\documentstyle[12pt,epsfig,axodraw]{article}

\newcommand{\be}{\begin{equation}}
\newcommand{\ee}{\end{equation}}
\def\aprle{\buildrel < \over {_{\sim}}}
\def\aprge{\buildrel > \over {_{\sim}}}
\def\um{1/2}
\def\sq{1/\sqrt{2}}
\begin{document}
\topmargin 0pt
\oddsidemargin=-0.4truecm
\evensidemargin=-0.4truecm
\renewcommand{\thefootnote}{\fnsymbol{footnote}}
\newpage
\setcounter{page}{1}
\begin{titlepage}     
\vspace*{-2.0cm}
\begin{flushright}
FISIST/1-2000/CFIF \\
hep-ph/0001264
\end{flushright}
\vspace*{0.5cm}
\begin{center}
{\Large \bf Neutrino physics} 
\footnote{Lectures given at Trieste Summer School in Particle Physics, 
June 7 -- July 9, 1999.} \\ 
\vspace{1.0cm}

{\large E. Kh. Akhmedov\footnote{On leave from National Research Centre 
Kurchatov Institute, Moscow 123182, Russia. 
E-mail: akhmedov@gtae2.ist.utl.pt} }\\
\vspace{0.05cm}
{\em Centro de F\'\i sica das Interac\c c\~oes Fundamentais (CFIF)} \\
{\em Departamento de F\'\i sica, Instituto Superior T\'ecnico }\\
{\em Av. Rovisco Pais, P-1049-001, Lisboa, Portugal }\\
\end{center}
\vglue 0.8truecm
\begin{abstract}
In the present lectures the following topics 
are considered: general properties of neutrinos, neutrino mass phenomenology 
(Dirac and Majorana masses), neutrino masses in the simplest extensions of 
the standard model (including the seesaw mechanism), neutrino oscillations in 
vacuum, neutrino oscillations in matter (the \mbox{MSW} effect) in 2- and
3-flavour schemes, implications of CP, T and CPT symmetries for neutrino 
oscillations, double beta decay, solar neutrino oscillations and the solar
neutrino problem, and atmospheric neutrinos. We also give a short overview of 
the results of the accelerator and reactor neutrino experiments and of future 
projects. Finally, we discuss how the available experimental data on neutrino 
masses and lepton mixing can be summarized in the phenomenologically
allowed forms of the neutrino mass matrix.


\end{abstract}
\end{titlepage}
\renewcommand{\thefootnote}{\arabic{footnote}}
\setcounter{footnote}{0}
\newpage
\section{Introduction 
\label{intro}}

In June of 1998 a very important event in neutrino physics occured -- the 
Super-Kamiokande collaboration reported a strong evidence for neutrino 
oscillations in their atmospheric neutrino data \cite{SK1}. This was not a 
thunderstorm in a clear sky: the evidence for oscillating neutrinos has been 
mounting during the last two decades in the solar neutrino experiments 
\cite{SNP}, and previous data on atmospheric neutrinos also gave indications 
in favour of neutrino oscillations \cite{Mann}. In addition, the LSND  
Collaboration has reported 
an evidence for $\bar{\nu}_\mu \leftrightarrow \bar{\nu}_e$ and $\nu_\mu 
\leftrightarrow \nu_e$ oscillations in their accelerator neutrino experiment 
\cite{LSND}. However, it was the Super-Kamiokande experiment that, for the
first time, not only showed with a high statistics the deficit of the 
detected neutrino flux compared to expectations, but also demonstrated that 
this deficit depends on the neutrino pathlength and energy in the way it
is expected to depend in the case of neutrino oscillations. Since neutrinos 
are massless in the standard model of electroweak interactions, the
evidence for neutrino oscillations (and therefore for neutrino mass) is 
the first strong evidence for physics beyond the standard model. 

Neutrino physics is a very active field now, both experimentally and 
theoretically. The Sudbury Neutrino Observatory (SNO) experiment was put 
into operation in 1999; it should add a very important information to 
our knowledge of solar neutrinos. The SNO experiment and the forthcoming 
Borexino experiment will complement the already existing data of Homestake, 
Gallex, SAGE, Kamiokande and Super-Kamiokande experiments and this will 
hopefully lead to the resolution of the long-standing solar neutrino problem. 
The first long baseline terrestrial neutrino experiment, K2K, has already
started taking data, and several more experiments -- KamLAND, MINOS,
MiniBooNE and CERN -- Gran Sasso experiments are either scheduled to start 
in a very near future or in advanced stage of planning. There are also very 
wide discussions of possible next generation long baseline experiments with 
muon storage rings \cite{nuoscind}. All these experiments are designed to 
probe a wide range of the neutrino mass squared differences 
and lepton mixing angles, and possibly CP-violating effects in neutrino
oscillations. They may also be able to test the fascinating possibility of
matter enhancement of neutrino oscillations -- the MSW effect \cite{MSW}. 

On the theoretical side, there have been many analyses of the available
experimental data and predictions for the forthcoming experiments made 
within various scenarios for neutrino properties, studies of neutrino-related 
processes in stars and supernovae, phenomenological studies of the allowed 
structures of neutrino mass matrices and developments of particle physics 
models capable of producing the requisite mass matrices. 

Why are we so much interested in neutrinos? Neutrinos play a very important 
role in various branches of subatomic physics as well as in astrophysics
and cosmology. The smallness of neutrino mass is very likely related to
existence of new, yet unexplored mass scales in particle
physics. These scales are so high that their direct experimental study 
may not ever be possible. Neutrinos can provide us with a very valuable, 
though indirect, information about these mass scales and the new physics 
related to these scales. In fact, they may even hold a clue to the general 
problem of the fermion mass generation. 

There are intricate relationships between neutrino physics and nuclear 
physics. The existence of neutrinos was first theoretically suggested to
explain apparent energy nonconservation and wrong spin-statistics relations 
in nuclear beta decay; inverse beta decay processes were used for the 
first experimental detections of reactor antineutrinos and solar neutrinos.  
The chiral nature of neutrinos, parity nonconservation and $V-A$ structure
of weak interactions were also established in nuclear physics experiments 
\cite{history}. Neutrino-nuclear reactions serve as a very clean probe
of nuclear and nucleon structure; on the other hand, nuclear physics
provides us with the knowledge of the reaction cross sections which is
very important for calculating the solar neutrino fluxes and neutrino 
detection rates. 

Neutrinos play a very important role in astrophysics and cosmology. They carry 
away up to $99\%$ of the energy released in the type II supernova explosions 
and therefore dominate the supernova energetics. Neutrino reactions play a 
crucial role in the mechanism of supernova explosions. 
Neutrinos are copiously produced in thermonuclear reactions which  
occur in the stellar interior and in particular in our sun. Solar neutrinos 
carry information about the core of the sun which is unaccessible to direct 
optical observations. The detection of solar neutrinos has confirmed the 
hypothesis that the sun is powered by thermonuclear reactions. 
At the same time, the sun and supernovae give us a possibility of studying
neutrino properties over extremely long baselines 
and probe the neutrino mass differences as small as $10^{-5}$ eV 
or even smaller, beyond the reach of the terrestrial neutrino experiments. 

The big bang nucleosynthesis depends sensitively on neutrino interactions and 
on the number of light neutrino species. Neutrinos of the mass $m_\nu \sim
1$ eV could constitute the so-called hot dark matter and may be important for 
galaxy formation. Neutrinos may also play an important role in baryogenesis: 
the observed excess of baryons over antibaryons in the universe may be related 
to decays of heavy Majorana neutrinos. 

In the present lecture notes a number of issues pertaining to neutrino 
physics are considered. These include general properties of neutrinos, 
neutrino mass phenomenology, neutrino masses in the simplest extensions of 
the standard model (including the seesaw mechanism of neutrino mass
generation), neutrino oscillations in vacuum,  
neutrino oscillations in matter (the MSW effect), implications of CP, T and 
CPT symmetries for neutrino oscillations, solar neutrino oscillations and 
the solar neutrino problem, and atmospheric neutrinos. We also give a 
short overview of the results of the accelerator and reactor neutrino 
experiments and of future projects. Finally, we discuss how the available 
experimental data on neutrino masses and lepton mixing can be summarized in 
the phenomenologically allowed forms of the neutrino mass matrix. Except for 
the simplest extensions of the standard models, we do not discuss the models 
of neutrino mass since they were covered in the lectures of K.S. Babu at 
this school \cite{Babu} (see also \cite{numass} for recent reviews).
Astrophysical implications are limited to the 
solar neutrino problem; other issues of neutrino astrophysics were discussed 
in the lectures of T.P. Walker \cite{Walker}, and additional information can
be found, e.g. in \cite{raffelt,sarkar,raffelt2}. We do not discuss neutrino 
electromagnetic properties and their implications because of the lack
of space. Some material on these issues can be found in \cite{BilPe,PM,Akh1}. 
Informative reviews and monographs on neutrino physics include (but are 
not limited to) \cite{BilPe,PM,GR,BGG,KP,BV}. 

\section{General properties of neutrinos
\label{genpro}}
What do we know about neutrinos? Neutrinos are electrically neutral particles 
of spin 1/2. There are at least three species (or flavours) of 
very light neutrinos, $\nu_e$, $\nu_\mu$ and $\nu_\tau$, which are
left handed,  
and their antiparticles $\bar{\nu}_e$, $\bar{\nu}_\mu$ and $\bar{\nu}_\tau$, 
which are right handed. 
Electron-type neutrinos and antineutrinos are produced in nuclear $\beta^\pm$ 
decay, 
\begin{eqnarray}
A(Z,N)\to A(Z+1, N-1)+e^- +\bar{\nu}_e\,,
\nonumber \\  
A(Z,N)\to A(Z-1, N+1)+e^+ +\nu_e\,,
\label{beta}
\end{eqnarray}
and in particular in the neutron decay process $n\to p+e^- +\bar{\nu}_e$. 
They are also produced in muon decays $\mu^\pm\to e^\pm +\bar{\nu}_\mu
(\nu_\mu)+\nu_e(\bar{\nu}_e)$, and as a subdominant mode, in pion decays 
$\pi^\pm \to e^\pm +\nu_e(\bar{\nu}_e)$ and in some other decays and 
reactions. 
The elementary processes responsible for the nuclear beta decays or pion
decays are actually the quark transitions $u\to d+e^+ +\nu_e$ and 
$d\to u+e^- +\bar{\nu}_e$. Muon neutrinos and antineutrinos are produced
in muon decays, pion decays $\pi^\pm \to \mu^\pm + \nu_\mu(\bar{\nu}_\mu)$ 
and some other processes. Neutrinos of the third type, tau,  
are expected to be produced in $\tau^\pm$ decays. They have not been 
experimentally detected yet
\footnote{The DONUT experiment at Fermilab is looking for $\nu_\tau$ and at 
present has several candidate events.}, 
but there is no doubt in their existence. 
Neutrinos of each flavour participate in reactions in which the charged lepton 
of the corresponding type are involved; these reactions are mediated by 
$W^\pm$ bosons. Thus, these so-called charged current reactions involve the 
processes $W^\pm \to l_a^\pm + \nu_a(\bar{\nu}_a)$ where $a=e,~\mu$ or
$\tau$, or related processes. Neutrinos can also participate in neutral 
current reactions mediated by $Z^0$ bosons; these are elastic or quasielastic 
scattering processes and decays $Z^0 \to \nu_a \bar{\nu}_a$. 

The latter process allowed us to count the number of light neutrino species 
that have the usual electroweak interactions.  Indeed, neutrinos from the 
$Z^0$ decays are not detected, and therefore the difference between the 
measured total width of the $Z^0$ boson and the sum of its partial widths
of decay into quarks and charged leptons, the so-called invisible width, 
$\Gamma_{inv}=\Gamma_{tot}-\Gamma_{vis}=498 \pm 4.2$ MeV, should be due to 
the decay into $\nu\bar{\nu}$ pairs. Taking into 
account that the partial width of $Z^0$ decay into one $\nu\bar{\nu}$ pair 
$\Gamma_{\nu\bar{\nu}}=166.9$ MeV one finds the number of the light active 
neutrino species \cite{PDG}:
\be
N_\nu=\frac{\Gamma_{inv}}{\Gamma_{\nu\bar{\nu}}}=2.994 \pm 0.012\,,
\label{nnu}
\ee
in a very good agreement with the existence of the three neutrino flavours. 
There are also indirect limits on the number of light ($m< 1$ MeV)
neutrino species (including possible electroweak singlet, i.e. ``sterile'' 
neutrinos) coming from big bang nucleosynthesis:
\be
N_\nu < 3.3\,,
\ee
though this limit is less reliable than the laboratory one (\ref{nnu}), 
and probably four neutrino species can still be tolerated \cite{sarkar}. 

\begin{figure}[htb] 
\hbox to \hsize{\hfil\epsfxsize=12cm\epsfbox{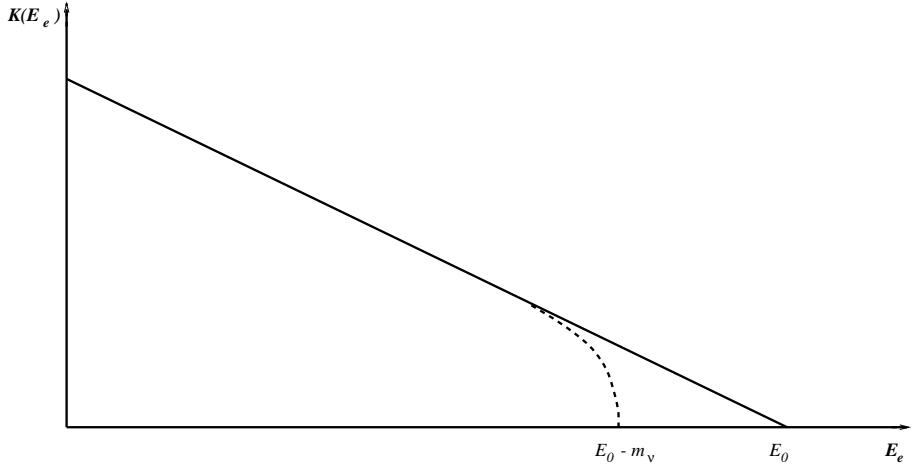}\hfil}
\caption{\small Kurie plots for $m_\nu=0$ (solid line) and 
$m_\nu \ne 0$ (dashed line) 
}
\label{kurie} 
\end{figure}

What do we know about the neutrino mass? Direct kinematic searches of
neutrino mass produced only the upper limits \cite{PDG}: 
\be
m_{\nu_1}< 2.5~\mbox{eV at 95\% c.l. (Troitsk)}\,; < 6~\mbox{eV at 95\% c.l. 
(Mainz)}\,; 
\label{m1}
\ee
\be
m_{\nu_2}< 170~\mbox{keV at 90\% c.l. (PSI}; ~\pi^+\to\mu^++\nu_\mu) \,; 
\label{m2}
\ee
\be
m_{\nu_3}< 18.2~\mbox{MeV at 95\% c.l. (ALEPH}; ~\tau^-\to 5\pi+\nu_\tau) 
\label{m3}
\ee
Here $\nu_1$, $\nu_2$ and $\nu_3$ are the primary mass components of 
$\nu_e$, $\nu_\mu$ and $\nu_\tau$, respectively. 
%
%
The limits in eqs. (\ref{m2}) and (\ref{m3}) come from
the comparison of the total energy release with the energy of decay products.  
The limits in eq. (\ref{m1}) are obtained from the tritium beta decay
experiments and are based on the analyses of the so-called Kurie plot. 
The electron spectrum in the allowed $\beta$ decay is 
\begin{eqnarray}
N_e(E_e) dE_e & \propto & F(Z,E_e)\sqrt{E_e^2-m_e^2}\,E_e(E_0-E_e)^2\, dE_e
\,,\quad (m_\nu=0)\,; 
\label{sp1} \\
N_e(E_e) dE_e & \propto & F(Z,E_e)\sqrt{E_e^2-m_e^2}\,E_e (E_0-E_e) 
\sqrt{(E_0-E_e)^2-m_\nu^2}\, dE_e\,, \quad (m_\nu\neq 0)\,. 
\label{sp2} 
\end{eqnarray}
Here $F(Z,E_e)$ is the well known function which takes into account
the interaction between the emitted electron and the nucleus in the final
state and $E_0$ is the energy release. Thus, the shape of the $\beta$
spectrum should depend on whether or not neutrinos have a mass. 
It follows from these equations that the plot of the Kurie function
$K(E_e)\equiv [N_e(E_e)/(F(Z,E_e)p_e E_e)]^{1/2}$ versus $E_e$ should be a
straight line when $m_\nu=0$ but should have a different shape close to the
endpoint of the spectrum when $m_\nu \ne 0$ (fig. \ref{kurie}). Analyses
of the Kurie plots for beta decay of tritium which has a very low energy 
release give the results of eq. (\ref{m1}). However, all the
experiments showed some excess of the number of electrons near the 
endpoint of the spectrum rather than a deficiency that is expected if 
$m_\nu\ne 0$. This excess is most likely due to unknown systematic 
effects. For this reason the conservative upper limit on the lightest 
neutrino mass recommended by the Particle Data Group \cite{PDG} is 
\be
m_1 < 15 ~\mbox{eV}
\label{m1a}
\ee
rather than the limits in (\ref{m1}).

Are neutrinos Dirac particles (like quarks or charged leptons) or Majorana
particles? In other words, can neutrinos be their own antiparticles? 
It is known experimentally that 
neutrinos emitted in $\beta^-$ decay (which we call electron antineutrinos) 
cannot be captured in reactions which are caused by electron neutrinos; 
for example, the reaction $\bar{\nu}_e + ^{37}$Cl$ \to ^{37}$Ar$+e^-$ does 
not occur, whereas the reaction $\nu_e + ^{37}$Cl$ \to ^{37}$Ar$+e^-$ 
does, and in fact was used in the first experiment to detect the solar 
neutrinos. Does this mean that neutrinos are different from antineutrinos,
i.e. cannot be Majorana particles? Not necessarily. The reason is that the
weak interactions that are responsible for neutrino interactions are chiral 
($V-A$): in the Cl-Ar reaction mentioned above, only neutrinos of left handed 
chirality can be detected. The particles which we call $\bar{\nu}_e$ are 
right handed, i.e. have the ``wrong chirality'', and so cannot participate
in the Cl-Ar reaction (this is called ``chiral prohibition''). If neutrinos 
are massive, the chirality (i.e. $1\pm \gamma_5$) is not a good quantum number,
and an electron antineutrino which was produced right handed can develop a 
small left handed component, so that the reaction $\bar{\nu}_e + ^{37}$Cl$
\to ^{37}$Ar$+e^-$ can become possible. However, the ``wrong chirality'' 
admixture is of the order $m_\nu/E$, i.e. extremely small due to the
smallness of the neutrino mass compared to typical beta decay energies.
Therefore the probability of the $\bar{\nu}_e + ^{37}$Cl$ \to ^{37}$Ar$+e^-$ 
process should be suppressed by the factor $(m_\nu/E)^2$, 
which can be the reason for its non-observation. 

\begin{figure}[htb] 
\hbox to \hsize{\hfil\epsfxsize=12cm\epsfbox{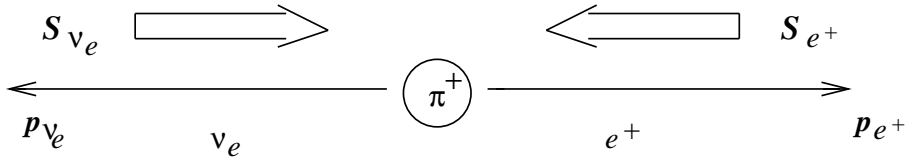}\hfil}
\caption{\small Charged pion decay }
\label{pidecay} 
\end{figure}

On of the most outstanding manifestations of the chiral prohibition rule  
are the decays of charged pions. They also show that one should carefully
discriminate between the chirality ($1\pm \gamma_5$) and helicity, which
is the projection of the spin of the particle on its momentum. 
In the limit of massless fermions they coincide, but for massive fermions 
they do not (see Problem 1 in the next section). Let us assume that 
neutrinos are massless and consider the decay $\pi^+ \to l^+ + \nu_l$ of 
pions at rest, where $l^+$ is $\mu^+$ or $e^+$ and $\nu_l$ is the 
corresponding neutrino. 
The $V-A$ structure of weak interactions requires the emitted $\nu_l$ to be 
of left handed chirality. For $m_\nu=0$ this also means that it has the left 
handed (or negative) helicity with its spin antiparallel to its momentum.
Conservation of total angular momentum then requires $l^+$ to be of negative 
helicity, too (see fig. \ref{pidecay}). However, the $l^+$ are antiparticles, 
and due to the $V-A$ structure of weak interactions they must be produced in 
the states of right handed chirality. Therefore the amplitude of the process 
must be proportional to the admixture of right handed chirality in left handed 
(negative) helicity of the charged lepton, i.e. to its mass: $A(\pi^+ \to
l^+ + \nu_l)\propto m_l$. We therefore expect
\be
R_\pi\equiv \frac{\Gamma(\pi^+ \to e^+ +\nu_e)}{\Gamma(\pi^+ \to \mu^+
+\nu_\mu)}
=\left(\frac{m_e}{m_\mu}\right)^2 \left(\frac{m_\pi^2-m_e}
{m_\pi^2-m_\mu^2}\right)^2 = 1.28 \times 10^{-4}\,.
\label{Rpi}
\ee
where we have taken into account the difference of the phase space factors
for the decays into $\mu^+$ and $e^+$. Experimentally we have 
\be
R_\pi =(1.230 \pm 0.004)\times 10^{-4}\,,
\ee
in a very good agreement with the theoretical prediction; the 4\% 
difference is actually due to the fact that (\ref{Rpi}) does not
include radiation corrections.  
 
Neutrinos are elusive particles, and neutrino experiments 
are very difficult. This is because the interactions of neutrinos are 
mediated by heavy $W^\pm$ and $Z^0$ bosons and so at low energies
they are very weak.  The mean free path of a 1 MeV neutrino in lead is 
about 1 light year! Therefore neutrino detection requires very large 
detectors and/or very intense neutrino beams. 

What are the main sources of neutrinos? One of the strongest neutrino sources 
is our sun. It emits about $2\times 10^{38}$ electron neutrinos per second, 
leading to the neutrino flux at the surface of the earth of $\sim 6\times 
10^{10}$ $cm^{-2} s^{-1}$ in the energy range $E \le 0.42$ MeV and $\sim 
5\times 10^{6}$ $cm^{-2} s^{-1}$ in the energy range 0.8 MeV $\aprle E \le 15$ 
MeV. Nuclear power plants are a powerful source of electron antineutrinos.
A 3 GW plant emits about $7.7\times 10^{20}$ $\bar{\nu}_e$ of the energy 
$\sim$ a few MeV per second and creates a flux of $\sim 6\times 10^{11}$ 
$cm^{-2} s^{-1}$ at 100 $m$. At remote locations the averaged $\bar{\nu}_e$ 
number density created by all nuclear power plants on the earth is about  
$10^{6} - 10^{7}$ $cm^{-3}$. The relic neutrinos, i.e. the neutrinos left
over from the early epochs of the evolution of the universe, have a number 
density of about 110 $cm^{-3}$ for each neutrino species and a black-body 
spectrum with the average energy of about $5\times 10^{-4}$ eV. Natural 
radioactivity of the earth results in the flux of $\sim 6\times 10^{6}$ 
$cm^{-2} s^{-1}$ (or number density of $\sim 2\times 10^{-4}$ $cm^{-3}$) of 
neutrinos of the energy $E\aprle 1$ MeV. The flux of atmospheric neutrinos at 
the earth's surface is $\sim 10^{-1}$ $cm^{-2} s^{-1}$. An important 
tool for studying neutrino properties are particle accelerators which produce 
neutrino beams of the energy ranging typically between 30 MeV and 30 GeV. 
Finally, a very rare and short-lived but yet a very important source of 
neutrinos are type II supernovae. They emit $\sim 6\times 10^{58}$ neutrinos 
and antineutrinos of all flavours over the time interval of about 10 $s$
and with the typical energies $E\aprle 30$ MeV. Observation of the neutrino 
burst from the supernova 1987A allowed us to obtain important constraints 
on neutrino properties.

\section{Neutrino mass phenomenology: Weyl, Dirac and Majorana neutrinos 
\label{WDM}}
A massless neutrino (or any other fermion) is described by a two-component
Weyl spinor field. It has a certain chirality, $L$ or $R$. The chirality 
projector operators $P_{L,R}$ are defined as 
\be
P_L = \frac{1-\gamma_5}{2}\,, \quad\quad
P_R = \frac{1+\gamma_5}{2}\,, 
\ee
and have the following properties:
\be
P_L^2=P_L\,,\quad P_R^2=P_R\,, \quad P_L P_R=P_R P_L=0\,, \quad P_L+P_R=1\,.
\label{rel1}
\ee
The terms ``left handed'' and ``right handed'' originate from the fact that 
for relativistic particles chirality almost coincides with helicity
defined as the projection of the spin of the particle on its momentum. 
The corresponding projection operators are
\be
P_{\pm} = \frac{1}{2}\left(1 \mp\frac{\mbox{\boldmath $\sigma$}{\bf p}}
{|\bf p|}\right)\,.
\ee
They satisfy relations similar to (\ref{rel1}). 
For a free fermion, helicity is conserved but chirality in general is not; 
it is only conserved in the limit $m=0$ when it coincides with helicity.
However, for relativistic particles chirality is nearly conserved and 
the description in terms of chiral states is useful.

\noindent
{\it Problem 1. Using plane wave solutions of the Dirac equation show that 
for positive energy solutions, in the limit $m=0$, eigenstates of L (R) 
chirality coincide with the eigenstates of negative (positive) helicity.}

For our discussion we will need the particle - antiparticle conjugation 
operator $\hat{C}$. Its action on a fermion field $\psi$ is defined
through 
\be
\hat{C}: \; \psi \to \psi^c = C \bar{\psi}^T\,, \quad\quad 
C=i\gamma_2 \gamma_0\,.
\label{C}
\ee
The matrix $C$ has the properties
\be
C^\dagger=C^T=C^{-1}=-C\,, \quad\quad C\gamma_\mu C^{-1}=-\gamma_\mu^T\,.
\label{C1}
\ee
Some useful relations based on these properties are 
\be
(\psi^c)^c=\psi\,,\quad 
\overline{\psi^c}=\psi^T C\,,\quad
\overline{\psi_1} \psi_2^c=\overline{\psi_2^c} \psi_1\,,\quad
\overline{\psi_1} A \psi_2 =\overline{\psi_2^c} (C A^T C^{-1})\psi_1^c\,,
\label{C2}
\ee
where $\psi$, $\psi_1$, $\psi_2$ are 4-component fermion fields and 
$A$ is an arbitrary $4\times 4$ matrix. Using the commutation properties
of the Dirac $\gamma$ matrices it is easy to see that, acting on a chiral 
field, $\hat{C}$ flips its chirality: 
\be
\hat{C}: \; \psi_L\to(\psi_L)^c=(\psi^c)_R\,, \quad\quad
\psi_R\to (\psi_R)^c=(\psi^c)_L\,,
\label{C3}
\ee
i.e. the antiparticle of a left handed fermion is right handed. 

\noindent
{\it Problem 2. Prove eqs. (\ref{C2}) and (\ref{C3}).}

The particle - antiparticle conjugation operation $\hat{C}$ must not be 
confused with the charge conjugation operation 
C which, by
definition, flips all the charge-like quantum numbers of a field (electric 
charge, baryon number, lepton number, etc.) but leaves all the other 
quantum numbers (e.g., chirality) intact. In particular, charge conjugation 
would take a left handed neutrino into a left handed antineutrino that does 
not exist, which is a consequence of the C-noninvariance 
of weak interactions. At the same time, particle - antiparticle conjugation 
converts a left handed neutrino into right handed antineutrino which does 
exist and is the antiparticle of the left handed neutrino. 

We are now ready to discuss the Dirac and Majorana mass terms. 
For a massive fermion, the mass term in the Lagrangian has the form
\be
-{\cal L}_m=m\bar{\psi}\psi=\overline{(\psi_L+\psi_R)}(\psi_L+\psi_R)
=\overline{\psi_L}\psi_R+\overline{\psi_R}\psi_L
\,,
\label{mass}
\ee
Thus, the mass terms couple the left handed and right handed components of 
the fermion field, and therefore a massive field must have both components:
\be
\psi=\psi_L+\psi_R\,.
\label{psi}
\ee
Now, there are essentially two possibilities. First, the right handed
component of a massive field can be completely independent of the left 
handed one; in this case we have a Dirac field. 
Second, the right handed field can be 
just a $\hat{C}$ - conjugate of the left handed one: $\psi_R=
(\psi_L)^c=(\psi^c)_R$, or 
\be
\psi=\psi_L+\eta (\psi^c)_R=\psi_L+\eta (\psi_L)^c\,.
\label{psi1}
\ee
where we have included the phase factor $\eta=e^{i\varphi}$ with an 
arbitrary phase $\varphi$. 
In this case we have a Majorana field; one can construct
it with just one Weyl field. From (\ref{psi1}) it immediately follows that
the $\hat{C}$ - conjugate field coincides with itself up to a phase factor: 
\be
\psi^c= \eta^* \psi\,.
\label{C4}
\ee
This means that particles described by Majorana fields are genuinely
neutral, i.e. coincide with their antiparticles. Thus, Majorana particles 
are fermionic analogs of photons and $\pi^0$ mesons. To construct a massive 
Dirac field, one needs two independent 2-component Weyl fields, $\psi_L$ and 
$\psi_R$; together with their $\hat{C}$-conjugates, $(\psi_L)^c=\psi^c_{\,R}$ 
and $(\psi_R)^c=\psi^c_{\,L}$, this gives four degrees of freedom. In
contrast with this, a Majorana fermion has only two degrees of freedom, 
$\psi_L$ and $(\psi_L)^c=\psi^c_{\,R}$. 

For massive fermions, particle - antiparticle conjugation $\hat{C}$ 
and charge conjugation C coincide. For Dirac fermions we have 
\begin{eqnarray}
\hat{C}: & & \psi=\psi_L+\psi_R\to(\psi)^c=(\psi_L)^c+(\psi_R)^c=
(\psi^c)_R + (\psi^c)_L\,, \nonumber \\
{\rm C}: & &
\psi=\psi_L+\psi_R\to\tilde{\psi}=\tilde{\psi}_L+\tilde{\psi}_R 
\equiv (\psi^c)_L + (\psi^c)_R\,,
\label{C5}
\end{eqnarray}
where tilde means charge conjugation. For Majorana neutrinos, both 
particle - antiparticle conjugation and charge conjugation leave the field 
unchanged because it does not have any charges
\footnote{There may be, however, some differences in the phase factors, 
see \cite{Kayser}.}. 
As we have already pointed 
out, particle - antiparticle and charge conjugations are not equivalent
when acting on chiral fields. 

For $n$ fermion species (flavours), the Majorana mass term can be written as 
\be
-{\cal L}_m=\frac{1}{2}\left[\overline{(\psi_L)^c}M \psi_L+
\overline{\psi_L}M^\dagger (\psi_L)^c\right]=\frac{1}{2}\left[\psi_L^T C M 
\psi_L+ \overline{\psi_L}C M^\dagger \overline{\psi_L}^T\right ]=
\frac{1}{2}\left[\psi_L^T C M \psi_L+ h.c.\right] \,,
\label{maj1}
\ee
where $\psi=(\psi_1,...,\psi_n)^T$ is a vector in the flavour space and 
$M$ is a $n\times n$ matrix. Using the anticommutation property of the 
fermion fields and eq. (\ref{C1}), it is easy to show that the matrix 
$M$ must be symmetric: $M_{ij}=M_{ji}$. It is interesting to note that 
in classical field theory it would be antisymmetric, $M_{ij}=-M_{ji}$. 
This, in particular, means that in the case of just one neutrino species 
the Majorana mass vanishes identically in classical field theory. It is,
therefore, an essentially quantum quantity. 
Kinematically, Dirac and Majorana masses are indistinguishable: they lead 
to the same relation between energy, momentum and mass of the particle,
$E=\sqrt{{\bf p}^2+m^2}$. 

{}From eq. (\ref{maj1}) a very important difference between the Dirac and
Majorana mass terms follows. The Dirac mass terms $\bar{\psi}\psi$ are 
invariant with respect to the $U(1)$ transformations 
\be
\psi \to e^{i\alpha}\psi\,,\quad\quad
\bar{\psi} \to \bar{\psi}e^{-i\alpha}\,,
\ee
i.e. they conserve the corresponding charges (electric charge, lepton or
baryon number, etc.). It follows from (\ref{maj1}) that the Majorana mass 
terms break all the charges  
that the field $\psi$ has by two units. This, in particular, 
means that, since the electric charge is exactly conserved, no charged 
particle can have Majorana mass. Therefore, out of all known fermions, 
only neutrinos can be Majorana particles. If neutrinos have Majorana
masses, the total lepton number is not conserved, while it is conserved if 
neutrinos are Dirac particles. 

\section{Neutrino masses in the standard model and slightly beyond 
\label{numass}}
In the standard model of electroweak interactions, all quarks and charged 
fermions get their masses through the Yukawa couplings with the Higgs
field $H=(H^+, \,H^0)^T$:
\be
-{\cal L}_Y=h^u_{ij} \overline{Q}_{Li} u_{Rj}\tilde{H} +h^d_{ij} 
\overline{Q}_{Li} d_{Rj} H + f^e_{ij} \overline{l}_{Li} e_{Rj} H + h.c.
\label{yuk}
\ee
where $Q_{Li}$ and $l_{Li}$ are left handed quark and lepton doublets, 
$u_{Ri}$, $d_{Ri}$ and $e_{Ri}$ are $SU(2)_L$ - singlet right-handed fields 
of up-type quarks, down-type quarks and charged leptons respectively, 
$\tilde{H}\equiv i\tau_2 H^*$, $\tau_a$ being the isospin Pauli matrices, 
and $i,\,j$ are the generation indices. 
After the electroweak symmetry is broken by a nonzero vacuum expectation
value (VEV) $v$ of the Higgs field, the Yukawa terms in (\ref{yuk}) yield
the mass matrices of quarks and charged leptons 
\be
(m_u)_{ij}=h^u_{ij}v\,,\quad 
(m_d)_{ij}=h^d_{ij}v\,,\quad (m_e)_{ij}=f^e_{ij}v\,. 
\label{mqe}
\ee

Neutrinos are massless in the minimal standard model. They cannot have Dirac 
masses because there are no $SU(2)_L$ - singlet (``sterile'') right-handed 
neutrinos $\nu_R$ in the standard model. Can neutrinos have Majorana 
masses? The answer is no. The reason for this is rather subtle, and it is 
worth discussing it in some detail.

The Majorana mass term should be of the form $\nu_L^T C \nu_L$ [see 
eq. (\ref{maj1})]. Since $\nu_L$ has the weak isospin projection $I_3=1/2$, 
the Majorana mass term has $I_3=1$, i.e. it is a component of the
isotriplet operator $l^T C\, i\tau_2 \mbox{\boldmath $\tau$} \,l \sim
(3,\, -2)$.
Therefore, in order to introduce the Majorana mass in a gauge invariant way 
so as not to spoil the renormalizability of the standard model, one would 
need an isotriplet Higgs field $\mbox{\boldmath $\Delta$}\sim (3, \,2)$:  
\be
-{\cal L}_{Yuk}^\Delta = f_\Delta (l^T C\, i\tau_2 \mbox{\boldmath $\tau$}
\,l)\mbox{\boldmath $\Delta$} + h.c.
\label{maj2}
\ee 
\noindent
{\it Problem 3. Show that for any SU(2) spinor $\chi$, $\chi^T i\tau_2$ 
has the same transformation properties as $\chi^\dagger$, and in particular 
that both $\chi_1^\dagger \chi_2$ and $\chi_1^T i\tau_2 \chi_2$ are invariants 
while $\chi_1^\dagger \mbox{\boldmath $\tau$} \chi_2$ and $\chi_1^T i\tau_2 
\mbox{\boldmath $\tau$} \chi_2$ transform as vectors. Hint: use the property 
$\tau_2 \mbox{\boldmath $\tau$}^T \tau_2=-\mbox{\boldmath $\tau$}$ of the 
Pauli matrices. }

\noindent
When the electrically neutral component of $\mbox{\boldmath $\Delta$}$ 
develops a VEV, Majorana neutrino mass is  
generated. However, such a Higgs does not exist in the standard model. 
Can one construct a composite triplet Higgs operator out of two Higgs 
doublets? Yes, the operator $H^T i\tau_2 \mbox{\boldmath $\tau$} 
\,H \sim (3,\, 2)$, i.e. has the correct quantum numbers. However, the term 
$(l^T C\, i\tau_2 \mbox{\boldmath $\tau$} \,l)(H^T i\tau_2 \mbox{\boldmath 
$\tau$} \,H )$ has the dimension $d=5$, i.e. it cannot enter in the Lagrangian 
of a renormalizable model at the fundamental level. Can the operator 
\be
\frac{f}{M}(l^T C\, i\tau_2 \mbox{\boldmath $\tau$} \,l)(H^T i\tau_2 
\mbox{\boldmath $\tau$} \,H )
\label{maj3}
\ee
be generated as an effective operator at some higher loop level? If so, it
would produce the Majorana mass term for neutrinos $m_L\simeq f v^2/M$ 
when the Higgs field develops a nonvanishing VEV, with $M$ being the
characteristic mass scale of the particles in
the loop. In principle, this is possible. However, in the standard model this 
does not happen because the total lepton number $L$ (more precisely, the 
difference of the baryon and lepton numbers $B-L$) is exactly conserved. 

Let us discuss this point in more detail. In the standard model, lepton
and baryon numbers are conserved at the perturbative level due to
accidental symmetries of the Lagrangian. 
These symmetries are called accidental because they are not imposed on the 
Lagrangian but are just the consequence of the particle content of the 
standard model, its gauge invariance, renormalizability and Lorentz 
invariance. Indeed, consider, for example, the lepton number violating 
operators $l l$ where for simplicity the isospin structure has been suppressed.
Such operators have isotriplet and isosinglet components: $(2,-1)\otimes(2,-1)
\sim (3,-2)\oplus (1,-2)$, both of them with hypercharge $-2$. Thus, we need 
a hypercharge 2 field (or product of the fields) to obtain a gauge 
invariant expression. One can, e.g., try $\bar{e}_R \,l l$ which is
the gauge invariant $\Delta L=1$ operator. However it is a product of three
fermion fields and therefore is neither Lorentz invariant nor renormalizable. 
The lepton number violating ($\Delta L=2$) operator in eq. (\ref{maj3}) is
both gauge and Lorentz invariant, but, as we have already mentioned, it is a  
$d=5$ nonrenormalizable operator. One can construct, e.g., a baryon number 
and lepton number violating ($\Delta B=\Delta L=1$) gauge and Lorentz
invariant operator 
\be
\varepsilon_{\alpha\beta\gamma}(u_R^{\alpha \,T} C d_R^{\beta})
(u_R^{\gamma\, T} C e_R)\,,
\ee
where $\alpha,\beta,\gamma$ are the $SU(3)_c$ indices; however it is a $d=6$ 
nonrenormalizable operator. Similarly, one can consider all other possible 
$L$- and $B$-violating operators that can be constructed with the 
standard-model fields and make sure that all of them either violate gauge
or Lorentz invariance, or are nonrenormalizable. Therefore $L$ and $B$ are 
automatic symmetries of the standard model at the perturbative level.  

Nonperturbatively, both $B$ and $L$ are violated by the electroweak sphalerons 
since these symmetries are anomalous. This violation is tiny and of no 
practical consequences at low energy and temperatures but may be very 
important, e.g., in the early universe \cite{RS}. The triangle anomalies 
of the baryon and lepton number currents are equal to each other in the 
standard model, so that $B-L$ is anomaly free and therefore conserved exactly. 
Since the operators of the type (\ref{maj3}) which could produce Majorana
mass terms for neutrinos break not only $L$ but also $B-L$ by two units, they 
cannot be induced in the standard model even nonperturbatively. Thus, 
neutrinos are exactly massless in the minimal standard model. 

It is interesting to note that if the lepton number (or $B-L$) conservation 
was not exact, neutrino Majorana mass term could be generated even in the 
standard model framework. It has been speculated that quantum gravity
should respect only gauge symmetries while violating all the global
symmetries, such as baryon or lepton numbers. If so, it could induce the 
operators of the type (\ref{maj3}) in the standard model with the mass 
scale $M$ of the order of the Planck mass $M_{\rm Pl}\simeq 1.2\times 
10^{19}$ GeV \cite{Planckscale}. This would give neutrino Majorana masses 
$m\sim v^2/M_{\rm Pl} \sim 10^{-5}$ eV, which is exactly of the right
order of magnitude to account for the solar neutrino deficiency through
the vacuum neutrino oscillations. 

How can we extend the standard model so as to accommodate nonzero neutrino 
mass? One can either extend the Higgs content of the model, or the fermion 
content, or both, or enlarge the gauge group (which also requires extended 
particle content). One possibility would be to introduce the isotriplet 
Higgs field $\mbox{\boldmath $\Delta$}$ discussed above. However, such a 
theory would not be very attractive. The VEV $u$ of the triplet Higgs would 
modify the masses of the $W$ and $Z$ bosons at the tree level, and in order 
not to run into a contradiction with the experimental value of the parameter 
$\rho=m_W^2/\cos^2\theta_W m_Z^2$ it should be $\aprle$ a few GeV. The Yukawa 
coupling (\ref{maj2}) conserves the lepton number provided that  one assigns 
the lepton charge $-2$ to the isotriplet Higgs.  If the Higgs potential is 
also $L$-conserving, the VEV $u$ would break the lepton number spontaneously
and therefore would give rise to the corresponding massless neutral Goldstone 
boson $J$ which is called ``triplet Majoron'' \cite{tripl} (in fact, due to 
the electroweak symmetry breaking, the Majoron is a linear combination of 
the isodublet and isotriplet fields with the dominant triplet contribution). 
In this case even more stringent constraint on the VEV of the triplet 
Higgs results from astrophysics, $u < 5$ keV. Such small a value of 
$u$ means that the massive counterpart of the Majoron, $\rho$, should be 
very light and can be produced in the decays $Z^0\to J+\rho$. Such a decay 
would add to the invisible width of the $Z^0$ boson a quantity which is 
equal to the width of decay into two extra neutrino species, in sharp 
contradiction with eq. (\ref{nnu}). Therefore the triplet Majoron model 
is ruled out now. If one allows the $L$-breaking terms in the Higgs
potential (e.g., $\mu (H^T i\tau_2 \mbox{\boldmath $\tau$} \,H )
\mbox{\boldmath $\Delta$}^\dagger+ h.c.)$, lepton number would be broken
explicitly and no Majoron appears. However, in the standard model, it is
highly unnatural to have the triplet Higgs VEV $u \ll v$ -- there is no
symmetry or dynamical mechanism to protect it, and its expected value is 
$u\sim v$, which would lead to a grossly wrong value of the $\rho$ parameter. 

It is interesting to note that isotriplet Higgs fields may be quite natural 
in the framework of extended gauge theories. For example, the Higgs content 
of the most popular version of the left-right symmetric theories \cite{LR} 
based on the gauge group $SU(2)_L\times SU(2)_R\times U(1)_{B-L}$ includes a 
bi-doublet $\Phi$, which gives the usual Dirac masses $m_D$ of fermions, and 
the fields $\mbox{\boldmath $\Delta$}_L$ and $\mbox{\boldmath $\Delta$}_R$
which are triplets of $SU(2)_L$ and $SU(2)_R$ respectively. Their VEVs 
$u_L$ and $u_R$ produce the Majorana mass terms for the left-handed and 
right-handed neutrinos, $\nu_L$ and $\nu_R$. The VEV $u_R$ determines the 
mass scale at which parity is spontaneously broken, and is expected to be 
very large, thereby explaining non-observation of the right-handed ($V+A$) 
interactions at low energies. Right handed neutrinos are also very heavy 
in this model, while the masses of $\nu_L$ are naturally very small due to 
the seesaw mechanism \cite{seesaw}: $m_L\sim m_D^2/M_R$ (we shall discuss
this mechanism in detail in sec. \ref{ssmech}). The minimization of 
the Higgs potential leads to $u_L \sim v^2/u_R$, i.e. the smallness of
$u_L$ is a natural consequence of the largeness of $u_R$, and a kind of the 
seesaw mechanism for the Higgs VEVs operates. 


Another possible extension of the standard model is just to add three 
$SU(2)_L$ singlet neutrinos $\nu_R$, one per fermion generation. In a sense, 
this is natural since it restores the quark-lepton symmetry: in the 
minimal standard model all left handed (isodoublet) quarks and charged 
leptons have right handed (isosinglet) counterparts, while neutrinos do not  
\footnote{This argument, however, has to be taken with some caution: the
existence of the right handed counterparts of charged left handed fermions
is dictated by the requirement of the electroweak anomaly cancellation 
whereas this argument does not apply to right handed neutrinos, 
see 
below.}.  

Right handed neutrinos can have the usual Yukawa couplings to lepton
doublets. Since they have zero isospin and are electrically neutral, their 
hypercharge is also zero, i.e. they 
are electroweak singlets. Therefore they can have ``bare'' Majorana
mass terms which are invariant with respect to $SU(2)_L\times U(1)_Y$. 
The part of the Lagrangian that is relevant for the lepton mass generation
is therefore  
\be
-{\cal L}_{Y+M}=f^e_{ij} \overline{l}_{Li} e_{Rj} H + f^\nu_{ij}\, 
\overline{l}_{Li} \nu_{Rj} \,\tilde{H} +\frac{1}{2} M_{ij}\, \nu_{Ri}^T 
C \nu_{Rj} + h.c.
\label{yuk1}
\ee
The second term here yields the Dirac neutrino mass matrix $(m_D)_{ij}= 
f^\nu_{ij} v$. If one assigns the lepton charge $L=+1$ to all the $\nu_R$ 
fields, the first two terms on the r.h.s. of eq. (\ref{yuk1}) conserve 
the total lepton number $L$ [although not the individual lepton numbers, 
or lepton ``flavours'' $L_i$ ($i=e, \mu,\tau$)]. However, the third  term 
breaks $L$ by two units. Alternatively, if one assigns zero lepton charge 
to $\nu_R$, the third term in (\ref{yuk1}) conserves $L$, but then the
second term breaks it. Therefore the introduction of $\nu_R$ in the 
standard model leads to a qualitatively new situation: lepton number is no 
longer an automatic symmetry of the Lagrangian following from gauge and 
Lorentz invariance and renormalizability. 

Since $\nu_R$ are electroweak singlets, they do not contribute to the 
electroweak anomaly, and their number is not fixed by the requirement of the 
anomaly cancellation. In particular, their number need not coincide with the 
number of generations $N_g$: it may as well be smaller or even larger than 
$N_g$. There are, however, astrophysical and cosmological constraints on 
their number that depend on their masses and mixing parameters 
\cite{Walker,raffelt,sarkar}. 

\noindent
{\it Problem 4. Assuming that there are two right-handed neutrinos per 
fermion generation and that the Majorana mass terms are absent, write down 
the most general neutrino mass term and analyze it. What would be the
neutrino mass spectrum in this case? What would happen if there were instead 
two right-handed charged leptons per fermion generation? }

\section{General Dirac + Majorana case 
\label{D+M}}
We shall consider now the most general neutrino mass term for the case of
$n$ species of left handed and right handed neutrinos. It includes not only 
the Dirac mass $m_D$ and Majorana mass $m_R$ for $\nu_R$ but also the 
Majorana mass $m_L$ for left handed neutrinos. 
%
%
The neutrino mass term can be written as 
\be
-{\cal L}_m= \frac{1}{2} \nu_{L}^T\, C\, m_L\, \nu_{L}+{\overline \nu_{L}}
\,m_D^*\, \nu_{R}+ \frac{1}{2}\nu_{R}^T\, C \,m_R^*\,\nu_{R} + h.c. =
\frac{1}{2}\, n_L^T \,C {\cal M}\, n_L +h.c. \,. 
\label{Lm1}
\ee
Here $n_L=(\nu_L,\,(\nu_R)^c)=(\nu_L,\,\nu^c_{\,L})$ is the vector of $2n$ 
left handed fields (which we have written here as a line rather than column 
to save space),  $m_L$ and $m_R$ are complex symmetric $n\times n$
matrices, $m_D$ is a complex $n\times n$ matrix, and we have introduced the 
Dirac and right handed Majorana mass matrices through their complex 
conjugates for simplicity of the further notation. The matrix ${\cal M}$ has 
the form 
\be
{\cal M}=\left(\begin{array}{cc}
m_L      & m_D     \\
m_D^T      & m_R    
\end{array} \right)\,,
\label{mL1}
\ee
and in deriving eq. (\ref{Lm1}) we have used the relations 
\be
(\nu_R^T \,C m^*\, \nu_R)^\dagger=(\nu^c)_L^T \,C m\,(\nu^c)_L\,,\quad
\overline{\nu_R}\, m\, \nu_L=(\nu^c)_L^T \,C\,m\, \nu_L=\nu_L^T \,C\,m^T\,
(\nu^c)_L\,.
\label{rel2}
\ee
\noindent
{\it Problem 5. Prove these relations using properties (\ref{C1}) and
(\ref{C2}) of the matrix $C$.} 

It is instructive to consider first the simple one-generation case in
which $m_L$, $m_R$ and $m_D$ are just numbers, and ${\cal M}$ is a 
$2\times 2$ matrix. 
For simplicity we shall 
assume all the mass parameters to be real. 
The matrix ${\cal M}$ can be
diagonalized by the transformation 
$U^T{\cal M} U={\cal M}_d$ where $U$ is an orthogonal matrix and 
${\cal M}_d=diag(m_1,\, m_2)$. 
%
%
%
We introduce the fields $\chi_L$ through $n_L=U \chi_L$, or 
\be
n_L=\left(\begin{array}{c}
\nu_L \\
\nu^c_{\,L}   
\end{array} \right)=
\left(\begin{array}{cc}
\cos\theta      & \sin\theta      \\
-\sin\theta     & \cos\theta    
\end{array} \right)
\left(\begin{array}{c}
\chi_{1L} \\
\chi_{2L}   
\end{array} \right)\,. 
\label{mL2}
\ee
Here $\chi_{1L}$ and $\chi_{2L}$ are the left handed components of
neutrino mass eigenstates.  
%
The mixing angle $\theta$ is given by  
\be
\tan 2\theta=\frac{2 m_D}{m_R-m_L}\,,
\label{theta}
\ee
and the neutrino mass eigenvalues are 
\be
m_{1,2}=\frac{m_R+m_L}{2}\mp \sqrt{\left(\frac{m_R-m_L}{2}\right)^2
+m_D^2}\,.
\label{m1m2}
\ee
They are real but can be of either sign. The mass term can now be rewritten 
as 
\begin{eqnarray}
-{\cal L}_m &=& \frac{1}{2}\, n_L^T \,C {\cal M}\, n_L+h.c.= \frac{1}{2}\,
\chi_L^T \,C {\cal M}_d\, \chi_L +h.c. \nonumber \\ 
&=& \frac{1}{2}(m_1\, \chi_{1L}^T \,C\,\chi_{1L}+m_2\, 
\chi_{2L}^T \,C\,\chi_{2L})+h.c. =\frac{1}{2} (\, |m_1|\,
\overline{\chi}_1\chi_{1}+|m_2|\, \overline{\chi}_2\chi_{2}\,)\,. 
\label{mL3}
\end{eqnarray}
Here we have defined 
\be
\chi_1=\chi_{1L}+\eta_1(\chi_{1L})^c\,,\quad\quad
\chi_2=\chi_{2L}+\eta_2(\chi_{2L})^c\,. 
\label{chi}
\ee
with $\eta_{1,2}=1$ or $-1$ for $m_{1,2}>0$ or $<0$ respectively. It follows 
immediately from eq.~(\ref{chi}) that the mass eigenstates $\chi_1$ and
$\chi_2$ are Majorana neutrinos. The relative signs of the mass
eigenvalues ($\eta_1$ and $\eta_2$) determine the relative CP parities of 
$\chi_1$ and $\chi_2$; physical masses $|m_1|$ and $|m_2|$ are positive, as 
they should. 
The fact that the mass eigenstates in the case of the most general Dirac + 
Majorana mass term are Majorana particles should not be surprising. We
have four neutrino degrees of freedom, $\nu_L$, $\nu_R$ and their $\hat{C}$- 
conjugates $(\nu_L)^c=\nu^c_{\,R}$ and $(\nu_R)^c=\nu^c_{\,L}$. The mass 
matrix has two different eigenvalues, so there are two massive neutrino 
fields. Each of them corresponds to two degrees of freedom, therefore they
have to be Majorana particles. Analogously, in the case of $n$ generations, 
the most general mass term (\ref{Lm1}) leads to $2n$ massive Majorana 
neutrinos. 

It is instructive to consider some limiting cases. In the limit of no
Majorana masses ($m_L=m_R=0$), pure Dirac case has to be recovered. 
In this case the mass matrix (\ref{mL1}) has the form 
\be
{\cal M}=\left(\begin{array}{cc}
0      & m     \\
m      & 0    
\end{array} \right)\,.
\label{mL4}
\ee
This matrix corresponds to a conserved lepton number $L_{\nu_L}-
L_{\nu^c_{\,L}}=L_{\nu_L}+L_{\nu_{R}}$ which can be identified with the 
total lepton number $L$. Thus, the lepton number is conserved in this 
limiting case, as expected. Let us now check that the usual Dirac mass
term is recovered. 

The matrix ${\cal M}$ in (\ref{mL4}) is diagonalized by the rotation 
(\ref{mL2}) with $\theta=45^\circ$, and its eigenvalues are $-m$ and $m$. 
This means that we have two Majorana neutrinos that have the same mass, 
opposite CP parities and are maximally mixed. We now want to demonstrate
that this is equivalent to having one Dirac neutrino of mass $m$. We have 
$\eta_2=-\eta_1=1$; from eqs.~(\ref{mL2}) and (\ref{chi}) it then follows 
$\chi_1+\chi_2=\sqrt{2}(\nu_L+\nu_R)$, $\chi_1-\chi_2=-\sqrt{2}(\nu^c_{\,L}+
\nu^c_{\,R})=-(\chi_1+\chi_2)^c$. This gives 
\be
\frac{1}{2}\,m\,(\overline{\chi}_1\chi_1 +\overline{\chi}_2\chi_2)=
\frac{1}{4}\,m\,[\overline{(\chi_1+\chi_2)}(\chi_1+\chi_2) +
[\overline{(\chi_1-\chi_2)}(\chi_1-\chi_2)]=m\,
\bar{\nu}_D \nu_D\,,
\label{check1}
\ee
where
\be
\nu_D\equiv \nu_L +\nu_R\,.
\ee
The counting of the degrees of freedom also shows that we must have a Dirac 
neutrino in this case -- there are four degrees of freedom and just one 
physical mass. 
Thus, two maximally mixed degenerate Majorana neutrinos of opposite CP 
parities merge to form a Dirac neutrino.  It is easy to see that in this case 
their contributions into the probability amplitude of the neutrinoless double 
beta decay $2\beta 0\nu$ (which is only possible for Majorana neutrinos)
exactly cancel, see eq. (\ref{ampl}) in sec. \ref{2beta}. 

If the Majorana mass parameters $m_L$ and $m_R$ do not vanish but are
small compared to $m_D$, the resulting pair of Majorana neutrinos will 
be quasi-degenerate with almost maximal mixing and opposite CP parities. 
The physical neutrino masses in this case are $|m_{1,2}|\simeq m_D \pm 
(m_L+m_R)/2$. Such a pair in many respects behaves as a Dirac neutrino and 
therefore sometimes is called a ``quasi-Dirac neutrino''. In particular,
its contribution to the $2\beta 0\nu$ decay amplitude (\ref{ampl}) is
proportional to the mass difference $(m_L+m_R)/2$ which is much smaller than 
the mass of each component. 

\section{Seesaw mechanism of neutrino mass generation 
\label{ssmech}}
The natural mass scale in the standard model is the electroweak scale
which is of the order of 
$v\simeq 174$ GeV. The smallness of, e.g., the electron mass $m_e\simeq 0.511$ 
MeV is not explained; however, it is easily accommodated in the standard
model through the proper choice of the corresponding Yukawa coupling, 
$f_e\simeq 3\times 10^{-6}$. At the same time, similar explanation of the 
smallness of the electron neutrino mass, $m_{\nu_e}\aprle 5$ eV, would
require the Yukawa coupling $f_{\nu_e}\aprle 3\times 10^{-11}$. Does this 
pose any problem? If we are willing to accept a very small Yukawa coupling
of the electron, why should not we accept small neutrino Yukawa couplings 
as well? After all, $10^{-11}$ may be as good (or as bad) as $10^{-6}$. 

The problem is that, except for neutrinos, the masses of all the fermions 
in each of the three generations are within 1 - 2 orders of magnitude 
of each other. The inclusion of neutrinos leads to huge disparities of
the fermion masses {\it within} each generation. Therefore, if a future 
more complete theory explains why there is a large mass hierarchy between 
generations, it would still remain to be explained why neutrinos are so 
light compared to the other fermions of the same generation. 

The seesaw mechanism \cite{seesaw} provides a very simple and 
attractive explanation of the smallness of neutrino mass. It relates 
the smallness of $m_\nu$ with the existence of very large mass scales. Although 
the seesaw mechanism is most natural in the framework of the grand unified 
theories (such as $SO(10)$) or left-right symmetric models, it also 
operates in the standard model extended to include the right handed 
(``sterile'') neutrinos $\nu_R$. This is the model 
discussed in the end of sec. \ref{numass}, and the relevant part of the
Lagrangian is given by eq.~(\ref{yuk1}). We shall now consider the seesaw 
mechanism 
in some detail.

The most general mass term for $n$ generations of left handed and right
handed neutrinos is written in eq. (\ref{Lm1}) with the mass matrix given
in eq. (\ref{mL1}).  
Notice that in the standard model there is no Majorana mass term for 
left handed neutrinos since there are no triplet Higgs scalars; however, 
$m_L$ is different from zero in some extensions of the standard model, 
so we shall keep it for generality.  
The right handed neutrino $\nu_R$ is an electroweak singlet and so its
mass is not protected by the electroweak symmetry. One can therefore
expect it to be very large, possibly at the Planck scale or at some 
intermediate scale $M_I\sim \sqrt{v M_{\rm Pl}}\sim 10^{10} - 10^{12}$ GeV 
which may be relevant for the physics of parity breaking. 

Let us first consider the limit $m_L\ll m_D\ll m_R$ of the simple 
one-generation case discussed in the previous section. In this limit  
\be
\theta\simeq \frac{m_D}{m_R}\ll 1\,, \quad \quad  
m_1\simeq m_L-\frac{m_D^2}{m_R}\,,\quad \quad m_2 \simeq m_R\,,
\label{ss1}
\ee
\be 
\chi_1\simeq \nu_L+\eta_1(\nu_L)^c \,, \quad\quad
\chi_2\simeq (\nu_R)^c +\eta_2 \nu_R \,.
\label{ss2}
\ee
Thus we have a very light Majorana mass eigenstate $\chi_1$ predominantly 
composed of $\nu_L$ and a heavy eigenstate $\chi_2$ mainly composed of 
$\nu_R$. The admixture of the singlet neutrino state $\nu_R$ in $\chi_1$
and that of the usual neutrinos $\nu_L$ in $\chi_2$ are of the order of
$m_D/m_R$. As follows from eq. (\ref{ss1}), it is 
$\nu_R$ being heavy that makes $\nu_L$ light
\footnote{This is the commonly used jargon: in fact, $\nu_L$ and $\nu_R$,
being chiral fields, do not have any mass but rather are components of some 
massive fields. It would be more correct to say that the fact that $\chi_2$ 
made predominantly of $\nu_R$ and its $\hat{C}$-conjugate $(\nu_R)^c$ is 
heavy explains the lightness of $\chi_1$ made predominantly of $\nu_L$ and
$(\nu_L)^c$, but this is just too lengthy.}. 

\begin{figure}[h]
\begin{center}
\begin{picture}(359,80)(-5,-60)
\ArrowLine(40,-40)(100,-40)
\DashLine(100,-40)(100,5){5}
\Text(100,15)[c]{$\langle H\rangle$}
\Text(60,-50)[l]{\small $\nu_L$}
\Text(100,-50)[c]{\small $m_D$}
\ArrowLine(100,-40)(160,-40)
\Text(130,-50)[l]{\small $\nu_R$}
\Text(160,-40)[c]{$\times$}
\Text(160,-50)[c]{\small $M_R$}
\ArrowLine(220,-40)(160,-40)
\Text(190,-50)[l]{\small $\nu_R$}
\DashLine(220,-40)(220,5){5}
\Text(220,15)[c]{$\langle H\rangle$}
\Text(220,-50)[c]{\small $m_D$}
\ArrowLine(280,-40)(220,-40)
\Text(250,-50)[l]{\small $\nu_L$}
\end{picture}
\end{center}
\vspace{-0.6cm}
\caption{\small Seesaw mechanism of $m_L$ generation}
\label{seesawdiag}
\end{figure}
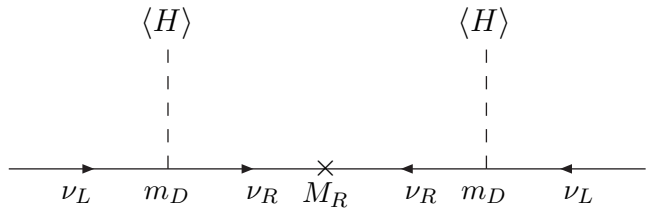

Consider now the full $n$-generation case. We want to block-diagonalize
the matrix ${\cal M}$ in eq. (\ref{mL2}) so as to decouple light and heavy 
neutrino degrees of freedom: 
\be
n_L=U \chi_L\,,\quad\quad
U^T {\cal M}\,U \,= \,
U^T \left(\begin{array}{cc}
m_L & m_D  \\
m_D^T & M_R    
\end{array}\right)\,U\,=\,
 \left(\begin{array}{cc}
\tilde{m}_L & 0  \\
0           & \tilde{M}_R    
\end{array}\right)\,,
\label{ss3}
\ee
where $U$ is a unitary $2n\times 2n$ matrix, and we have changed the
notation $m_R \to M_R$. We shall be looking for the matrix $U$ of the 
form 
\be
U=
 \left(\begin{array}{cc}
1  &  \rho \\
-\rho^\dagger & 1    
\end{array}\right)\,,\quad\quad U^\dagger U = 1 +{\cal O}(\rho^2)\,,
\label{U}
\ee
where the elements are $n\times n$ matrices, and $\rho$ will be treated as 
a perturbation. 
We shall neglect for simplicity possible CP violation in the leptonic sector 
and take $m_L$, $m_D$ and $M_R$ to be real matrices (effects of CP violation 
in neutrino oscillations will be discussed in sec. \ref{impCPT}). The matrix 
$\rho$ can then also be chosen to be real. Block-diagonalization of 
${\cal M}$ gives 
\be
\rho\simeq m_D M_R^{-1}\,,\quad\quad 
\tilde{m}_L \simeq m_L - m_D M_R^{-1} m_D^T\,,\quad\quad ~
\tilde{M}_R\simeq M_R\,.
\label{ss4}
\ee
These relations generalize those of eq. (\ref{ss1}) to the case of $n$ 
generations. The diagonalization of the effective mass matrix $\tilde{m}_L$  
yields $n$ light Majorana neutrinos which are predominantly composed of
the usual (``active'') neutrinos $\nu_L$ with very small ($\sim m_D/M_R$)
admixture of ``sterile'' neutrinos $\nu_R$; diagonalization of $\tilde{M}_R$ 
produces $n$ heavy Majorana neutrinos which are mainly composed of $\nu_R$. 
It is important that the active neutrinos get Majorana masses $\tilde{m}_L$ 
even if they have no ``direct'' masses, i.e. $m_L=0$, as it is in the standard 
model. The masses of active neutrinos are then of the order of $m_D^2/M_R$. 
Generation of the effective Majorana mass of light neutrinos is 
diagrammatically illustrated in fig. \ref{seesawdiag}.  
It is interesting that with the largest Dirac mass eigenvalue of the order
of the electroweak scale, $m_D\sim 200$ GeV, the right handed scale 
$M_R\sim 10^{15}$ GeV which is close 
to the typical GUT scales, and assuming that the direct mass term $m_L\aprle 
m_D^2/M_R$, one obtains the mass of the heaviest of the light neutrinos 
$m_\nu\sim (10^{-2} - 10^{-1})$ eV, which is just of the right order of 
magnitude for the neutrino oscillation solution of the atmospheric neutrino 
anomaly. 

\noindent
{\it Problem 6. Perform the approximate block diagonalization of the matrix  
${\cal M}$ and verify eq.~(\ref{ss4}).} 

As was pointed out before, in general the number $k$ of ``sterile'' neutrino 
species need not coincide with the number $n$ of the ``active'' ones. The
seesaw mechanism works 
when $k\ne n$ as well, and the 
formulas of eqs. (\ref{ss3}) - (\ref{ss4}) are still valid. The  
difference is that now the matrices $m_D$ and $\rho$ are rectangular 
$n\times k$ matrices ($n$ lines, $k$ columns) rather than square matrices.
The matrices $m_L$ and $M_R$ are square matrices of dimension $n\times n$ 
and $k\times k$ respectively; the same is true for the unit matrices in the 
left upper  and right lower corners of the matrix $U$ in (\ref{U}). 

In deriving eq. (\ref{ss4}) we have assumed that the matrix $\rho$ can be 
considered as a small parameter, i.e. that $M_R \gg m_D$. In what sense one 
can consider one matrix to be much larger than another? Does that mean
that we have to require that all the elements of  $M_R$ be much larger
than all the elements of $m_D$? Obviously, this is not correct. One can,  
for example, consider the matrix ${\cal M}$ in the basis of $\nu_R$ in
which $M_R$ has been diagonalized; in this case all off-diagonal elements 
of $M_R$ are zero and yet the seesaw approximation works perfectly well.
On the contrary, if one chooses the matrix $M_R$ with all its elements 
equal to each other, the standard seesaw mechanism fails, no matter how
large the elements of $M_R$. This is just because $det(M_R)=0$ in this case, 
so that the matrix $M_R^{-1}$ does not exist. The true criterion of the 
applicability of the seesaw approximation is that all eigenvalues of $M_R$ 
be large compared to all eigenvalues of $m_D$. 

\noindent
{\it Problem 7. Consider the seesaw mechanism for three $\nu_L$ and
three $\nu_R$ species in the case when one of the eigenvalues of $M_R$ is 
zero. What will be the neutrino mass spectrum in this case? Hint: go to 
the $\nu_R$ basis where $M_R$ is diagonal and include the line and the 
column that contain zero eigenvalue into a redefined matrix $m_L$. }

\section{Neutrino oscillations in vacuum 
\label{vac}}
Neutrino oscillations are the most sensitive probe of neutrino mass. Solar 
and supernova neutrino experiments may be able to discover the neutrino mass 
as small as $10^{-5}$ eV or even smaller, far beyond the reach of the direct 
kinematic search experiments. The idea of neutrino oscillations was first 
introduced by Pontecorvo \cite{Pontecorvo}. The essence of this effect is
very simple, and examples of the oscillation phenomena can be found in 
nearly every textbook on quantum mechanics. Consider, for example, a 
two-level quantum system. If the system is in one of its stationary states 
$|\Psi_i\rangle$ (eigenstates of the Hamiltonian), it will remain in this 
state, and the time evolution of the wave function is very simple -- it just 
picks up a phase factor: $|\Psi_i(t)\rangle=e^{-i E_i t} |\Psi_i(0)\rangle$. 
If, however, a state is prepared which is not one of the eigenstates 
of the Hamiltonian of the system, the probability to find the system in 
this state will oscillate in time with the frequency $\omega_{21}=E_2-E_1$
where $E_1$ and $E_2$ are the eigenenergies of the system. 

In the case of neutrino oscillations, neutrinos are produced by the 
charged-current weak interactions and therefore are weak-eigenstate neutrinos 
$\nu_e$, $\nu_\mu$ or $\nu_\tau$. However, the neutrino mass matrix in this 
(flavour) basis is in general not diagonal. This means that the mass 
eigenstate neutrinos $\nu_1$, $\nu_2$ and $\nu_3$ (the states that diagonalize 
the neutrino mass matrix, i.e. the free propagation eigenstates) are in 
general different from the flavour eigenstates. Therefore the probability of 
finding a neutrino created in a given flavour state to be in the same state 
(or any other flavour state) oscillates with time. 

We shall consider neutrino oscillations in the case of Dirac neutrino mass
term and then will comment on the Majorana and Dirac + Majorana cases. 
The part of the Lagrangian that describes the lepton masses and charged
current interactions is    
\be
-{\cal L}_{W+m}=\frac{g}{\sqrt{2}}\,\overline{e}_{aL}'\,\gamma^\mu\,\nu_{aL}'
\, W^-_\mu + (m_l)_{ab} \overline{e}_{aL}' e_{bR}' + (m_D)_{ab}\, 
\overline{\nu}_{aL}' \nu_{bR}' + h.c. 
\label{yuk2}
\ee
Here the primes are used to denote the flavour eigenstate fields. It follows 
from this expression that the individual lepton flavours $L_e$, $L_\mu$ and 
$L_\tau$ are not conserved when the Dirac neutrino mass term is present,
but the total lepton number $L=L_e+L_\mu+L_\tau$ is still conserved. 

The mass matrix of charged leptons $m_l$ and the neutrino mass matrix $m_D$ 
are general complex matrices which can be diagonalized by bi-unitary 
transformations. Let us write 
\be
e_L'=V_L \,e_L\,, \quad\quad e_R'=V_R \,e_R\, \quad\quad
\nu_L'=U_L\,\nu_L\,, \quad\quad \nu_R'=U_R\,\nu_R\,,
\label{trans} 
\ee
and choose the unitary matrices 
$V_L$, $V_R$, $U_L$ and $U_R$ so that they diagonalize the mass matrices of 
charged leptons and neutrinos:
\be
V_L^\dagger \,m_l\,V_R=(m_l)_{diag}\,,\quad\quad 
U_L^\dagger \,m_D\,U_R=(m_D)_{diag}\,. 
\label{diag}
\ee
The ``unprimed'' fields $e_{iL}$, $e_{iR}$, $\nu_{iL}$ and $\nu_{iR}$ are
then the components of the Dirac mass eigenstate fields $e_i=e_{iL}+e_{iR}$ 
and $\nu_i=\nu_{iL}+\nu_{iR}$. The Lagrangian in eq. (\ref{yuk2}) can be
written in the mass eigenstate basis as 
\be
-{\cal L}_{W+m}=\frac{g}{\sqrt{2}}\, \overline{e_i}\,\gamma^\mu\, (V_L^\dagger 
U_L)_{ij}\,\nu_{Lj}\, W^-_\mu + m_{li} \overline{e}_{Li} e_{Ri} + m_{Di}\, 
\overline{\nu}_{Li} \nu_{Ri} + h.c. \,,
\label{yuk3}
\ee
where $m_{li}$ are the charged lepton masses 
and $m_{Di}$ are the neutrino masses. The matrix $U=V_L^\dagger U_L$ is
called the lepton mixing matrix, or Maki-Nakagawa-Sakata (MNS) matrix
\cite{MNS}. It is the leptonic analog of the CKM mixing matrix. It relates a 
neutrino flavour eigenstate $|\nu'_a\rangle$ produced or absorbed alongside 
with the corresponding charged lepton, to the mass eigenstates $|\nu_i\rangle$:
\be
|\nu'_a\rangle
=U^*_{ai}\,|\nu_i\rangle\,,
\label{rel3}
\ee
In what follows, to simplify the notation we shall omit the primes and 
distinguish between the flavour and mass eigenstates just by using the 
indices from the beginning of the Latin alphabet for the former and from its 
middle for the latter.  Assume that at a time $t=0$ the flavour eigenstate 
$|\nu_a\rangle$ was produced. What is the probability to find the neutrino 
in a state $|\nu_b\rangle$ at a later time $t$? Time evolution of the flavour 
eigenstates is not simple, therefore it is more convenient to follow the 
evolution of the system in the mass eigenstate basis (we shall discuss
evolution in the flavour basis in sec. \ref{matter}). The initial state at
$t=0$ is $|\nu(0)\rangle=|\nu_a\rangle =U^*_{aj}|\nu_j\rangle$; the neutrino 
state at a later time $t$ is then 
\be
|\nu(t)\rangle=U^*_{aj}\,e^{-iE_j t}|\nu_j\rangle \,.
\label{evol1}
\ee
The probability amplitude of finding the neutrino at the time $t$ in a
flavour state $|\nu_b\rangle$ is
\be
A(\nu_a\to \nu_b;\,t)=
\langle \nu_b|\nu(t)\rangle = U^*_{aj}\,e^{-iE_j t}\,\langle \nu_b|\nu_j
\rangle = U_{bi} U^*_{aj}\,e^{-iE_j t}\,\langle \nu_i|\nu_j\rangle = 
U_{bj}\,e^{-iE_j t}\, U^*_{aj} \,.
\label{evol2}
\ee
As usual, the sum over all intermediate states $j$ is implied. The last 
expression here has a very simple physical meaning. The factor $U^*_{aj}=
U^\dagger_{ja}$ is the amplitude of transformation of the initial flavour 
eigenstate neutrino $\nu_a$ into a mass eigenstate one $\nu_j$; the factor
$e^{-iE_j t}$ is just the propagator describing the time evolution of the 
mass eigenstate neutrino $\nu_j$ in the energy representation, and finally
the factor $U_{bj}$ converts the time-evolved mass eigenstate $\nu_j$
into the flavour eigenstate $\nu_b$. The neutrino oscillation probability, 
i.e. the probability of the transformation of a flavour eigenstate neutrino 
$\nu_a$ into another one $\nu_b$, is then 
\be
P(\nu_a\to \nu_b;\,t)=
|A(\nu_a\to \nu_b;\,t)|^2=|U_{bj}\,e^{-iE_j t}\, U^*_{aj}|^2 \,.
\label{evol3}
\ee

We have discussed neutrino oscillations in the case of Dirac neutrinos.
What happens if neutrinos have a Majorana mass term rather than the Dirac 
one? Eq. (\ref{yuk2}) now has to be modified: the term $(m_D)_{ab}\, 
\overline{\nu}_{aL}' \nu_{bR}' +h.c.$ has to be replaced by $(m_M)_{ab}\, 
\overline{\nu_{aL}^c}' \nu_{bR}' + h.c.=(m_M)_{ab}\, \nu_{aL}'^T\,C\,\nu_{bR}' 
+ h.c.$ This mass term breaks not only the individual lepton flavours but 
also the total lepton number. 
The symmetric Majorana mass matrix $(m_M)_{ab}$ is diagonalized by 
the transformation $U_L^T\,m_M\,U_L=(m_M)_{diag}$, so one can again use the 
field transformations (\ref{trans}). Therefore the structure of the charged 
current interactions is the same as in the case of the Dirac neutrinos, and 
the diagonalization of the neutrino mass matrix in the case of the $n$ fermion 
generations again gives $n$ mass eigenstates. Thus the oscillation 
probabilities in the case of the Majorana mass term are the same as in the 
case of the Dirac mass term. This, in particular, means that one cannot 
distinguish between Dirac and Majorana neutrinos by studying neutrino 
oscillations. Essentially, this is because the total lepton number is not 
violated by the neutrino flavour oscillations. 

The situation is different in the case when the neutrino mass term is of the 
most general Dirac + Majorana (D + M) form (which, like in the pure Dirac
case, requires the existence of the electroweak singlet neutrinos $\nu_R$).  
In particular, in the case of $n$ $\nu_L$ and $n$ $\nu_R$ species, the 
neutrino mass matrix has the dimension $2n\times 2n$, leading to $2n$
massive Majorana neutrino states. The total lepton number conservation is
violated by the Majorana mass term. Unlike in the pure Dirac and pure
Majorana cases, in this D + M case one can have a new type of neutrino 
oscillations: in addition to the usual flavour oscillations $\nu_a \to
\nu_b$, oscillations into ``sterile'' states, $\nu_a \to \nu^c_b$, can occur. 
These oscillations violate the total lepton number $L$. Thus, the D + M case 
can in principle be distinguished from the pure Dirac and Majorana cases 
in neutrino oscillations experiments. For more detailed discussion of the 
D + M case see, e.g., \cite{BGG}. 

\subsection{2 flavour case 
\label{2fl}}
\begin{figure}[htb]
\setlength{\unitlength}{1cm}
\begin{center}
\epsfig{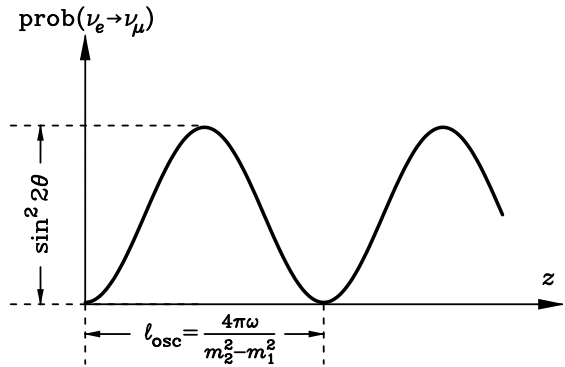}
\end{center}
\vspace{-0.5cm} 
\caption{\small Pattern of two--flavor neutrino oscillations (neutrino
energy $\omega$, distance $z$). From \cite{raffelt2}.}
\label{osc}
\end{figure}

Let us now consider neutrino oscillations in a simple case of just two
neutrino species, $\nu_e$ and $\nu_\mu$. The lepton mixing matrix $U$ can
be written as 
\be
U=\left(\begin{array}{cc}
c & s \\
-s & c \end{array} \right)\,,
\label{U2}
\ee
where 
$c=\cos\theta_0$, $s=\sin\theta_0$, $\theta_0$ being the mixing angle. 
The neutrino mass and flavour eigenstates are therefore related through 
\begin{eqnarray}
|\nu_e\rangle &=& c\,|\nu_1 \rangle+s\,|\nu_2 \rangle\,,\nonumber \\
|\nu_\mu\rangle &=& \!\!\!\! -s\,|\nu_1 \rangle+c\,|\nu_2 \rangle\,.
\label{mix1}
\end{eqnarray} 
%
Substituting (\ref{U2}) into (\ref{evol3}) and taking into account that for 
relativistic neutrinos of the momentum $p$, 
\be
E_i=\sqrt{p^2+m_i^2}\simeq p+\frac{m_i^2}{2p}\simeq p+\frac{m_i^2}{2E}\,, 
\ee
we find the transition probabilities  
\be
P(\nu_e\to\nu_\mu ;\,t)=P(\nu_\mu\to\nu_e;\,t)=
\sin^2 2\theta_0 \,\sin^2 \left (\frac{\Delta m^2}{4E}t\right)\,. 
\label{prob1}
\ee
Here $\Delta m^2=m_2^2-m_1^2$. 
The survival probabilities are  $P(\nu_e\to\nu_e;\,t)=P(\nu_\mu\to\nu_\mu; 
\,t)=1-P(\nu_e\to\nu_e;\,t)$. 
It is convenient to rewrite the transition probability in terms of the
distance $L$ travelled by neutrinos. 
For relativistic neutrinos $L \simeq t$, and one has 
\be
P(\nu_e\to\nu_\mu ;\,L)=
\sin^2 2\theta_0\,\sin^2 \left(\pi \frac{L}{l_{osc}}\right)\,,
\label{prob2}
\ee
where $l_{osc}$ is the oscillation length defined as  
\be
l_{osc}=\frac{4\pi E}{\Delta m^2}\,\simeq \,2.48\;m\,\frac{E\,\mbox{(MeV)}}
{\Delta m^2\, (\mbox{eV}^2)}=\,2.48\;km\,\frac{E\,\mbox{(GeV)}}
{\Delta m^2\, (\mbox{eV}^2)}
\label{losc}
\ee
It is equal to the distance between any two closest minima or maxima of
the transition probability (see fig. \ref{osc}). 
Notice that $l_{osc}$ is inversely proportional to 
the energy difference of the neutrino mass eigenstates: $l_{osc}=
2\pi/(E_2-E_1)$. Another convenient form of the expression for the 
transition probability is 
\be
P(\nu_e\to\nu_\mu ;\,L)=
\sin^2 2\theta_0\,\sin^2 \left(1.27 \Delta m^2\frac{L}{E}\right)\,,
\label{prob3}
\ee
where $L$ is in $m$ and $E$ in MeV or $L$ is in $km$ and $E$ in GeV. 

Let us discuss the probability of neutrino oscillations (\ref{prob2}) (see 
fig. \ref{osc}). It has two factors. The first one ($\sin^2 2\theta_0$) 
does not depend on the distance travelled by neutrinos; 
it describes the amplitude (or depth) of the neutrino oscillations. The 
amplitude is maximal (equal to one) when the mixing angle $\theta_0=45^\circ$, 
which corresponds
to maximal mixing. When $\theta_0$ is close to zero or $90^\circ$, flavour 
eigenstates are nearly aligned with mass eigenstates, which corresponds to
small mixing. In this case the oscillation amplitude is small. The second 
factor oscillates with time or distance $L$ travelled by neutrinos. The
phase of the sine (the oscillation phase) is proportional to the energy
difference of the mass eigenstates $\Delta m^2/2E$ and to the distance $L$. 
In order to have an appreciable transition probability, it is not enough to
have large mixing: in addition, the oscillation phase should be not too small. 
When the oscillation phase is very large, the transition probability 
undergoes fast oscillations. Averaging over small energy intervals 
corresponding to the finite energy resolution of the detector, or over small 
variations of the distance between the neutrino production and detection 
points corresponding to the finite sizes of the neutrino source and 
detector, results then in averaging out the neutrino oscillations. The
observed transition probability in this case is
\be
\overline{P(\nu_e\to \nu_\mu)}=\overline{P(\nu_\mu\to \nu_e)}
=\frac{1}{2}\, \sin^2 2\theta_0\,.
\label{aver}
\ee

In our discussion we have been assuming that the initially produced neutrino 
state has a certain momentum. Since it is a mixture of different mass 
eigenstates, its energy is not well defined in this case. We then consider
the time evolution of the neutrino state. Alternatively, one could assume that 
the neutrino is produced in a state of a certain energy but not well defined 
momentum, and consider the evolution of the neutrino system in space. For
relativistic neutrinos the result will be the same. Which description is more 
correct? Does the initially produced state have a certain energy or a certain 
momentum? In general, since the production process takes finite time and is 
localized in space, neither neutrino energy nor momentum is well defined;
moreover, the transition probability depends on both the neutrino production 
and detection processes. Fortunately, for relativistic neutrinos these 
subtleties are unimportant, and the simple description in terms of the states 
with certain momentum or energy gives the correct answer 
\footnote{Uncertainties of neutrino energy and momentum may be important for 
the question of coherence of the neutrino state in the case of the
oscillations over very long baselines $L$, e.g. for oscillations
of solar or supernova neutrinos on their way to the earth. For
discussion see, e.g., \cite{MS-Uspekhi}.}. 

\subsection{3 flavour case 
\label{3fl}}
Consider now the case of three neutrino flavours. The neutrino flavour
eigenstate and mass eigenstate fields are related through 
\be
\left(\begin{array}{c}
\nu_{eL} \\
\nu_{\mu L}\\   
\nu_{\tau L}
\end{array} \right)=
\left(\begin{array}{ccc}
U_{e1} & U_{e2} & U_{e3}  \\
U_{\mu 1} & U_{\mu 2} & U_{\mu 3}  \\
U_{\tau 1} & U_{\tau 2} & U_{\tau 3}  
\end{array} \right)
\left(\begin{array}{c}
\nu_{1L} \\
\nu_{2L}\\   
\nu_{3L}
\end{array} \right)\,. 
\label{mix2}
\ee
In general, in the case of Dirac neutrinos the lepton mixing matrix $U$ in
(\ref{mix2}) depends on three mixing angles $\theta_{12}$, $\theta_{13}$ and 
$\theta_{23}$ and one CP-violating phase $\delta$ (in the case of Majorana 
neutrinos there are two additional, so-called Majorana phases, see discussions 
in sections \ref{impCPT} and \ref{3flmatt}). It is convenient to use the
parametrization of the matrix $U$ which coincides with the standard 
parametrization of the quark mixing matrix \cite{PDG}:
\be
U =
\left(\begin{array}{ccc}
c_{12}\,c_{13}   & s_{12}\, c_{13}    & s_{13}\, e^{-i\delta} \\
-s_{12}\,c_{23}-c_{12}\,s_{23}\,s_{13}\,e^{i\delta} &
c_{12}\,c_{23}-s_{12}\,s_{23}\,s_{13}\,e^{i\delta} &
s_{23}\,c_{13}   \\
s_{12}\,s_{23}-c_{12}\,c_{23}\,s_{13}\,e^{i\delta} &
-c_{12}\,s_{23}-s_{12}\,c_{23}\,s_{13}\,e^{i\delta} &
c_{23}\,c_{13}
\end{array}
\right)\,.
\label{U3}
\ee
Here $c_{ij}=\cos \theta_{ij}$, $s_{ij}=
\sin\theta_{ij}$. The probabilities of oscillations between various 
flavour states are given by the general expression (\ref{evol3}). Unlike in 
the two-flavour case, they in general do not have a simple form. There are, 
however, several practically important limiting cases in which one can obtain 
very simple approximate expressions for the oscillation probabilities in terms 
of the 2-flavour ones. Assume first that the neutrino mass squared differences 
$\Delta m_{ij}^2=m_i^2-m_j^2$ have a hierarchy
\be
|\Delta m_{21}^2| \ll |\Delta m_{31}^2| \simeq |\Delta m_{32}^2|\,.
\label{hier}
\ee
This means that either $m_1\ll(\aprle)\, m_2 \ll m_3$ (direct hierarchy)
or $m_3 \ll m_1\approx m_2$ (inverted mass hierarchy). These cases are of 
practical interest since the solar neutrino data indicate that one needs a 
small mass squared difference $\Delta m_\odot^2 \sim 10^{-5}$ eV$^2$ for the 
solution of the solar neutrino problem through the matter-enhanced neutrino 
oscillations (or $\Delta m_\odot^2\sim 10^{-10}$ eV$^2$ for the vacuum 
oscillations solution) whereas the explanation of the atmospheric neutrino 
experiments through the neutrino oscillations requires $\Delta m_{atm}^2 
\sim 10^{-3}$ eV$^2$, much larger than $\Delta m_\odot^2$. 
Consider first the oscillations over the baselines $L$ for which 
\be
\frac{\Delta m_{21}^2}{2E}L\ll 1\,. 
\label{cond1} 
\ee
This case is relevant for atmospheric, reactor and accelerator neutrino 
experiments. It follows from (\ref{cond1}) that the oscillations due to the 
small mass difference $\Delta m_{21}^2$ are effectively frozen in this case, 
and one can consider the limit $\Delta m_{21}^2\to 0$. 
The probability of $\nu_a\to\nu_b$ oscillations then takes a very simple
form: 
\be
P(\nu_a\to\nu_b;\,L)=4 |U_{a3}|^2\,|U_{b3}|^2\,\sin^2 \left(\frac{\Delta 
m_{31}^2}{4E}\,L\right)\,. 
\label{prob4}
\ee 
It resembles the 2-flavour oscillation probability. The probabilities 
of oscillations between $\nu_e$, $\nu_\mu$ and $\nu_\tau$ are 
\begin{eqnarray}
P(\nu_e\to\nu_\mu;\,L)=4 |U_{e3}|^2\,|U_{\mu 3}|^2\,\sin^2 
\left(\frac{\Delta m_{31}^2}{4E}\,L\right)=
s_{23}^2\,\sin^2 2\theta_{13}\,\sin^2\left(\frac{\Delta
m_{31}^2}{4E}\,L\right)\,, \label{prob5a} \\
P(\nu_e\to\nu_\tau;\,L)=4 |U_{e3}|^2\,|U_{\tau 3}|^2\,\sin^2
\left(\frac{\Delta m_{31}^2}{4E}\,L\right)=
c_{23}^2\,\sin^2 2\theta_{13}\,\sin^2\left(\frac{\Delta
m_{31}^2}{4E}\,L\right)\,, \label{prob5b} \\
P(\nu_\mu\to\nu_\tau;\,L)=4 |U_{\mu 3}|^2\,|U_{\tau 3}|^2\,\sin^2
\left(\frac{\Delta m_{31}^2}{4E}\,L\right)= c_{13}^4\,\sin^2 2\theta_{23}\,
\sin^2\left(\frac{\Delta m_{31}^2}{4E}\,L\right)\,,  
\label{prob5c}
\end{eqnarray} 
with $P(\nu_b\to \nu_a;\,L)=P(\nu_a\to \nu_b;\,L)$. 
They depend only on the elements of the third
column of the lepton mixing matrix $U$ and one mass squared difference. 
The survival probability for electron neutrinos takes a particularly  
simple form  
\be
P(\nu_e\to\nu_e;\,L)=1-\sin^2 2\theta_{13}\,\sin^2\left(\frac{\Delta m_{31}^2}
{4E}\,L\right)\,,
\label{prob6}
\ee 
i.e. it coincides with the $\nu_e$ survival probability in the 2-flavour
case with the mass squared difference $\Delta m^2=\Delta m_{31}^2$ and 
mixing angle $\theta_0=\theta_{13}$.

Consider now another limiting case, which is relevant for the solar neutrino 
oscillations and also for very long baseline reactor experiments (such as 
KamLAND, see sec. \ref{reacc}). We shall be again assuming the hierarchy 
(\ref{hier}) and in addition  
\be
\frac{\Delta m_{31}^2}{2E} L\simeq \frac{\Delta m_{32}^2}{2E}L\gg 1\,.
\label{cond2} 
\ee
whereas the condition (\ref{cond1}) is no longer necessary. 
In this case the oscillations due to the mass squared differences 
$\Delta m_{31}$ and $\Delta m_{32}$ are very fast and lead to an averaged 
effect; the $\nu_e$ survival probability is then 
\be
P(\nu_e\to \nu_e) \simeq c_{13}^4 P + s_{13}^4\,,
\label{prob7}
\ee
where $P$ is the $\nu_e$ survival probability in the 2-flavour 
case with the mass squared difference $\Delta m^2=\Delta m_{21}^2$ and 
mixing angle $\theta_0=\theta_{12}$. In the case of neutrino oscillations 
in vacuum one has  
\be
P=1-\sin^2 2\theta_{12}\,
\sin^2\left(\frac{\Delta m_{21}^2}{4E}\,L\right)\,. 
\label{prob8}
\ee
Finally, consider the limit $U_{e3}=0$ (the results will be also approximately 
valid for $|U_{e3}|\ll 1$). In this case one obtains 
\begin{eqnarray}
& & P(\nu_e \to \nu_\mu;\,L)=c_{23}^2 \sin^2 2\theta_{12}\,
\sin^2 \Delta_{21} \,, 
\label{prob9} \\
& & P(\nu_e \to \nu_\tau;\,L)=s_{23}^2 \sin^2 2\theta_{12}\,
\sin^2 \Delta_{21}\,,
\label{prob10} \\
& & P(\nu_\mu \to \nu_\tau;\,L)=\sin^2 2\theta_{23} 
\,(-s_{12}^2\,c_{12}^2\sin^2 \Delta_{21}   
+s_{12}^2\,\sin^2 \Delta_{31} + c_{12}^2\sin^2 \Delta_{32} )\,, 
\label{prob11}
\end{eqnarray}
where $\Delta_{ij}\equiv (\Delta m_{ij}^2/4E)L$, and no assumption about the 
hierarchy of the mass squared differences has been made. Notice that the 
limiting cases that we have discussed here are not mutually excluding,
i.e. have some overlap with each other. 

In general, when considering the propagation of solar neutrinos in the sun 
or in the earth, one should take into account matter effects on neutrino 
oscillations. The same is true for the terrestrial atmospheric and long 
baseline accelerator neutrino oscillation experiments in which the neutrino 
trajectories or their significant portions go through the matter of the 
earth. Matter effects on $\nu_\mu \leftrightarrow \nu_\tau$ oscillations are 
relatively small (they vanish in the 2-flavour approximation), but they may 
be quite appreciable for $\nu_e \leftrightarrow \nu_\mu$ and $\nu_e 
\leftrightarrow \nu_\tau$ oscillations. We shall discuss these effects in 
sec. \ref{matter}. As we shall see, 
eq. (\ref{prob7}) remains valid in the case of neutrino 
oscillations in matter as well, but the two-flavour probability (\ref{prob8}) 
has to be modified. Eqs. (\ref{prob5a}) - (\ref{prob5c}), (\ref{prob6}) and 
(\ref{prob9}) - (\ref{prob11}) are also modified in the case of neutrino
oscillations in matter (see sec. \ref{3flmatt}). 

\noindent
{\it Problem 8. Assuming that conditions (\ref{hier}) and (\ref{cond1})
are satisfied derive eq. (\ref{prob4}).}

\noindent
{\it Problem 9. Derive 
eqs. (\ref{prob7}) - (\ref{prob8}) assuming 
(\ref{hier}) and (\ref{cond2}).}

\noindent
{\it Problem 10. Assuming $U_{e3}=0$ derive eqs. (\ref{prob9}) 
- (\ref{prob11}).} 

\noindent
{\it Problem 11. Imagine a world described by the standard model
supplemented 
by three right handed neutrinos $\nu_R$, but in which the charged leptons
are massless (Yukawa couplings $f_{ij}^e$ in eq. (\ref{yuk1}) vanish).
Will neutrinos oscillate in such a world?}

\noindent
{\it Problem 12. Do neutrinos produced in the decay $Z^0\to \nu \bar{\nu}$ 
oscillate? (When you come to an answer, compare it with that in \cite{SZ}).} 

\subsection{Implications of CP, T and CPT symmetries 
\label{impCPT}}

We shall now consider the consequences of CP, T and CPT symmetries for 
neutrino oscillations. As we discussed in sec. \ref{WDM}, charge conjugation
operation C is not well defined for neutrinos as it would convert a left 
handed neutrino into a non-existent left handed antineutrino. CP, however,
is well defined: it converts a left handed neutrino $\nu_L$ into a right
handed antineutrino which is the antiparticle of $\nu_L$. Thus, CP 
essentially acts as the particle - antiparticle conjugation. If CP is 
conserved, the probabilities of oscillations between particles and their
antiparticles coincide:
\be
{\rm CP:}
\quad \quad \quad 
P(\nu_a \to \nu_b;\,t)=P(\bar{\nu}_a \to \bar{\nu}_b;\,t)\,.
\label{CP1}
\ee
In quantum field theory, the same field operator that 
annihilates a particle also creates its antiparticle,  whereas its Hermitean 
conjugate does the opposite. Therefore the action of the particle --  
antiparticle conjugation on the lepton mixing matrix $U$ amounts to 
$U\to U^*$. This means that CP is only conserved in the leptonic sector if
the mixing matrix $U$ is real or can be made real by a rephasing of the lepton 
fields. 

In general, a unitary $n\times n$ matrix depends on $n(n-1)/2$ angles 
and $n(n+1)/2$ phases. In the Dirac case, $2n-1$ phases can be removed by
a proper rephasing of the left handed fields, leaving 
$n(n+1)/2-(2n-1)=(n-1)(n-2)/2$ physical phases (the rephasing of the left
handed field leaves the lepton mass terms unchanged since the phases can be 
absorbed into the corresponding rephasing of the right-handed fields). 
Thus, in the Dirac case CP non-conservation is only possible in the case of 
$n\ge 3$ generations. In the Majorana case there is less freedom to rephase 
the fields since the Majorana mass terms are of the form $\nu_L \nu_L + h.c.$ 
rather than of the form $\bar{\nu}_R \nu_L +h.c.$ and so the phases of 
neutrino fields cannot be absorbed. 
Therefore in the Majorana case only $n$ phases can be removed, leaving 
$n(n+1)/2-n=n(n-1)/2$ physical phases. Out of these phases, $(n-1)(n-2)/2$ are 
the usual, Dirac-type phases while the remaining $n-1$ are specific for the 
Majorana case, so called Majorana phases. Majorana phases do not lead to any 
observable effects for neutrino oscillations \cite{Majorosc}, and we shall 
not consider them here.

CPT transformation can be considered as a combined action of CP, which
results in the interchange between paricles and antiparticles, and time
reversal, which interchanges the initial and final states. Therefore under 
CPT transformation the oscillation probability $P(\nu_a\to\nu_b;\,t)$ goes 
into $P(\bar{\nu}_b\to \bar{\nu}_a;\,t)$. One the other hand, from 
eq. (\ref{evol2}) it follows that the CPT transformation $U\to U^*$, $t\to
-t$ transforms the oscillation amplitude $A(\nu_a\to\nu_b;\,t)$ into its 
complex conjugate. Therefore the oscillation probabilities are invariant with 
respect to CPT, i.e. the following equality holds:
\be
{\rm CPT:}
\quad \quad \quad 
P(\nu_a \to \nu_b;\,t)=P(\bar{\nu}_b \to \bar{\nu}_a;\,t)\,. 
\label{CPT}
\ee
CPT invariance implies, in particular, that the survival probabilities for
neutrinos and antineutrinos are the same: $P(\nu_a\to \nu_a;\,t)=
P(\bar{\nu}_a\to \bar{\nu}_a;\,t)$. 

Finally, time reversal T interchanges the initial and final states, so if 
T is conserved one has 
\be
{\rm T:}
\quad \quad \quad 
P(\nu_a \to \nu_b;\,t)=P(\nu_b \to \nu_a;\,t)\,. 
\label{T}
\ee
{}From CPT invariance it follows that CP conservation is equivalent to T
conservation; indeed, it is easy to see that (\ref{CP1}) and 
(\ref{CPT}) lead to (\ref{T}). 

Let us now consider effects of CP-violation on neutrino oscillations.  
If CP is not conserved, the oscillation probabilities for neutrinos are 
different from those for antineutrinos. As we pointed out before, this is
only possible if the lepton mixing matrix $U$ is essentially complex, i.e. 
has unremovable phases. For three lepton generations, there is only one such 
phase $\delta$, and so there should be only one CP-odd oscillation asymmetry.  
Let us denote the CP-odd asymmetries as  
\be
\Delta P_{ab}\equiv P(\nu_a \to \nu_b;\,t)-P(\bar{\nu}_a \to \bar{\nu}_b;
\,t)\,.
\label{asymm1}
\ee
{}From CPT invariance one finds $\Delta P_{ab}=-\Delta P_{ba}$. Using the 
parametrization (\ref{U3}) of the mixing matrix $U$ it is not difficult 
to find 
\begin{eqnarray}
\Delta P_{e\mu}=\Delta P_{\mu \tau}=\Delta P_{\tau e}=4 s_{12}\,c_{12}\,
s_{13}\,c_{13}^2\,s_{23}\,c_{23}\,\sin\delta 
\quad \quad \quad \quad \quad \quad \quad \quad \quad \quad \quad 
\nonumber \\
\times \left[\sin\left(\frac{\Delta m_{12}^2}{2E}t 
\right)+ \sin\left(\frac{\Delta m_{23}^2}{2E}t\right)+ \sin\left(
\frac{\Delta m_{31}^2}{2E}t\right)\right]\,.
\label{asymm2}
\end{eqnarray}

\noindent
{\it Problem 13. Derive eq. (\ref{asymm2}) using parametrization
(\ref{U3}) and unitarity property of the lepton mixing matrix $U$.}

\noindent
This expression has several interesting features. First, as expected, it 
vanishes in the limit $\delta=0$. Second, it vanishes if any of the mixing 
angles $\theta_{12}$, $\theta_{13}$ or $\theta_{23}$ is zero or $90^\circ$. 
In particular, oscillation probabilities (\ref{prob9})-(\ref{prob11})
obtained in the limit $\theta_{13}=0$ are CP-symmetric. Third, since the 
mass squared differences satisfy the 
relation $\Delta m_{12}^2+\Delta m_{23}^2+\Delta m_{31}^2=0$, the CP-odd 
asymmetry (\ref{asymm2}) vanishes if even one of $\Delta m_{ij}^2$ 
is zero. We have already encountered this situation -- the transition 
probabilities (\ref{prob4}) - (\ref{prob6}). 
derived in the limit $\Delta m_{21}^2=0$ depend only on the absolute values 
of $U_{ai}$ and are therefore CP-invariant. 
The relation between $\Delta m_{ij}^2$ also means that in the
limit of the small arguments of the sines (small oscillation phases) in
eq. (\ref{asymm2}) the probability asymmetry is cubic (more precisely,  
tri-linear) in small phases rather than linear. Another important difference 
between the usual oscillation probabilities and the CP-odd asymmetry 
(\ref{asymm2}) is that while the former contain squared sines of the 
oscillation phases and therefore oscillate near non-zero average values,
the latter is linear in these sines and so oscillates around zero. This means 
that in the case of very large oscillation phases, when the averaging
regime sets in, the CP-odd asymmetry of neutrino oscillations averages to 
zero. It is for this reason that the oscillation probability (\ref{prob7})  
is CP-invariant. 

It follows from the above discussion that the experimental observation of 
CP violation effect in neutrino oscillations is a very difficult task.  The 
CP-odd probability asymmetry is suppressed if any one of the three lepton 
mixing angles is small, and we know from the CHOOZ reactor neutrino experiment 
that the mixing angle $\theta_{13}$ is small (see sec. \ref{reacc} below
for a discussion of the reactor and accelerator data). The atmospheric neutrino 
experiments indicate that the mixing angle $\theta_{23}$ is rather large, but 
the value of $\theta_{12}$ is largely unknown: the present solar neutrino 
data allow both large and small values for this mixing angle. The hierarchy 
(\ref{hier}) of the mass squared differences which follows from the solar
and atmospheric neutrino observations further hinders experimental searches 
of CP violation in the neutrino oscillation experiments. In addition, matter 
effects on neutrino oscillations may mimic CP violation and so make the 
searches of the genuine CP violation even more difficult (see sec. 
\ref{3flmatt}). Still, the presently allowed values of neutrino parameters
do not exclude a possibility of observation of CP violation effects in the 
future neutrino experiments. Discovering CP-nonconservation in the leptonic 
sector would be of great importance, so the goal is worth pursuing. 

\noindent
{\it Problem 14. Using {\rm CPT} invariance and the unitarity condition 
$P(\nu_e\to\nu_e)+P(\nu_e\to\nu_\mu)+P(\nu_e\to\nu_\tau)=1$ (and similar 
conditions for $\nu_\mu$ and $\nu_\tau$) show that there is only one 
independent {\rm CP}-odd oscillation asymmetry in the case of three
neutrino flavours.} 

\noindent
{\it Problem 15. Apply the same approach to the four-flavour case and find
the number of independent {\rm CP}-odd oscillation asymmetries in that case. 
Compare it with the number of the physical Dirac-type phases in the mixing 
matrix $U$ in the case of four generations.}

\section{Neutrino oscillations in matter 
\label{matter}}

Neutrino oscillations in matter may differ from the oscillations in vacuum
in a very significant way. The most striking manifestation of the matter 
effects on neutrino oscillations is the resonance enhancement of the 
oscillation probability -- the Mikheyev - Smirnov - Wolfenstein (MSW) effect 
\cite{MSW,KuoPa}. In vacuum, the oscillation probability cannot exceed 
$\sin^2 2\theta_0$, and for small mixing angles it is always small. Matter can 
enhance neutrino mixing, and the probabilities of neutrino oscillations in 
matter can be large (close to unity) even if the mixing angle in vacuum is
very small. Matter enhanced neutrino oscillations provide a very elegant
solution of the solar neutrino problem; matter effects on oscillations of 
solar and atmospheric neutrinos inside the earth can be quite important. 
Neutrino oscillations in supernovae and in the early universe may also be  
strongly affected by matter. 

How does the matter affect neutrino propagation? Neutrinos can be absorbed 
by the matter constituents, or scattered off them, changing their momentum
and energy. However the probabilities of these processes, being proportional 
to the square of the Fermi constant $G_F$,  
are typically very small. Neutrinos can also experience forward 
scattering, an elastic scattering in which their momentum is not changed. 
This process is coherent, and it creates mean potentials $V_a$ for neutrinos 
which are proportional to the number densities of the scatterers. These 
potentials are of the first order in $G_F$, but one could still expect them 
to be too small and of no practical interest. This expectation, however, 
would be wrong. To assess the importance of matter effects on neutrino 
oscillations, one has to compare the matter-induced potentials of neutrinos 
$V_a$ with the characteristic neutrino kinetic energy differences 
$\Delta m^2/2E$. Although the potentials $V_a$ are typically very small, 
so are $\Delta m^2/2E$; if $V_a$ are comparable to or larger than 
$\Delta m^2/2E$, matter can strongly affect neutrino oscillations.

\subsection{Evolution equation 
\label{eveq}}
\begin{figure}[h]
\begin{center}
\begin{picture}(359,80)(-5,-40)
\ArrowLine(40,50)(70,25)
\ArrowLine(40,-50)(70,-25)
\Photon(70,25)(70, -25)3 4
\Text(95,0)[r]{\small$W^\pm$}
\Text(95,35)[r]{\small$e$}
\Text(100,-35)[r]{\small$\nu_e$}
\Text(40,-35)[l]{\small$e$}
\Text(40,35)[l]{\small$\nu_e$}
\ArrowLine(70,25)(100,50)
\ArrowLine(70,-25)(100,-50)
\ArrowLine(240,50)(270,25)
\ArrowLine(240,-50)(270,-25)
\Photon(270,25)(270, -25)3 4
\Text(290,0)[r]{\small$Z^0$}
\Text(325,35)[r]{\small$\nu_{e, \mu,\tau}$}
\Text(325,-35)[r]{\small$p, n, e$}
\Text(220,-35)[l]{\small$p,n,e$}
\Text(220,35)[l]{\small$\nu_{e,\mu,\tau}$}
\ArrowLine(270,25)(300,50)
\ArrowLine(270,-25)(300,-50)
\end{picture}
\end{center}
\caption{\small Neutrino scattering diagrams \label{WZ}}
\end{figure}
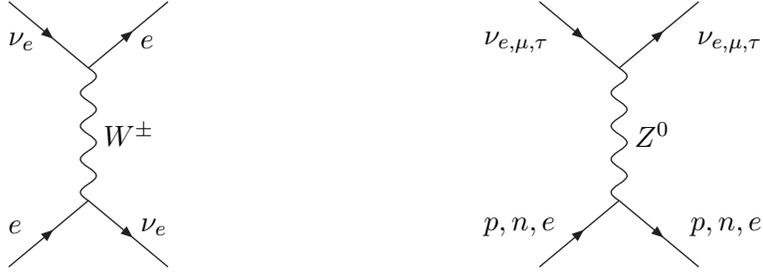

We shall now consider neutrino oscillations in matter in some detail. 
Neutrinos of all three flavours -- $\nu_e$, $\nu_\mu$ and $\nu_\tau$ -- 
interact with the electrons, protons and neutrons of matter through neutral 
current (NC) interaction mediated by $Z^0$ bosons. Electron neutrinos in 
addition have charged current (CC) interactions with the electrons of the 
medium, which are mediated by the $W^\pm$ exchange (see fig. \ref{WZ}). 

Let us consider the CC interactions. At low neutrino energies, they
are described by the effective Hamiltonian 
\be
H_{CC}=\frac{G_F}{\sqrt{2}}\,[\bar{e}\gamma_\mu (1-\gamma_5)\nu_e]
[\bar{\nu}_e\gamma^\mu (1-\gamma_5)e]
=\frac{G_F}{\sqrt{2}}\,[\bar{e}\gamma_\mu (1-\gamma_5)e]
[\bar{\nu}_e\gamma^\mu (1-\gamma_5)\nu_e]\,,
\ee
where we have used the Fierz transformation. In order to obtain the coherent 
forward scattering contribution to the energy of $\nu_e$ in matter (i.e. the 
matter-induced potential for $\nu_e$) we fix the variables corresponding to 
$\nu_e$ and integrate over all the variables that correspond to the electron:  
\be
H_{eff}(\nu_e)=
\langle H_{CC}\rangle_{electron}\equiv \bar{\nu}_e V_e \nu_e\,.
\label{Heff}
\ee
Furthermore, we have 
\be
\langle \bar{e}\gamma_0 e \rangle=\langle e^\dagger e \rangle=N_e\,,
\quad\quad \langle \bar{e}\mbox{\boldmath $\gamma$} 
e\rangle=\langle {\bf v}_e\rangle \,, \quad 
\langle \bar{e}\gamma_0\gamma_5 e \rangle=\langle \frac{\mbox{\boldmath 
$\sigma$}_e{\bf p}_e}{E_e}\rangle\,, \quad 
\langle \bar{e}\mbox{\boldmath $\gamma$} \gamma_5 e \rangle=\langle 
\mbox{\boldmath $\sigma$}_e \rangle\,,
\label{terms}
\ee
where $N_e$ is the electron number density. For 
unpolarized medium of zero total momentum only the first term survives,
and we obtain 
\be
(V_e)_{CC}\equiv V_{CC} = \sqrt{2}\, G_F N_e\,. 
\label{Ve}
\ee
Analogously, one can find the NC contributions $V_{NC}$ to the 
matter-induced neutrino potentials. Since NC interaction are flavour 
independent, these contributions are the same for neutrinos of all three 
flavours. In an electrically neutral medium, the number densities of protons 
and electrons coincide, and the corresponding contributions to $V_{NC}$ 
cancel. The direct calculation of the contribution due to the NC scattering 
of neutrinos off neutrons gives $(V_a)_{NC}=-G_F N_n/\sqrt{2}$, where 
$N_n$ is the neutron number density. Together with eq.~(\ref{Ve}) this
gives 
\be
V_e = \sqrt{2}\, G_F\,\left(N_e-\frac{N_n}{2}\right)\,,\quad\quad 
V_\mu = V_\tau=\sqrt{2}\, G_F\,\left(-\frac{N_n}{2}\right)\,.
\label{Va}
\ee
For antineutrinos, one has to replace $V_a \to -V_a$. 

Let us now consider the evolution of a system of oscillating neutrinos in 
matter. In vacuum, the evolution is most easily followed in the mass
eigenstate basis. In matter it is more convenient to do that in the 
flavour basis because the effective potentials of neutrinos are diagonal
in this basis. Consider first the two-flavour case. As usual, we write 
$\nu_{fl}=U\nu_m$ where $\nu_{fl}$ and $\nu_m$ are two-component vectors
of neutrino fields in the flavour and mass eigenstate bases and the matrix $U$ 
is given by eq. (\ref{U2}). In the absence of matter, the evolution equation 
in the mass eigenstate basis is $i(d/dt)|\nu_m\rangle=H_m|\nu_m\rangle$, 
where $H_m=diag(E_1,\,E_2)$. 
This gives the 
evolution equation in the flavour basis: 
$i(d/dt)|\nu_{fl}\rangle=H_{fl}|\nu_{fl}\rangle = 
U H_m U^\dagger |\nu_{fl}\rangle$. 
For relativistic neutrinos $E_i\simeq p+m_i^2/2E$, and we thus obtain 
\be
i\frac{d}{dt}\left(\begin{array}{c}
\nu_e \\
\nu_\mu   
\end{array} \right)=
\left(\begin{array}{cc}
\left(p+\frac{m_1^2+m_2^2}{4E}\right)-\frac{\Delta m^2}{4E}\cos 2\theta_0 & 
\frac{\Delta m^2}{4E}\sin 2\theta_0 \\
\frac{\Delta m^2}{4E}\sin 2\theta_0 &
\left(p+\frac{m_1^2+m_2^2}{4E}\right)+\frac{\Delta m^2}{4E}\cos 2\theta_0
\end{array} \right)
\left(\begin{array}{c}
\nu_{e} \\
\nu_{\mu}   
\end{array} \right)\,. 
\label{ev1}
\ee
Here $\nu_e$ and $\nu_\mu$ stand for time dependent amplitudes of
finding the electron and muon neutrino respectively. The expressions in
the brackets in the diagonal elements of the effective Hamiltonian in 
eq. (\ref{ev1}) coincide.   
They can only modify the common phase of the neutrino states and therefore
have no effect on neutrino oscillations which depend on the phase differences. 
For this reason one can omit these terms. The evolution equation describing 
neutrino oscillations in vacuum in the flavour basis then takes the form  
\be
i\frac{d}{dt}\left(\begin{array}{c}
\nu_e \\
\nu_\mu   
\end{array} \right)=
\left(\begin{array}{cc}
-\frac{\Delta m^2}{4E}\cos 2\theta_0 & 
\frac{\Delta m^2}{4E}\sin 2\theta_0 \\
\frac{\Delta m^2}{4E}\sin 2\theta_0 &
\frac{\Delta m^2}{4E}\cos 2\theta_0
\end{array} \right)
\left(\begin{array}{c}
\nu_{e} \\
\nu_{\mu}   
\end{array} \right)\,. 
\label{ev2}
\ee

We now proceed to derive the neutrino evolution equation in matter. To do 
that, one has to add the matter-induced potentials $V_e$ and $V_\mu$ to the 
diagonal elements of the effective Hamiltonian $H_{fl}$ in eq. (\ref{ev2}). 
Notice that $V_e$ and $V_\mu$ contain a common term due to NC interactions. 
As we already know, such common terms in the diagonal elements are of no 
consequence for neutrino oscillations; we can therefore omit them 
\footnote{The NC contribution does, however, affect the oscillations between 
the usual and ``sterile'' (electroweak singlet) neutrinos in matter.}. 
This gives 
\be
i\frac{d}{dt}\left(\begin{array}{c}
\nu_e \\
\nu_\mu   
\end{array} \right)=
\left(\begin{array}{lc}
-\frac{\Delta m^2}{4E}\cos 2\theta_0 + \sqrt{2}\, G_F N_e ~~ & 
\frac{\Delta m^2}{4E}\sin 2\theta_0 \\
~~~\frac{\Delta m^2}{4E}\sin 2\theta_0 &
\frac{\Delta m^2}{4E}\cos 2\theta_0
\end{array} \right)
\left(\begin{array}{c}
\nu_{e} \\
\nu_{\mu}   
\end{array} \right)\,. 
\label{ev3}
\ee
This is the evolution equation which describes $\nu_e \leftrightarrow \nu_\mu$ 
oscillations in matter. The equation for $\nu_e \leftrightarrow \nu_\tau$ 
oscillations has the same form. 
In the two-flavour 
approximation, $\nu_\mu \leftrightarrow \nu_\tau$ oscillations are not
modified in matter since $V_\mu=V_\tau$; however, in the full 3-flavour 
framework matter does influence the $\nu_\mu \leftrightarrow \nu_\tau$ 
oscillations because of the mixing with $\nu_e$, see sec. \ref{3flmatt}.  

\subsection{Constant density case 
\label{const}}
Let us now consider the evolution equation (\ref{ev3}). In general, the 
electron number density $N_e$ depends on the coordinate along the 
neutrino trajectory or, in our description in eq.~(\ref{ev3}), on $t$. 
We shall first consider a simple case of constant matter density and
chemical composition, i.e. $N_e=const$. Diagonalization of the effective 
Hamiltonian in (\ref{ev3}) gives the following neutrino eigenstates 
in matter: 
\begin{eqnarray}
\nu_A &=& \nu_e \,\cos\theta + \nu_\mu \, \sin\theta 
\,, \nonumber \\
\nu_B &=& \!\!\!\! -\nu_e \,\sin\theta  + \nu_\mu \,\cos\theta \,, 
\label{eigenst}
\end{eqnarray}
where the mixing angle $\theta$ is given by
\be
\tan 2\theta=\frac{\frac{\Delta m^2}{2E}\sin 2\theta_0}{\frac{\Delta
m^2}{2E}\cos 2\theta_0-\sqrt{2}\,G_F \,N_e}\,.
\label{mix3}
\ee
It is different from the vacuum mixing angle $\theta_0$ and therefore the
matter eigenstates $\nu_A$ and $\nu_B$ do not coincide with mass 
eigenstates $\nu_1$ and $\nu_2$. 
The difference of neutrino eigenenergies in matter is 
\be
E_A-E_B=\sqrt{\left(\frac{\Delta m^2}{2E}\cos 2\theta_0-\sqrt{2}\,G_F\,N_e 
\right)^2+\left(\frac{\Delta m^2}{2E}\right)^2 \sin^2 2\theta_0}\;. 
\label{EAEB}
\ee
It is now easy to find the probability of $\nu_e\leftrightarrow \nu_\mu$ 
oscillations in matter: 
\be
P(\nu_e\to\nu_\mu ;\,L)=
\sin^2 2\theta\,\sin^2 \left(\pi \frac{L}{l_{m}}\right)\,,
\label{pr1}
\ee
where 
\be
l_m=\frac{2\pi}{E_A-E_B}=\frac{2\pi}{\sqrt{\left(\frac{\Delta m^2}{2E}
\cos 2\theta_0-\sqrt{2}\,G_F\,N_e \right)^2+\left(\frac{\Delta m^2}{2E}
\right)^2 \sin^2 2\theta_0}}\,. 
\label{lm}
\ee
It has exactly the same form as the probability of oscillations in vacuum 
(\ref{prob2}), except that the vacuum mixing angle $\theta_0$ and oscillation 
length $\l_{osc}$ are replaced by those in matter, $\theta$ and $\l_m$. 
In the limit of zero matter density $\theta=\theta_0$, $l_m=l_{osc}$, and
the vacuum oscillation probability is recovered. 

The oscillation amplitude 
\be
\sin^2 2\theta =\frac{\left(\frac{\Delta m^2}{2E}\right)^2\sin^2 2\theta_0}
{\left(\frac{\Delta m^2}{2E}\cos 2\theta_0-\sqrt{2}\,G_F \,N_e\right)^2
+\left(\frac{\Delta m^2}{2E}\right)^2\sin^2 2\theta_0}\, 
\label{amp}
\ee
has a typical resonance form, with the maximum value $\sin^2 2\theta=1$ 
achieved when the condition 
\be
\sqrt{2}\,G_F\,N_e =\frac{\Delta m^2}{2E}\cos 2\theta_0
\label{res1}
\ee
is satisfied. It is called the MSW resonance condition. 
{}From (\ref{mix3}) or (\ref{amp}) it follows that when (\ref{res1}) is 
fulfilled, mixing in matter is maximal ($\theta=45^\circ$), independently 
from the vacuum mixing angle $\theta_0$. Thus, the probability of neutrino 
flavour transition in matter can be large even if the vacuum mixing angle is 
very small! 

For the resonance enhancement of neutrino oscillations in matter to be 
possible, the r.h.s. of (\ref{res1}) must be positive:
\be
\Delta m^2 \cos 2\theta_0 = (m_2^2-m_1^2)(\cos^2\theta_0-\sin^2\theta_0)>0\,,
\label{cond3}
\ee
i.e. if $\nu_2$ is heavier than $\nu_1$, one needs $\cos^2 \theta_0 > 
\sin^2 \theta_0$, and vice versa. It follows from eq. (\ref{mix1}) 
that the condition (\ref{cond3}) is equivalent to the requirement that of 
the two mass eigenstates $\nu_1$ and $\nu_2$, the lower-mass one have a 
larger $\nu_e$ component. If one chooses the convention $\cos 2\theta_0>0$,
(as is usually done) then (\ref{cond3}) reduces to $\Delta m^2 \equiv 
\Delta m_{21}^2>0$. The resonance condition for antineutrinos is then 
$\Delta m^2<0$. Therefore, for a given sign of $\Delta m^2$, either
neutrinos or antineutrinos (but not both) can experience the resonantly 
enhanced oscillations in matter. 

\subsection{Adiabatic approximation 
\label{adiabatic}}
Let us now discuss the realistic case of matter of varying density. Typically, 
one deals with situations when a beam of non-monochromatic neutrinos (i.e.
neutrinos with some energy distribution) propagates in a medium with certain 
density profile. If $\Delta m^2$ is of the right order of magnitude, then for 
any value of the matter density (or a significant portion of them) there is a
value of neutrino energy for which the resonance condition (\ref{res1}) is 
satisfied. 
Conversely, every value (or a significant portion of the values) of neutrino 
energy $E$ 
``finds'' a value of the matter density for which the resonance condition 
(\ref{res1}) is satisfied. If the neutrino beam is monochromatic (as, e.g., 
in the case of the solar $^7$Be neutrinos), 
the resonance enhancement of neutrino oscillations is still possible if the
corresponding resonance density is within the density range of the matter 
in which neutrinos propagate. Thus, the MSW resonance condition does not 
involve any fine tuning. 

In general, for oscillations in a matter of an arbitrary non-uniform density, 
the evolution equation (\ref{ev3}) does not allow an analytic solution and 
has to be solved numerically. However, there is an important particular case 
in which one can get an illuminating approximate analytic solution. This
is the case of slowly (adiabatically) varying matter density. 

We shall start with a semi-quantitative discussion. Consider electron 
neutrinos born in a matter of a very high density, far above the MSW 
resonance one (e.g., in the core of the sun). We shall assume that the matter 
density is monotonically decreasing along the neutrino trajectory. From 
eq. (\ref{mix3}) it follows that the mixing angle in matter at the neutrino 
production point $\theta_i\simeq 90^\circ$, which means
that the neutrino mixing is strongly suppressed by matter. As neutrinos 
propagate towards the regions of smaller density, their mixing increases 
(mixing angle $\theta$ decreases); it becomes maximal at the resonance point, 
where $\theta=45^\circ$. As neutrinos propagate further towards smaller 
densities, their mixing angle continues decreasing; it reaches the value
$\theta_f=\theta_0$ at densities $N_e \ll (N_e)_{MSW}$, where $(N_e)_{MSW}$ 
is the resonance value of the electron number density given by 
eq. (\ref{res1}). 
{}From eq.~(\ref{eigenst}) it follows that at the production point, 
where $\theta\simeq 90^\circ$, the produced $\nu_e$ almost coincide with
the matter eigenstate neutrinos $\nu_B$. If the matter density changes
slowly enough (adiabatically) along the neutrino path, the neutrino system 
has enough time to ``adjust'' itself to the changing external conditions.
In this case the transitions between the matter eigenstates $\nu_A$ and 
$\nu_B$ are exponentially suppressed. This means that if the system was 
initially in the eigenstate $\nu_B$, it will remain in the same eigenstate. 
However, the flavour composition of this eigenstate changes as neutrino 
propagates in matter because the mixing angle $\theta$ that determines this
composition is the function of matter density. Since at the final point of
neutrino evolution $\theta\simeq \theta_0$, eq. (\ref{eigenst}) tells us 
that the matter eigenstate neutrino $\nu_B$ at this point has the component 
of originally produced $\nu_e$ with the weight $\sin^2 \theta_0$ and the 
component of $\nu_\mu$ with the weight $\cos^2 \theta_0$, i.e. the
transition probability is 
\be
P(\nu_e\to\nu_\mu)=\cos^2\theta_0\,.
\label{pr2}
\ee
This means that in the case of small vacuum mixing angle, one can have almost 
complete adiabatic conversion of $\nu_e$ to $\nu_\mu$! 

\begin{figure}[t] 
\hbox to \hsize{\hfil\epsfxsize=10cm\epsfbox{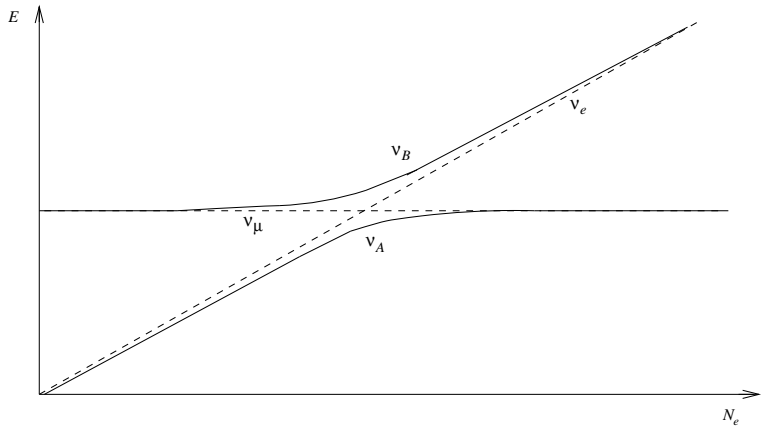}\hfil}
\caption{\small Neutrino energy levels in matter vs electron number
density $N_e$. Dashed line -- in the absence of mixing, solid line -- 
with mixing. }
\label{levcross} 
\end{figure}

This is illustrated by fig. \ref{levcross}  
which shows the energy levels of $\nu_A$ and $\nu_B$ along with those in 
the absence of mixing (i.e. of $\nu_e$ and $\nu_\mu$) as the function of
the electron number density. In the absence of mixing the energy levels cross 
at the MSW resonance point, but with nonvanishing mixing the levels ``repel'' 
each other, and the avoided level crossing results. If the probability of the 
transition between the two matter eigenstates is small, neutrinos produced as 
$\nu_e$ at high densities and propagating towards smaller densities follow 
the upper ($\nu_B$) branch and end up on the level that corresponds to 
$\nu_\mu$ at small $N_e$. This resonance conversion is similar to the well 
known Landau - Zener phenomenon in atomic and molecular physics. 

The expression (\ref{pr2}) for the conversion probability 
looks paradoxical: the smaller the vacuum mixing angle, the larger the 
probability that the initially produced $\nu_e$ will be converted into 
$\nu_\mu$ or vice versa. Does this mean that in the limit of vanishing
$\theta_0$ one can still have strong neutrino conversion? The answer, of
course, is no. The reason is simple: if $\theta_0$ becomes too small, the 
adiabaticity of the conversion gets broken, and 
eq. (\ref{pr2}) ceases to be valid. Similar situation takes place in the 
case of neutrino oscillations in a matter of constant density: if the MSW 
resonance condition is satisfied, the oscillation amplitude $\sin^2 2\theta_0
=1$, no matter how small $\theta_0$; however, in the limit $\theta_0 \to 0$ 
the phase of the second $\sin^2$ factor in eq. (\ref{pr1}) vanishes, and
no oscillations occur. 

We shall now turn to a more quantitative description of the neutrino 
conversion in the adiabatic regime, which will also allow us to establish
the domain of applicability of the adiabatic approximation. At any instant of 
time, the effective Hamiltonian $H_{fl}(t)$, which is given by the matrix on 
the r.h.s. of  eq. (\ref{ev3}), can be diagonalized by a unitary 
transformation 
\be
\nu_{fl}=\tilde{U}(t)\nu\,,\quad\quad
\tilde{U}(t)^\dagger H_{fl}(t) \tilde{U}(t)=\tilde{H}_d(t)=diag(E_A(t),\,
E_B(t))\,.
\label{diag2}
\ee
Here $\nu=(\nu_A(t),\,\nu_B(t))^T$ is the vector of the instantaneous 
matter eigenstates, $E_A(t)$ and $E_B(t)$ are instantaneous eigenvalues of 
$H_{fl}(t)$, and the matrix $\tilde{U}(t)$ has the form (\ref{U2}) except 
that the vacuum mixing angle $\theta_0$ has to be replaced by the mixing 
angle $\theta(t)$ given by eq.~(\ref{mix3}) with $N_e=N_e(t)$. The
evolution equation in the basis of the instantaneous eigenstates can
therefore be written as $i(d/dt)\nu=[\tilde{H}_d-i\tilde{U}^\dagger
(d\tilde{U}/dt)]  
\nu$, or 
\be
i\frac{d}{dt}\left(\begin{array}{c}
\nu_A \\
\nu_B   
\end{array} \right)=
\left(\begin{array}{cc}
E_A(t)  & -i\dot{\theta}(t) \\
i\dot{\theta}(t) & E_B(t)
\end{array} \right)
\left(\begin{array}{c}
\nu_A \\
\nu_B   
\end{array} \right)\,,
\label{ev4}
\ee
where $\dot{\theta}\equiv d\theta/dt$. Notice that the effective
Hamiltonian in this basis is not diagonal since the mixing angle $\theta$ 
is not constant, i.e. the matter eigenstate basis changes with time. 
If the off-diagonal terms are small, i.e. $|\dot{\theta}| \ll
|E_A-E_B|$,
%
the transitions between the instantaneous eigenstates $\nu_A$ and $\nu_B$ 
are suppressed. This corresponds to the adiabatic approximation. The
adiabaticity condition can be written as 
\be
\gamma^{-1}\equiv \frac{2|\dot{\theta}|}{|E_A-E_B|}=\frac{\sin 2\theta_0
\frac{\Delta m^2}{2E}}
{|E_A-E_B|^{3}}\,
|\dot{V}_{CC}| \ll 1\,, 
\label{adiab1}
\ee
where $E_A-E_B$ and $V_{CC}$ are given by eqs. (\ref{EAEB}) and (\ref{Ve}) 
respectively. The parameter $\gamma$ is called the adiabaticity parameter.
In the adiabatic limit, the effective Hamiltonian in (\ref{ev4}) is diagonal, 
and time evolution of the matter eigenstates is very simple -- they just 
receive phase factors. 

Suppose that at a time $t=t_i$ the electron neutrino was born:
\be
\nu(t_i)=\nu_e=\cos\theta_i \,\nu_A+\sin\theta_i\, \nu_B\,.
\label{init}
\ee
Then, in the adiabatic approximation, at a time $t_f$ we have 
\be
\nu(t_f)=
\cos\theta_i\, e^{-i\int_{t_i}^{t_f} E_A(t')dt'}\,\nu_A +\sin\theta_i\,
e^{-i\int_{t_i}^{t_f} E_B(t')dt'}\,\nu_B\,.
\label{fin}
\ee
Taking into account that at $t=t_f$ the mixing angle $\theta(t_f)\equiv 
\theta_f$ is different from $\theta_i$, one then finds 
\be
P(\nu_e\to\nu_\mu)=\frac{1}{2}-\frac{1}{2}\cos 2\theta_i\cos 2\theta_f-
\frac{1}{2}\sin 2\theta_i\sin 2\theta_f \cos \Phi\,,
\label{pr3}
\ee
where
\be
\Phi=\int_{t_i}^{t_f}(E_A-E_B) dt'\,.
\label{phi}
\ee
The second term in eq. (\ref{pr3}) is a smooth function of $t_f$, while
the third term oscillates with time. If the matter density at the neutrino 
production point is far above the MSW resonance one, then $\sin 2\theta_i 
\simeq 0$ and the third term is strongly suppressed. In this case the 
non-oscillatory neutrino conversion takes place, with the probability 
$P(\nu_e\to\nu_\mu)=\cos^2\theta_f$. If, in addition, the final point of 
neutrino evolution corresponds to a very low matter density, 
$\theta_f=\theta_0$ and eq. (\ref{pr2}) is recovered. Notice that 
eq. (\ref{pr3}) is symmetric with respect to the interchange of the initial 
and final evolution points $t_i \leftrightarrow t_f$. 

One can readily generalize this analysis to include possible transitions 
between the matter eigenstates $\nu_A$ and $\nu_B$ due to the violation of
the adiabaticity. Let $P'$ be the probability that the $\nu_A\leftrightarrow 
\nu_B$ transitions have occured in the course of the evolution of the
neutrino system (``hopping probability''). Then 
\be
\overline{P(\nu_e\to\nu_\mu)}\simeq \frac{1}{2}-\frac{1}{2}\cos
2\theta_i\cos 2\theta_f
(1-2P')\,,
\label{pr4}
\ee
where we have omitted oscillating terms which average to zero. In the case 
of very small vacuum mixing angle and $(N_e)_f\ll (N_E)_{MSW} \ll (N_e)_i$ 
(or $(N_e)_i\ll (N_E)_{MSW}\ll (N_e)_f$), it reduces to a very simple 
expression 
\be
\overline{P(\nu_e\to\nu_\mu)}\simeq 1-P'\,.
\label{pr5}
\ee
In the Landau-Zener approximation the hopping probability $P'$ can be 
written as
\be
P'\simeq e^{-\frac{\pi}{2}\gamma_r}\,,
\label{hop}
\ee
where $\gamma_r$ is the adiabaticity parameter 
taken at the MSW resonance point. In the adiabatic limit, $\gamma_r\gg 1$, 
one has $P'\simeq 0$ and eq. (\ref{pr3}) is recovered. In the non-adiabatic 
limit $\gamma_r \ll 1$, $P'\simeq 1$, and the survival and transition 
probabilities interchange; in that case the transition probability is small 
when $\theta_0$ is small. It should be noted that in the extreme non-adiabatic 
limit eq. (\ref{hop}) for the hopping probability is not quite correct and 
more accurate expressions should be used. 
It is important that the transition probability can be quite sizable even 
if the adiabaticity parameter is not very large: for $\gamma_r\simeq 1$ it
can be as large as about 80\%, and one can, e.g., consider the condition 
$\gamma_r>3$ (which can result in as large transition probabilities as 
99\%) as the adiabaticity criterion. 

Let us now discuss the adiabaticity condition (\ref{adiab1}) in more
detail. Since the difference of the eigenenergies reaches its minimum 
equal to $(\Delta m^2/2E)\sin 2\theta_0$ at the MSW resonance (see eqs. 
(\ref{EAEB}) and (\ref{res1}) and fig. \ref{levcross}), the adiabaticity is 
worst at the resonance. If the adiabaticity condition at the resonance is 
satisfied, it will then also be satisfied away from it 
\footnote{Assuming that $|\dot{\theta}|$ does not have strong maxima
outside the resonance region.}.  
The adiabaticity parameter at the MSW resonance is 
\be
\gamma_r=\left(\frac{\Delta m^2}{2E}\sin 2\theta_0\right)^2\frac{1}
{|\dot{V}_{CC}|_{res}}=\frac{\sin^2 2\theta_0}{\cos 2\theta_0}
\frac{\Delta m^2}{2E}\,L_\rho \,,
\label{adiab2}
\ee
where $L_\rho=|\dot{V}_{CC}/V_{CC}|_{res}^{-1}= |\dot{N}_{e}/
N_{e}|_{res}^{-1}$ is the density scale height at the resonance, i.e. the 
characteristic distance over which the electron number density varies 
significantly in the resonance region.

The adiabaticity condition has a simple physical meaning. 
Let us define the resonance width at half height $\Delta r$ as the spatial
width of the region where the amplitude of neutrino oscillations in matter
$\sin^2 2\theta\ge 1/2$. It is easy to obtain  
\be
\Delta r \simeq 2 \tan 2\theta_0\, L_\rho
\label{deltar}
\ee
{}From (\ref{lm}) we find that the oscillation length at the resonance is
$(l_m)_{res}=2\pi/|E_A-E_B|_{res}=(4\pi E/\Delta m^2)/\sin 2\theta_0$. 
Therefore the adiabaticity parameter at the resonance (\ref{adiab2}) can be 
written as 
\be
\gamma_r =\pi\frac{\Delta r}{(l_m)_{res}}\,,
\label{adiab3}
\ee
i.e. the adiabaticity condition $\gamma_r > 3$ is just the condition that 
at least one oscillation length fit into the resonance region. 

As follows from (\ref{res1}) and (\ref{adiab2}), both the MSW resonance
condition and the adiabaticity parameter at the resonance $\gamma_r$ depend 
on neutrino energy. Therefore the efficiency of the matter-enhanced neutrino 
flavour conversion is energy dependent. This should have important 
consequences, for example, for solar or supernova neutrinos, leading to
characteristic distortions of the energy spectra of detected neutrinos. 

\begin{figure}[htb] 
\hbox to \hsize{\hfil\epsfxsize=12cm\epsfbox{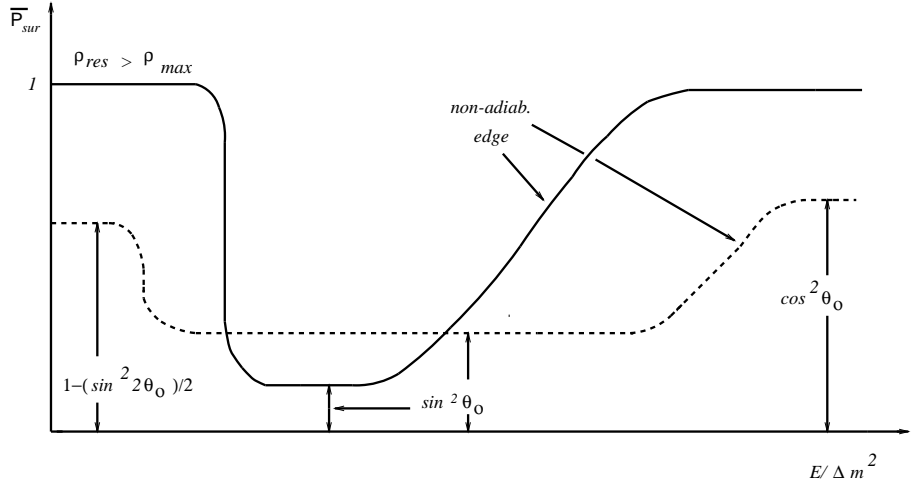}\hfil}
\caption{\small Averaged survival probability vs $E/\Delta m^2$ in 
the case of neutrino oscillations in a matter of monotonically decreasing  
density. Solid line - small $\theta_0$, dashed line -- large $\theta_0$. }
\label{bathtub} 
\end{figure}

For neutrino oscillations in a matter of density which decreases monotonically 
along the neutrino path (e.g., for neutrino oscillations in the sun), the 
average neutrino survival probability $P_S$ is schematically 
shown in fig. (\ref{bathtub}). It has a typical ``bathtub'' 
shape. In the case of small vacuum mixing angle, the low energy edge of the 
curve corresponds to no suppression (resonance density exceeds the maximal 
available matter density $(N_e)_{max}$ -- there is no MSW resonance). The
steep decrease of $P_S$ corresponds to the value of $E/\Delta m^2$ at which 
the MSW resonance condition (\ref{res1}) is satisfied at $N_e=(N_e)_{max}$. 
The bottom of the suppression bathtub corresponds to the adiabatic regime,
whereas the slow increase of $P_S$ at higher energies is due to the 
loss of adiabaticity. The high energy edge of the suppression curve 
corresponds to the non-adiabatic regime. In the case of larger vacuum
mixing angles, the suppression curve shows a similar behavior, but with 
smaller amplitudes of changes of the survival probability $P_S$ and a wider 
adiabatic region.

It should be noted that the fulfillment of the 
adiabaticity condition (\ref{adiab1}) is not, in general, sufficient for
the strong flavour conversion to take place: in addition, one has to
require that the total width (column density) $d$ of the matter traversed
by neutrinos be large enough. If matter density changes slowly along the
neutrino trajectory, the scale height $L_\rho$ is large, and so the 
adiabaticity parameter $\gamma_r$ and the resonance width $\Delta r$
given by eqs.~(\ref{adiab2}) and (\ref{deltar}) can formally be very
large, too; however, in reality the resonance width cannot 
exceed the total width of the matter slab traversed by neutrinos. For the 
transition probability to be appreciable, the column density $d$ must 
satisfy \cite{SmCec} 
\be
d\equiv \int_{t_i}^{t_f} N_e\, dt'\aprge \frac{1}{G_F \tan 2\theta_0}\,.
\label{cond4}
\ee

\noindent
{\it Problem 16. Derive eq. (\ref{pr3}). }  

\noindent
{\it Problem 17. Derive eq. (\ref{deltar}) for the MSW resonance width. }  

\noindent
{\it Problem 18. Assuming that 
the $\nu_A \leftrightarrow \nu_B$ transitions occur (with probability $P'$) 
in a narrow region around the MSW resonance and that the change of the
oscillation phases in that region is small, derive the non-adiabatic 
analog of eq. (\ref{pr3}). Show that upon averaging of the oscillating 
terms eq. (\ref{pr4}) is obtained.}  

\subsection{3 flavour case
\label{3flmatt}}
The evolution equation describing neutrino oscillations in matter in the case 
of three neutrino flavours $\nu_e$, $\nu_\mu$ and $\nu_\tau$ is
\be
i\frac{d}{dt}\left(\begin{array}{c}
\nu_e \\
\nu_\mu \\
\nu_\tau   
\end{array} \right)=
\left[ \frac{1}{2E} U
\left(\begin{array}{ccc}
m_1^2  &   0    & 0     \\
0      & m_2^2  & 0     \\
0      &   0    & m_3^2 
\end{array} \right) U^\dagger  +
\left(\begin{array}{ccc}
V_{CC}   &  ~ 0    &~~~ 0     \\
0                     &  ~ 0    &~~~ 0     \\
0                     &  ~ 0    &~~~ 0  
\end{array} \right) \right]
\left(\begin{array}{c}
\nu_{e} \\
\nu_{\mu}\\   
\nu_\tau
\end{array} \right)\,. 
\label{3f1}
\ee
Here $U$ is the 3-flavour vacuum mixing matrix defined in eq. (\ref{mix2}) 
or (\ref{U3}), and $V_{CC}$ was defined in eq. (\ref{Ve}). In general, it
is rather difficult to study this equation analytically, even in the 
adiabatic approximation. There are, however, several important particular
cases when the evolution of the 3-flavour neutrino system can be 
approximately expressed through the 2-flavour amplitudes. 
One such case is the case of the hierarchical mass squared differences 
(\ref{hier}). It is of particular interest because this is exactly the
type of the hierarchy which is suggested by the solar and atmospheric
neutrino experiments. We shall now discuss this case in some detail.

The mixing matrix $U$ in the three-flavour case can in general be written
as (see, e.g., \cite{Giunti})
\be 
U=V_{23} W_{13} V_{12} D\equiv V D\,,
\label{Unew}
\ee
where 
\be
V_{12}=\left(\begin{array}{ccc}
c_{12}    &  s_{12}  & 0     \\
-s_{12}   &  c_{12}    & 0     \\
0         &    0       & 1
\end{array} \right)\,,\quad
W_{13}=\left(\begin{array}{ccc}
c_{13}                  &    0       & s_{13}\,e^{-i\delta} \\
0                       &    1       & 0       \\
-s_{13}\,e^{i\delta}    &    0       & c_{13}
\end{array} \right)\,,\quad
V_{23}=\left(\begin{array}{ccc}
1         &    0       & 0        \\
0         &  c_{23}    & s_{23}     \\
0         & -s_{23}    & c_{23}
\end{array} \right)\,,\quad
\ee
and $D=diag(e^{-i\varphi_1},\, 1,\,e^{-i\varphi_2})$. The phase $\delta$ in 
$W_{13}$ is the usual Dirac-type CP violating phase, whereas the phases 
$\varphi_1$ and $\varphi_2$ are present only in the Majorana case. It 
immediately follows from eq. (\ref{3f1}) that the Majorana phases have no 
effect on neutrino oscillations, and therefore one can omit the factor $D$
and write $U=V$. In this case the matrix $U$ defined in (\ref{Unew}) coincides 
with the one in (\ref{U3}).  

We shall first consider the case (\ref{cond1}) which is relevant for
atmospheric, reactor and accelerator neutrino 
experiments. 
Subtracting the term $m_1^2$ from all the diagonal elements of the matrix 
$M^2=diag(m_1^2,\,m_2^2,\,m_3^2)$ and neglecting $\Delta m_{21}^2$
compared to $\Delta m_{31}^2$ we get $M^2 \to \tilde{M}^2= diag(0,\,0,\,
\Delta m_{31}^2)$. The new matrix $\tilde{M}^2$ commutes with the
rotation matrix $V_{12}$, so the evolution equation (\ref{3f1}) can be 
rewritten as $i(d/dt)\nu=[V_{23}W_{13}(\tilde{M}^2/2E) W_{13}^\dagger 
V_{23}^\dagger +V]\nu$, where $V$ is the second matrix on the r.h.s. of 
eq. (\ref{3f1}): $V=diag(V_{CC},\,0,\,0)$. Since $V$ commutes with $V_{23}$, 
it is convenient to go into a new basis defined through $\nu=V_{23} 
\tilde{\nu}$. The evolution equation in this basis is 
\be
i\frac{d}{dt}\left(\begin{array}{c}
\tilde{\nu}_1 \\
\tilde{\nu}_2 \\
\tilde{\nu}_3   
\end{array} \right)=
\left(\begin{array}{ccc}
\frac{\Delta m_{31}^2}{2E}s_{13}^2+V_{CC} &  0 
& \frac{\Delta m_{31}^2}{2E} s_{13} c_{13}    \\
0      & 0  & 0     \\
\frac{\Delta m_{31}^2}{2E} s_{13} c_{13} &  0 & 
\frac{\Delta m_{31}^2}{2E} c_{13}^2    \\
\end{array} \right) 
\left(\begin{array}{c}
\tilde{\nu}_1 \\
\tilde{\nu}_2 \\
\tilde{\nu}_3   
\end{array} \right)\,,
\label{3f2}
\ee
i.e. the problem is reduced to a 2-flavour problem of oscillations between 
$\tilde{\nu}_1=\nu_e$ and $\tilde{\nu}_3=s_{23}\,\nu_\mu+c_{23}\,\nu_\tau$. 
These oscillations are governed by the large mass squared difference 
$\Delta m_{31}^2$ and mixing angle $\theta_{13}$. The CP-violating phase 
$\delta$ is unobservable in the limit $\Delta m_{21}^2=0$ and so we
have omitted it. It is convenient to 
subtract the term $(\Delta m_{31}^2/2E+V_{CC})/2$ from all the diagonal
elements of the effective Hamiltonian in eq. (\ref{3f2}), so as to make
the Hamiltonian of the two-flavour system $(\tilde{\nu}_1,\,\tilde{\nu}_3)$  
traceless. The two-flavour evolution equation then takes the form 
\be
i\frac{d}{dt}\left(\begin{array}{c}
\tilde{\nu}_1 \\
\tilde{\nu}_3   
\end{array} \right)=
\left(\begin{array}{ll}
-\frac{\Delta m_{31}^2}{4E}\cos 2\theta_{13}+\frac{V_{CC}}{2}  
& ~~\frac{\Delta m_{31}^2}{4E} \sin 2\theta_{13}  \\
~~\frac{\Delta m_{31}^2}{4E} \sin 2\theta_{13} &   
~~\frac{\Delta m_{31}^2}{4E} \cos 2\theta_{13}-\frac{V_{CC}}{2}   \\
\end{array} \right) 
\left(\begin{array}{c}
\tilde{\nu}_1 \\
\tilde{\nu}_3   
\end{array} \right)\,.
\label{3f3}
\ee
For an arbitrary initial 
state $\tilde{\nu}(t_0)$ the solution at a time $t$ can be written as 
$\tilde{\nu}(t)=S(t,t_0) \tilde{\nu}(0)$, where $S(t,t_0)$ is the evolution 
matrix that corresponds to the effective Hamiltonian in eq. (\ref{3f3}). 
The unitary $2\times 2$ matrix $S(t,t_0)$ can be parametrized as 
\be
S(t,t_0)=Z-i\mbox{\boldmath $\tau$}{\bf W}\,,
\label{S}
\ee
where $\mbox{\boldmath $\tau$}$ are the Pauli matrices in the $(\tilde{\nu}_1, 
\tilde{\nu}_3)$ space, and the real parameters $Z$ and ${\bf W}=(W_1,\,W_2, 
\, W_3)$ satisfy $Z^2+{\bf W}^2=1$. The matrix $S(t,t_0)$ can be found for 
any matter density profile using the standard techniques developed for the 
2-flavour neutrino oscillations. In particular, for long-baseline accelerator 
experiments, neutrino path goes through the mantle of the earth where to 
a very good approximation the matter density can be considered to be constant. 
In this case the parameters $Z$ and ${\bf W}$ can be trivially found. For the 
atmospheric neutrino oscillations, in the case of neutrinos traversing both 
the mantle and the core of the earth, simple analytic expressions for the 
parameters $Z$ and ${\bf W}$ can be obtained in the step-function 
approximation to the density profile of the earth \cite{Akh2}. With known $Z$ 
and ${\bf W}$ it is easy to calculate the probabilities of neutrino 
oscillations:
\be
\!\!\!\! P(\nu_e\leftrightarrow \nu_\mu)=s_{23}^2\,P_2 \,,\quad\quad\quad
P(\nu_e\leftrightarrow\nu_\tau)=c_{23}^2\,P_2\,, \quad 
\label{3f4}
\ee
\be
P(\nu_\mu\to\nu_\tau)=
s_{23}^2\,c_{23}^2[2-P_2-2(Z\cos\Phi+W_3\sin\Phi)]\,,
\label{3f5}
\ee
where 
\be
\Phi=\frac{1}{2}\int_{t_0}^{t}\left(\frac{\Delta m_{31}^2}{2E}+V_{CC}\right) 
dt'\,, \quad\quad P_2=W_1^2+W_2^2\,,
\label{3f6}
\ee
$P_2$ being the probability of the $\tilde{\nu}_1\leftrightarrow 
\tilde{\nu}_3$ transitions described by eq. (\ref{3f3}). It is not difficult 
to take into account corrections due to nonvanishing $\Delta m_{21}^2$ 
\cite{ADLS} -- one just has to add the term $(-\Delta m_{21}^2/E)$ to the
integrand in eq. (\ref{3f6}). Eqs. (\ref{3f4}) - (\ref{3f6}) generalize the 
results (\ref{prob5a}) - (\ref{prob5c}) to the case of neutrino oscillations 
in matter.

\noindent
{\it Problem 19. Derive eqs. (\ref{3f4}) - (\ref{3f6}). For the case of
neutrino oscillations in a matter of constant density, find explicit
formulas for the parameters $Z$ and ${\bf W}$. Show that in the limit of 
vanishing matter density eqs. (\ref{prob5a}) - (\ref{prob5c}) are recovered.} 

In the limiting case (\ref{cond2}) which is relevant for the solar neutrino
oscillations, and assuming, as before, the hierarchy (\ref{hier}), one
finds that the $\nu_e$ survival probability is again given by 
eq. (\ref{prob7}). However in the presence of matter the two-flavour survival
probability $P$ in this expression is no longer given by eq. (\ref{prob8});  
it has to be calculated from the 2-flavour evolution equation in matter
of the type (\ref{ev3}) governed by the mass squared difference 
$\Delta m^2=\Delta m_{21}^2$ and mixing angle $\theta_0=\theta_{12}$,
except that the matter-induced potential for neutrinos $V_{CC}$ must be
replaced by the effective one $V_{eff}=c_{13}^2\,V_{CC}$ \cite{Lim}.

\noindent
{\it Problem 20. Assuming that $\Delta m_{31}^2/2E\simeq \Delta
m_{32}^2/2E\gg \Delta m_{21}^2/2E\sim V_{CC}$, $(\Delta m_{31}^2/2E)L \gg 1$, 
derive eq.~(\ref{prob7}) in the case of neutrino oscillations in matter and 
show that the survival probability $P$ in it has to be found from the
2-flavour evolution equation with $V_{eff}=c_{13}^2\,V_{CC}$. Hint: consider 
neutrino evolution in the basis $\nu'=V_{13}^\dagger V_{23}^\dagger \nu$
and take into account that the evolution of the state $\nu_3'$ decouples. }

{\it Problem 21. Derive expressions for 3-flavour oscillation probabilities  
in terms of the relevant 2-flavour evolution matrix in the case $U_{e3}=0$. 
Find explicit formulas for the case of the oscillations in a matter of
constant density. Show that in the limit of vanishing density 
eqs. (\ref{prob9})-(\ref{prob11}) are recovered.}

As we have already discussed, matter affects neutrino and antineutrino 
oscillations differently: for $\cos 2\theta_0>0$ and $\Delta m^2>0$, it can 
enhance oscillations of neutrinos but always suppresses oscillations of 
antineutrinos, whereas in the case $\Delta m^2<0$ the situation is opposite. 
This means that even in the absence of the CP-violating 
phase $\delta$ in the mixing matrix (\ref{U3}), the probabilities of 
neutrino oscillations are different from those of antineutrino oscillations, 
i.e. matter induces the CP violating effects. This is because the usual matter 
is itself CP (and also C and CPT) asymmetric since it consists of particles 
and not of antiparticles, or in general not of equal number of particles
and antiparticles. This CP violation is, however, ``macroscopic'' rather 
than microscopic; although it may be useful for experimental searches of 
matter effects on neutrino oscillations, it does not say anything about the 
CP violation in the leptonic sector at the fundamental level. 
Thus, matter effects can mimic the true CP violation and therefore may 
make it difficult to disentangle the genuine CP violation from the macroscopic 
one in the long baseline neutrino oscillation experiments \cite{CP}.  

There is, however, a circumstance that may in principle facilitate 
experimental searches of the genuine CP violation in the leptonic sector. In 
vacuum, CP-odd and T-odd asymmetries of the neutrino oscillation probabilities 
coincide; they both depend on the same CP- and T-violating parameter $\delta$. 
For neutrino oscillations in a medium, CP is violated by the matter but T
in general is not: it is only violated if the matter density profile is 
T-asymmetric, i.e. asymmetric with respect to the interchange of the
positions of the neutrino source and detector. A matter with T-symmetric 
density profile (and in particular, a matter of constant density) does not 
induce any T violation, and an experimental observation of a T violating 
asymmetry $P(\nu_a\to\nu_b)-P(\nu_b\to\nu_a)$ in such a matter would
directly measure the fundamental T (and CP) violating phase $\delta$.  

\vspace*{0.2cm}
Matter effects on propagating neutrinos are not limited to the MSW effect 
discussed in this section.  Matter can also modify significantly neutrino
decays in the case of unstable neutrinos \cite{decay}, and neutrino spin and 
spin-flavour precession in magnetic fields \cite{prec}. In particular, the
spin-flavour precession (simultaneous change of neutrino chirality and flavour 
due to interaction of neutrino flavour-off-diagonal magnetic moments with
magnetic fields) can be resonantly enhanced in matter; this phenomenon 
is similar to the MSW effect. 
Neutrino oscillations in matter can also be amplified even if the neutrino 
mixing angles both in vacuum {\it and} in matter are small, i.e. if there
is no MSW effect (no level crossing). This is possible if the change of matter 
density along the neutrino path and the change of the oscillation phase are 
``synchronized'' in a certain way \cite{ETC,Akh3,LS,ADLS}. In this case the 
oscillation probability can be strongly enhanced due to a cumulative build-up 
effect \cite{ETC,Akh3}. This phenomenon is of the same nature as the 
parametric resonance in mechanical or electromagnetic oscillations.  
It has been discovered recently that this parametric enhancement of neutrino 
oscillations can lead to observable effects in the oscillations of atmospheric 
\cite{LS,ADLS} and solar \cite{P,Akh2} neutrinos in the earth. 

\section{Double beta decay}
\label{2beta}

\begin{figure}[h]
\begin{center} 
\begin{picture}(349,80)(-15,-40)
\ArrowLine(-10,25)(20,25)
\ArrowLine(-10,-85)(20,-85)
\Photon(20,25)(20, 0)3 4
\Text(15,12)[r]{\small $W$}
\Text(17,-110)[c]{$(a)$}
\ArrowLine(20,0)(50,10)
\ArrowLine(50,-10)(20,0)
\ArrowLine(50,-50)(20,-60)
\ArrowLine(20,-60)(50,-70)
\Photon(20,-60)(20,-85)3 4
\Text(15,-72)[r]{\small$W$}
\Text(55,10)[l]{\small$e_L$}
\Text(55,-73)[l]{\small$e_L$}
\Text(55,-49)[l]{\small$\nu_{eL}$}
\Text(55,-10)[l]{\small$\nu_{eL}$}
\Text(45,35)[r]{\small$u_L$}
\Text(50,-93)[r]{\small$u_L$}
\Text(-10,-93)[l]{\small$d_L$}
\Text(-10,35)[l]{\small$d_L$}
\ArrowLine(20,25)(50,25)
\ArrowLine(20,-85)(50,-85)
\ArrowLine(120,25)(150,25)
\ArrowLine(120,-85)(150,-85)
\Photon(150,25)(150, 0)3 4
\Text(155,12)[l]{\small$W$}
\Text(152,-110)[c]{$(b)$}
\ArrowLine(150,0)(180,0)
\ArrowLine(150,-30)(150,0)
\Text(150.5,-30)[c]{$\times$}
\ArrowLine(150,-30)(150,-60)
\ArrowLine(150,-60)(180,-60)
\Photon(150,-60)(150,-85)3 4
\Text(155,-72)[l]{\small$W$}
\Text(185,0)[l]{\small$e_L$}
\Text(185,-60)[l]{\small$e_L$}
\Text(155,-30)[l]{\small$m_L$}
\Text(145,-15)[r]{\small$\nu_{eL}$}
\Text(145,-45)[r]{\small$\nu_{eL}$}
\Text(175,35)[r]{\small$u_L$}
\Text(180,-93)[r]{\small$u_L$}
\Text(120,-93)[l]{\small$d_L$}
\Text(120,35)[l]{\small$d_L$}
\ArrowLine(150,25)(180,25)
\ArrowLine(150,-85)(180,-85)
\ArrowLine(250,25)(280,25)
\ArrowLine(250,-85)(280,-85)
\Photon(280,25)(280, 0)3 4
\Text(285,12)[l]{\small$W_L$}
\Text(282,-110)[c]{$(c)$}
\ArrowLine(280,0)(310,0) 
\ArrowLine(280,-25)(280,0)
\Text(280.9,-21)[c]{$\times$}
\ArrowLine(280,-40)(280,-25)
\Text(280.9,-42)[c]{$\times$}
\ArrowLine(280,-40)(280,-60)
\ArrowLine(280,-60)(310,-60)
\Photon(280,-60)(280,-85)3 4
\Text(285,-72)[l]{\small$W_R$}
\Text(315,0)[l]{\small$e_L$}
\Text(315,-60)[l]{\small$e_R$}
\Text(285,-21)[l]{\small$m_D$}
\Text(285,-42)[l]{\small$M_R$}
\Text(275,-10)[r]{\small$\nu_{eL}$}
\Text(275,-30)[r]{\small$\nu_{eR}$}
\Text(275,-50)[r]{\small$\nu_{eR}$}
\Text(305,35)[r]{\small$u_L$}
\Text(310,-93)[r]{\small$u_R$}
\Text(250,-93)[l]{\small$d_R$}
\Text(250,35)[l]{\small$d_L$}
\ArrowLine(280,25)(310,25)
\ArrowLine(280,-85)(310,-85)
\end{picture}
\end{center}
\vspace{2.2cm}
\caption{\small $2\beta$ decay \label{2betafig}}
\end{figure}
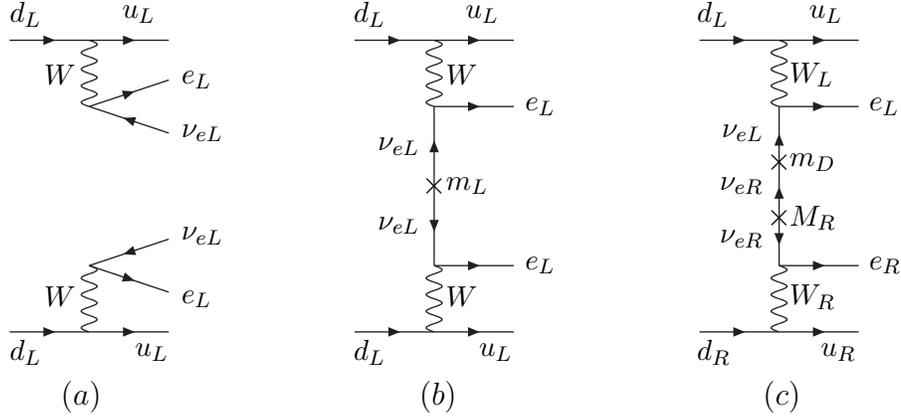

In some cases when the ordinary beta decay processes (\ref{beta}) are 
energetically forbidden, the double beta decay processes, in which a
nucleus $A(Z,N)$ is converted into an isobar with the electric charge 
differing by two units, may be allowed: 
\be
A(Z,N)\to A(Z\pm 2,N\mp 2)+2e^\mp +2\bar{\nu}_e(2\nu_e)\,.
\label{2beta2nu} 
\ee
In such decays two neutrons of the nucleus are simultaneously converted into  
two protons, or vice versa. At the fundamental (quark) level, these are  
transitions of two $d$ quarks into two $u$ quarks or vice versa (see 
fig. \ref{2betafig}a). Double beta decays are processes of the second order 
in weak interactions, and the corresponding decay rates are very low: typical 
lifetimes of the nuclei with respect to the $2\beta$ decay are $T\aprge  
10^{19}$ years. The processes (\ref{2beta2nu}) are called 
$2\beta 2\nu$ decays. Two-neutrino double beta decays with the emission of
two electrons ($2\beta^-$) were experimentally observed for a
number of isotopes with the half-lives in the range $10^{19} - 10^{21}$ 
years; there are very few candidate nuclei for $2\beta^+$ decay, and the 
experimental observation of this process is difficult because of the very 
small energy release ($Q$ values). 

If the lepton number is not conserved, the electron neutrino or antineutrino 
emitted in one of the elementary beta decay processes forming the $2\beta$
decay can be absorbed in another (fig. \ref{2betafig}b), leading to the 
neutrinoless double beta decay ($2\beta 0\nu$):
\be
A(Z,N)\to A(Z\pm 2,N\mp 2)+2e^\mp \,.
\label{2beta0nu} 
\ee
Such processes would have a very clear experimental 
signature: since the recoil energy of a daughter nucleus is negligibly small, 
the sum of the energies of the two electrons or positrons in the final state 
should be equal to the total energy release, i.e. should be represented by
a discrete energy line. Therefore $2\beta0\nu$ decays could serve as a 
sensitive probe of the lepton number violation. In some extended models 
exotic modes of $2\beta 0\nu$ decay are possible, e.g. decays with Majoron 
emission. In this case the sum of the energies of two electrons or
positrons is not a discrete line, but their spectrum is expected to be
different from that with no Majoron emission \cite{Klapdor}. 

Neutrinoless $2\beta$ decays break not only the lepton number; since the
absorbed $\nu_e$ or $\bar{\nu}_e$ has a ``wrong'' chirality, $2\beta 0\nu$ 
decays also break chirality conservation (this is another example of the 
chiral prohibition rule). Therefore $2\beta 0\nu$ decay is only possible 
if neutrinos have nonzero Majorana mass. Indeed, if $2\beta 0\nu$ decay is
mediated by the standard weak interactions, the amplitude of the process 
is proportional to neutrino mass (fig. \ref{2betafig}b):
\be
A(2\beta 0\nu) \propto \sum_i U_{ei}^2 \, m_i\equiv 
\langle m_{\nu_e}\rangle_{eff}\,.
\label{ampl}
\ee
Notice that this expression contains $U_{ei}^2$ rather than $|U_{ei}|^2$.  
If CP is conserved in the leptonic sector, the mixing matrix $U_{ai}^2$ 
can always be made real; however in this case the mass parameters $m_i$ 
in (\ref{ampl}) (eigenvalues of the neutrino mass matrix) can be of either
sign, their relative signs being related to the relative CP parities of
neutrinos. This means that in general significant cancellations between
various contributions to the sum in (\ref{ampl}) are possible. As we discussed 
in sec. \ref{D+M}, a pair of Majorana neutrinos with equal physical
masses $|m_i|$, opposite CP parities and maximal mixing is equivalent to 
a Dirac neutrino. It is easy to see that such a pair does not contribute 
to the amplitude in (\ref{ampl}). Analogously, the contribution of a 
quasi-Dirac neutrino is strongly suppressed.

Naively,  one could expect that the rates of $2\beta 0\nu$ decays to be
much higher than those of $2\beta 2\nu$ decays because the final state in
$2\beta 0\nu$ decays contains fewer particles than 
that of $2\beta 2\nu$ decays. In reality the situation is more complicated --  
the rates of $2\beta 0\nu$ decay are strongly suppressed by a very small 
factor $\langle m_{\nu_e}\rangle_{eff}^2/E_0^2$, where $E_0$ ($\sim$ a 
few MeV) is the typical energy release in $2\beta$ decay. Neutrinoless 
double beta decay was searched for experimentally but up to now 
have not been discovered. The experiments allowed to put the upper bound on 
the effective Majorana neutrino mass, the best limit coming from the 
Heidelberg -- Moscow experiment on $2\beta$ decay of $^{76}$Ge
\cite{Klapdor}:
\be
\langle m_{\nu_e}\rangle_{eff}< 0.2 - 0.6 ~\mbox{eV}\,,
\label{meff}
\ee
depending on the value of the nuclear matrix element which is not
precisely known. If the $2\beta 0\nu$ decay is discovered, it will be 
possible to infer the value of the effective Majorana neutrino mass 
$\langle m_{\nu_e}\rangle_{eff}$. As follows from (\ref{ampl}), this would
give the lower limit on the mass of the heaviest neutrino. 

In extended gauge models such as the left right symmetric (LR) or SUSY 
models, additional mechanisms of $2\beta 0\nu$ decay are possible. For the 
LR models a typical diagram is shown in fig. \ref{2betafig}c. It appears 
that no Majorana mass $m_L$ of $\nu_L$ is necessary in this 
case, i.e. $2\beta 0\nu$ decay can occur even if $m_L=0$. This, however, is 
incorrect: in all extended models in which $2\beta 0\nu$ decay occurs, 
the Majorana masses of $\nu_L$ are different from zero. In particular, in
the LR models in which the diagram \ref{2betafig}c exists, the seesaw
mechanism operates and so $m_L\ne 0$. It can be shown in a model-independent 
way that $2\beta 0\nu$ decay always means $m_L \ne 0$ \cite{Jose1}.

\section{The solar neutrino problem 
\label{SNP}}
It is generally believed that the stellar energy is provided by thermonuclear 
reactions. For so called main sequence stars (to which our sun belongs), 
the main result of these reactions is a fusion of hydrogen into
helium: 
\be
4 p + 2 e^- = {\rm ^4He} +2 \nu_e +26.73 ~\mbox{MeV}\,.
\label{fusion}
\ee
The main reaction chain in which the fusion process (\ref{fusion}) 
occurs in the sun is the $pp$ cycle (fig. \ref{ppcycle}), whereas the 
carbon-nitrogen-oxygen (CNO) cycle is responsible for less than 2\% of
solar energy. 
Neutrinos are emitted in six reactions of the solar $pp$ cycle 
(Table \ref{solar}). 
They are produced in nuclear beta decay or electron capture reactions, and 
their energy spectra are well known (see fig. \ref{spectrum}). Three of these 
reactions -- $pep$ and the electron capture into two different final states 
of $^7$Be -- produce monochromatic $\nu_e$ lines, whereas the neutrinos born 
in the other three reactions have continuous energy spectra. The fluxes of
the solar $\nu_e$ are calculated in the framework of the standard solar 
models. These models are based on the assumptions of local hydrostatic 
equilibrium, thermonuclear nature of solar energy, energy transfer in the sun 
by radiation and convection and the requirement that the present values of the 
solar radius, mass, luminosity and He/H ratio be reproduced as a result of 
the solar evolution. There are about 20 different solar models developed 
during the last decade by more than ten different groups of authors, all
based on the above mentioned principles. All these models except one agree 
with each other well (see fig. \ref{solmod}). 
A number of recent models also took into account diffusion of helium
and heavy elements. 
These models were able to reproduce with a very high accuracy the solar 
sound velocities inferred from the helioseismological measurements \cite{jnb}. 
This gives further credibility to the standard solar models. 

\begin{table}[b]
\caption{\label{solarreactions}Solar neutrino production in the $pp$
chains (from ref. \cite{raffelt2}).}  
\bigskip
\hbox to\hsize{\hss
\begin{tabular}{llrrr}
\hline
\hline
\noalign{\vskip4pt}
Name&Reaction&\multicolumn{1}{c}{$\langle E_\nu\rangle$}&
\multicolumn{1}{c}{$E_\nu^{\rm max}$}&
\multicolumn{1}{c}{Fractional}\\
&&\multicolumn{1}{c}{[MeV]}&
\multicolumn{1}{c}{[MeV]}&
\multicolumn{1}{c}{solar flux}\\
\noalign{\vskip4pt}   
\hline
$pp$ & $p+p \to {\rm D} + e^+ + \nu_e$&0.26& 0.42 &0.909\\
$ pep$ &$p + e^- + p \to {\rm D} + \nu_e$&1.44&---&$2\times10^{-3}$\\
$hep $ &${\rm ^3He} + p \to {\rm ^4He} + e^+ + \nu_e$&9.62& 18.77
&$2 \times 10^{-8}$\\
$ {\rm ^7Be}$ &${\rm ^7Be} + e^-\to  {\rm ^7Li} + \nu_e$&(90\%)
0.86&---&0.074\\
         &                                       &(10\%) 0.38&---&     \\
${\rm ^8B} $ &${\rm ^8B} \to  {\rm ^8Be}^{\ast} + e^+ + \nu_e$&6.71&
$\approx$ 15 &
$8.6 \times 10^{-5}$\\
\hline
\end{tabular}\hss}
\label{solar}
\end{table}
\begin{samepage}   
\begin{figure}
\setlength{\unitlength}{1cm}
\begin{center}
\vspace{5.2cm}
\epsfig{file=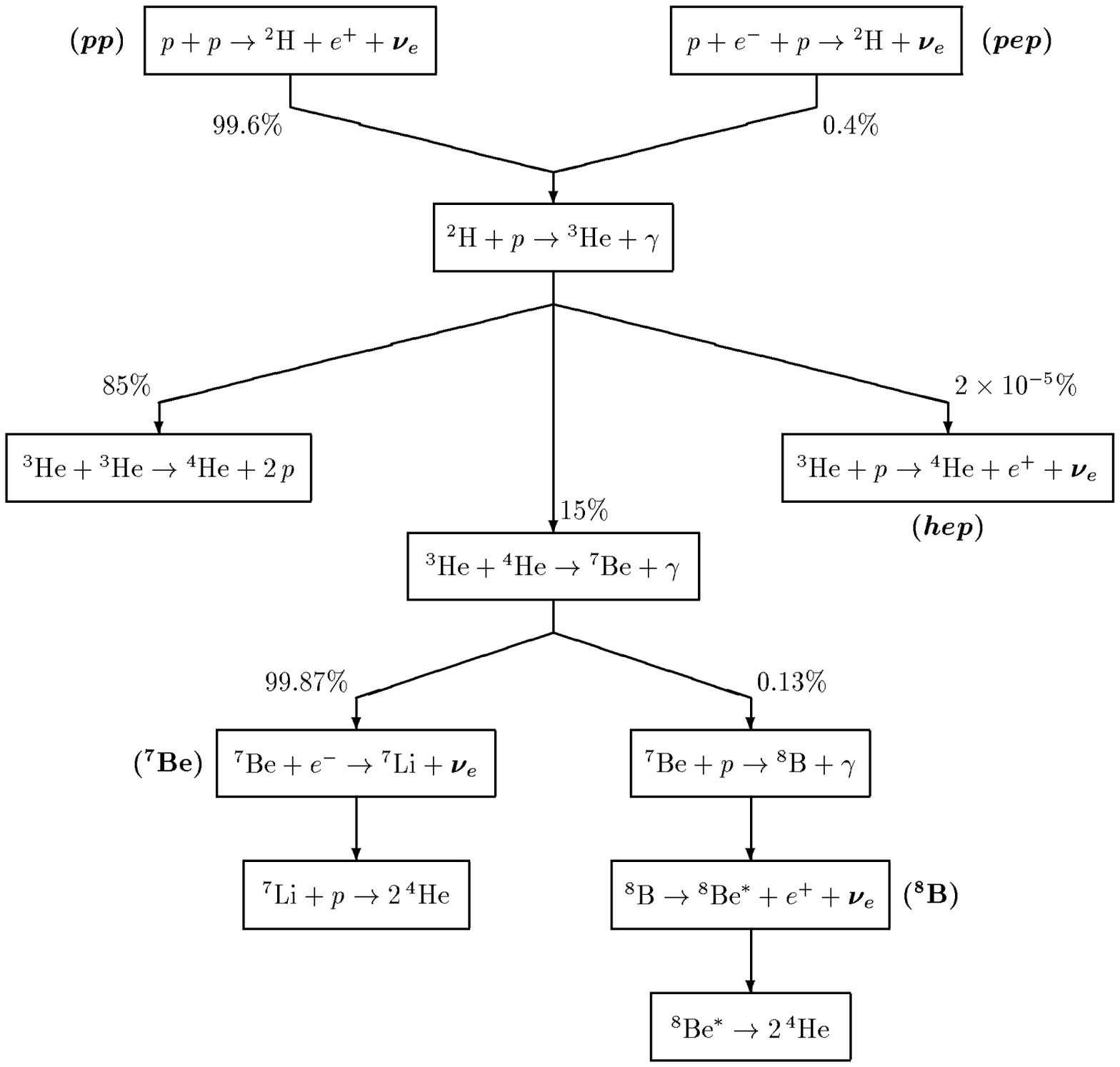,width=12cm}
\end{center}
\vspace{-7cm} 
\caption{\label{ppcycle}\small Energy generation in the Sun via the $pp$
chains (from ref.~\cite{BGG}).}
\begin{center}
\epsfig{file=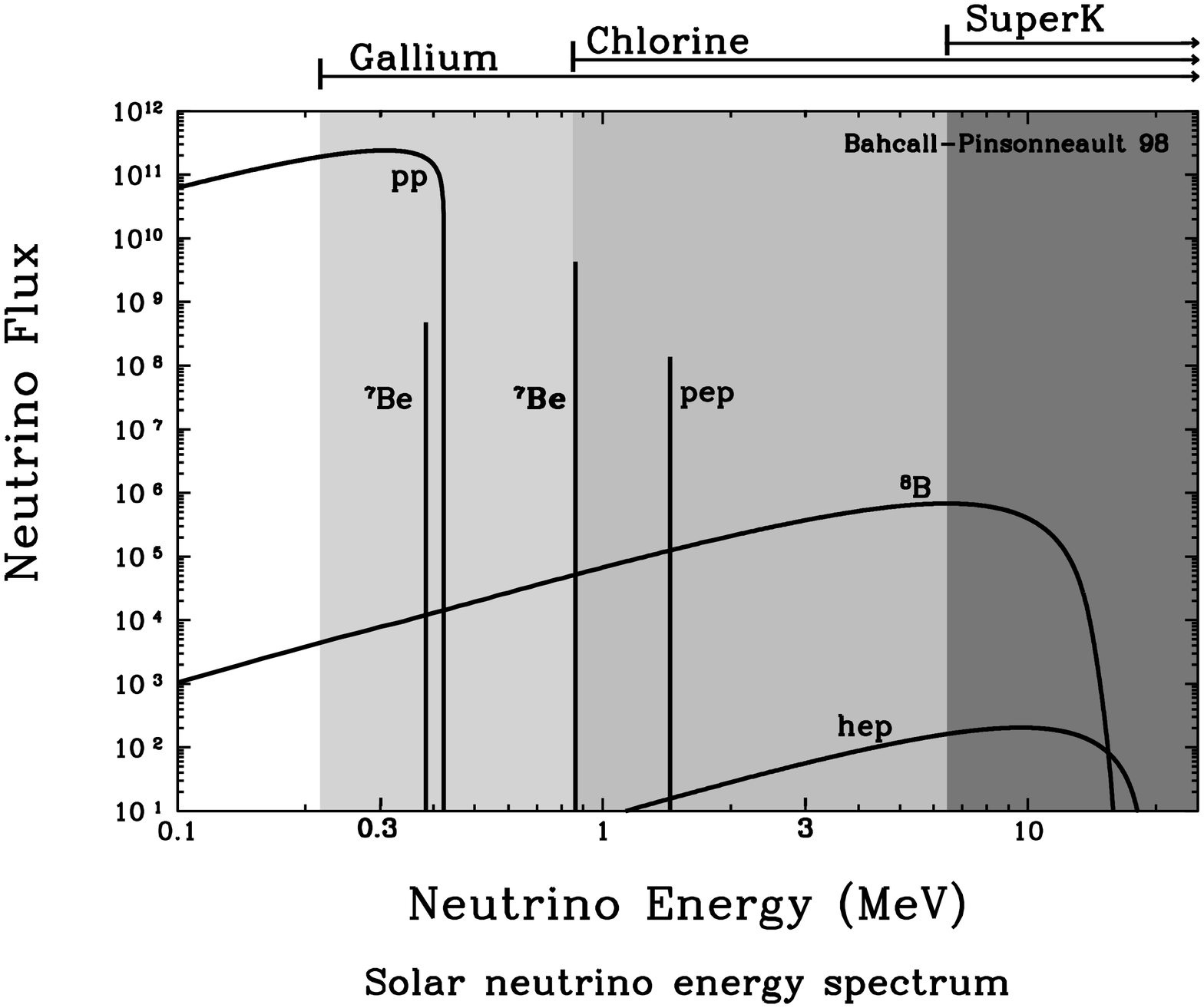,width=8.5cm,angle=270}
\end{center}
\caption{\label{spectrum}\small Solar neutrino spectrum and thresholds of
solar neutrino experiments as indicated above the figure
(from ref. \cite{jnb}).}
\end{figure}
\end{samepage}

The calculated fluxes of different components of the solar $\nu_e$ flux
depend not only on the general principles laid in the basis of the standard 
solar models, but also on a number of input parameters, such as the solar
opacity and various nuclear cross sections. These parameters are known to a 
varying degree of accuracy \cite{jnb}. The flux of the most abundant $pp$ 
neutrinos is closely related to the well known luminosity of the sun, and 
therefore has a very small uncertainty, $\sim 1\%$. On the other hand, the
flux of the energetic $^8$B neutrinos has a relatively large uncertainty, 
$\sim 20\%$. The flux of the most energetic $hep$ neutrinos is the least 
certain one -- it is only known to an accuracy of 1 - 2 orders of magnitude.
Fortunately, this flux is very small, and its contribution to the
detection rates of solar neutrinos is practically negligible. It can,
however, affect the high energy part of the solar neutrino spectrum. 

The solar neutrinos have been detected in five solar neutrino experiments. 
Historically, the first one was the Homestake
experiment of Davis and his collaborators. It is based on the reaction
\be
\nu_e+{\rm ^{37}Cl}\to {\rm ^{37}Ar}+e^-\,.
\label{Cl} 
\ee
The radioactive argon atoms produced in this reaction are extracted using
chemical methods and counted in proportional counters. 
The energy threshold of reaction (\ref{Cl}) is 0.814 MeV, so only 
the $^8$B and $^7$Be and $pep$ neutrinos are detected in the Homestake
experiment, the largest contribution coming from the $^8$B neutrinos (see 
figs. \ref{spectrum} and \ref{solexpdata}). Radiochemical techniques are also 
used in two other solar neutrino experiments -- SAGE and Gallex, in which
the reaction
\be
\nu_e+{\rm ^{71}Ga}\to {\rm ^{71}Ge}+e^-\,.
\label{Ga} 
\ee
is employed. The energy threshold of this reaction is 0.234 MeV, and so
the gallium experiments can also detect the lowest energy $pp$ neutrinos.
Since the flux of the $pp$ neutrinos is very large, they are expected to
give the main contribution to the event rates in the SAGE and Gallex
detectors (fig. \ref{solexpdata}). The remaining two experiments --
Kamiokande and its up-scaled version Super-Kamiokande -- are the water 
Cherenkov detectors and use the neutrino-electron scattering reaction 
\be
\nu_a+e^-\to \nu_a + e^-\,.
\label{nue} 
\ee
to detect solar neutrinos. 
This reaction has zero physical threshold, but one has to introduce energy
cuts to suppress the background. In the Kamiokande experiment solar
neutrinos with the energies $E>7.5$ MeV were detected, whereas the
threshold used by Super-Kamiokande is at present 5.5 MeV. 
With these energy cuts, the Kamiokande and Super-Kamiokande detection rates 
are only sensitive to the $^8$B component of the solar neutrino flux. 
Reaction (\ref{nue}) has a very interesting feature -- for neutrino
energies $E\gg m_e$ the angular distribution of the recoil electrons 
is forward-peaked, i.e. it points in the direction of the momentum of the 
incoming neutrino. The angular distributions of neutrinos detected in the
Kamiokande and Super-Kamiokande experiments have a prominent peak at
$180^\circ$ from the direction to the sun, which is a beautiful proof of 
the solar origin of these neutrinos. The origin of the neutrinos 
detected in the radiochemical experiments is unknown, and our belief that 
they come from the sun is based on the fact that no other sources of 
sufficient strength and energy are known.

\begin{samepage}   
\begin{figure}
\setlength{\unitlength}{1cm}
\begin{center}
\vspace{-0.4cm}
\epsfig{file=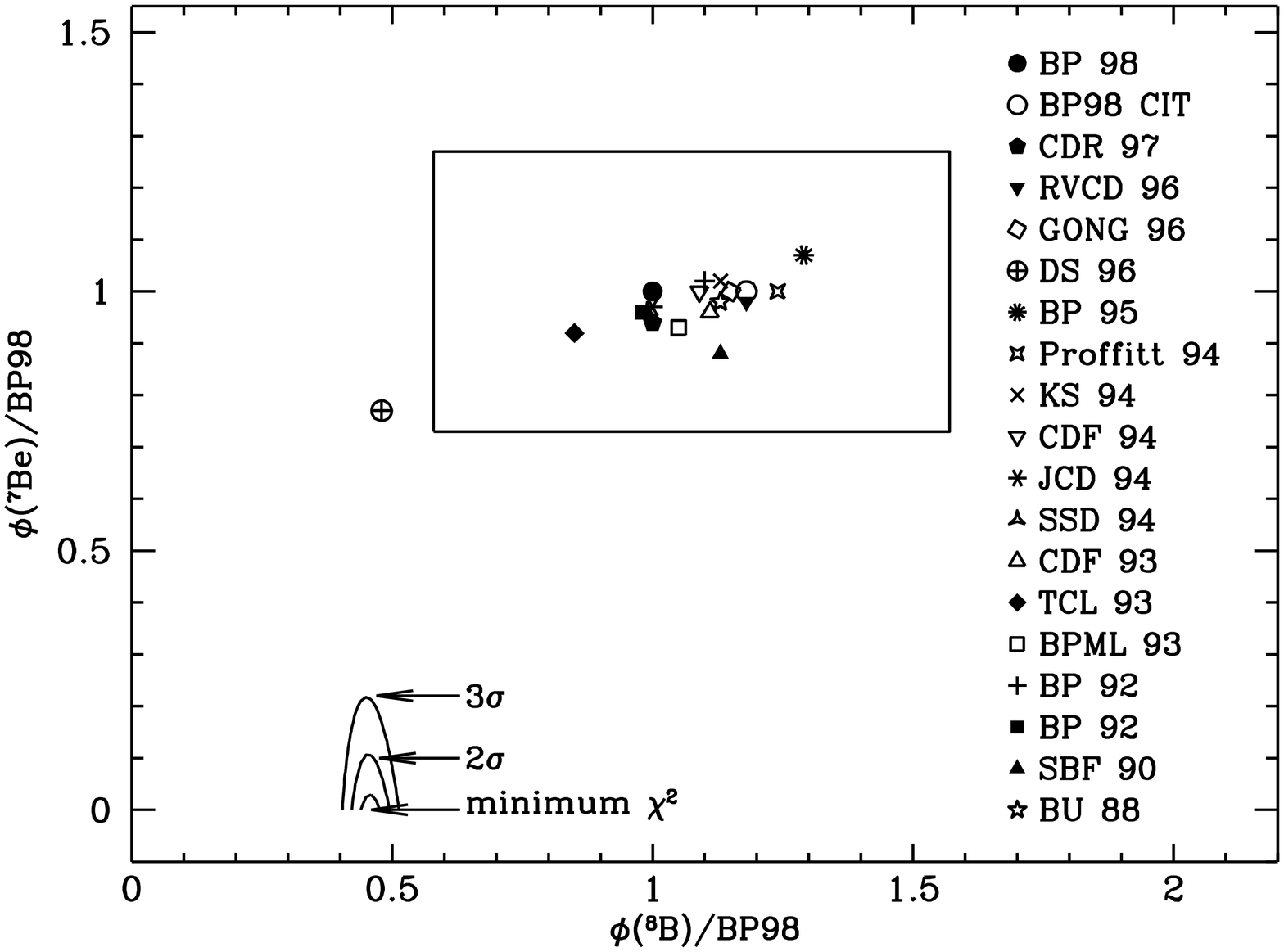,width=10cm}
\end{center}
\vspace{-0.5cm} 
\caption{\label{solmod} \small Predictions of standard solar models since
1988. The $^7$Be and $^8$B fluxes are normalized to the predictions of the 
BP98 model. The rectangular box defines the 3$\sigma$ error range of the 
BP98 model. The best-fit $^7$Be flux is negative 
(from ref.~\cite{jnb1}).}
\begin{center}
\vspace{-0.5cm}
\epsfig{file=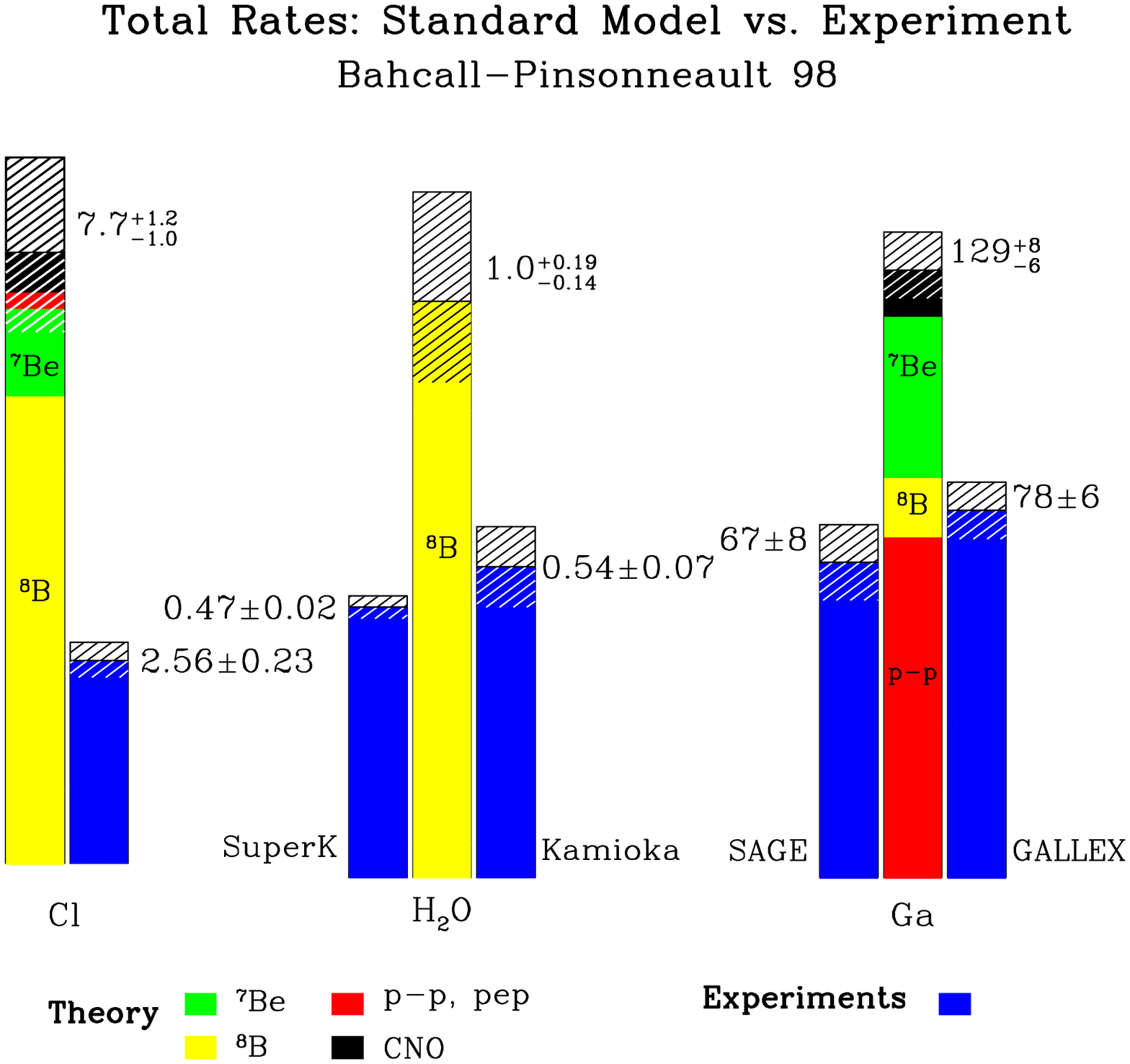,height=14.5cm,angle=270}
\end{center}
\vspace{-1.0cm}
\caption{\label{solexpdata}\small Solar neutrino measurements and
theoretical flux predictions. For Cl and Ga experiments, the units are SNU
(1 SNU = $10^{-36}$ captures per target atom per second); for H$_2$O 
experiments, the ratio data/exp. is shown. From ref. \cite{jnb}.}
\end{figure}
\end{samepage}

In all five solar neutrino experiments, fewer neutrinos than expected 
were detected, the degree of deficiency being different in the experiments
of different types (fig. \ref{solexpdata}). This is the essence of the
solar neutrino problem. The observed deficiency of solar neutrinos could be 
due to an insufficient knowledge of solar physics or an error in some input 
parameters (astrophysical solution), or due to experiment-related errors 
(such as miscalculated detection efficiency or cross section), or due to 
some unknown neutrino physics (particle physics solution of the solar 
neutrino problem). As we have already discussed, the standard solar models 
are well grounded, agree with each other and with helioseismological
observations. 
All the solar neutrino experiments but one (Homestake) have been calibrated, 
and their experimental responses   
were found to be in a very good agreement with expectations. The argon 
extraction efficiency of the Homestake detector was also checked by doping it 
with a known small number of radioactive argon atoms, but no calibration
has been carried out since no artificial source of neutrinos with a suitable 
energy spectrum exists. 

The solar neutrino problem is not just the problem of the deficit of the 
observed neutrino flux: results of different experiments seem
to be inconsistent with each other. 
%
%
In the absence of new neutrino physics, the energy spectra of the various 
components of the solar neutrino flux are given by the standard nuclear 
physics and well known, and only the total fluxes of these components may 
be different from those predicted by the standard solar models. One can 
then infer the flux of $^8$B neutrinos directly from the Kamiokande or 
Super-Kamiokande data, and use this flux to find the corresponding 
contribution to the Homestake detection rate. This contribution turns out to 
be larger than the total detection rate, so the best-fit contribution of the 
$^7$Be neutrinos to the Homestake detection rate is negative! 
In fact, if one assumes that the solar neutrino spectra are undistorted, one 
can demonstrate the existence of the solar neutrino problem without using 
any information about solar physics except the solar luminosity constraint, 
or even without this constraint, i.e. in a solar-model independent way. 
Therefore the probability of the astrophysical solution of the solar
neutrino problem is very low. 

Alternatively, one could assume that the observed deficit of solar neutrinos 
is due to some unknown experimental errors. However, in this case it is not 
sufficient to assume that one of the solar neutrino experiments is wrong; 
one would have to assume that the experiments of at least {\it two}
different types (chlorine, gallium and water Cherenkov) are wrong, which is 
very unlikely. The remaining possibility is that the spectra of solar 
neutrinos are distorted, which requires new neutrino physics. 

There are several possible particle-physics solutions of the solar neutrino 
problem, the most natural one being neutrino oscillations (for an alternative 
solution related to nonzero neutrino magnetic moments, see e.g. \cite{Akh1}). 
The neutrino oscillation solution has become even more plausible after the 
strong evidence for atmospheric neutrino oscillations was reported by the 
Super-Kamiokande Collaboration. Neutrino oscillations can convert a fraction 
of solar $\nu_e$ into $\nu_\mu$ or $\nu_\tau$ (or their combination).
Since the energy of solar neutrinos is smaller than the masses of muons
and tauons, these $\nu_\mu$ or $\nu_\tau$ cannot be detected in the CC
reactions of the type (\ref{Cl}) or (\ref{Ga}) and therefore are invisible 
in the chlorine and gallium experiments. They can scatter on electrons
through the NC interactions and therefore should contribute to the
detection rate in water Cherenkov detectors. However, the cross section of
the NC channel of reaction (\ref{nue}) is about a factor of 6 smaller 
than that of the CC channel, and so the deficit of the neutrino flux 
observed in the Kamiokande and Super-Kamiokande experiments can be 
explained. The probabilities of neutrino oscillations depend on neutrino
energy, and the distortion of the energy spectra of the experimentally 
detected solar neutrinos, which is necessary to reconcile the data of
different experiments, is readily obtained. 

\begin{figure}[h] 
\setlength{\unitlength}{1cm}
\hbox{\hfill
\epsfxsize=8.0cm\epsfbox{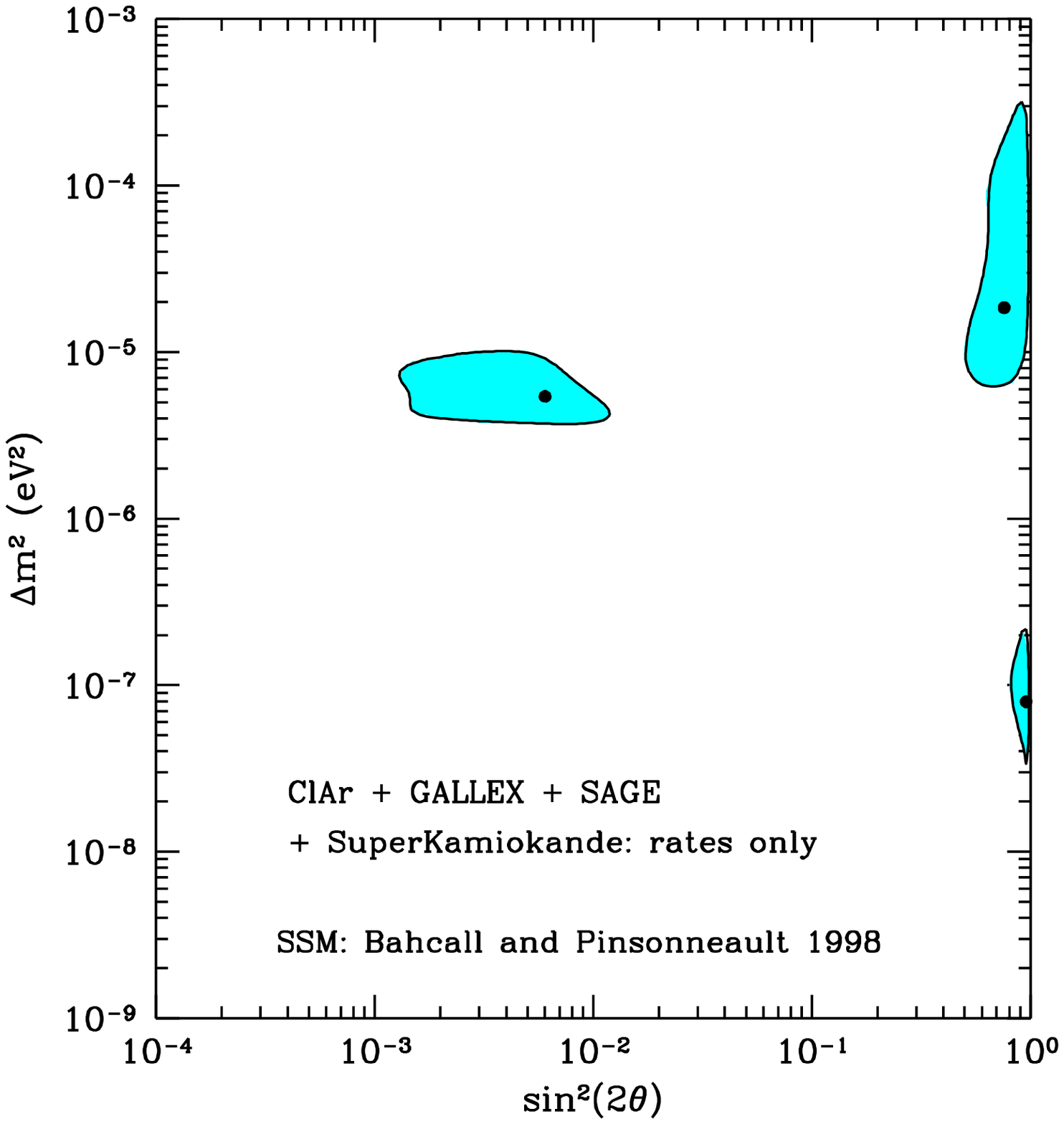}
\hspace{-1.5cm}
\epsfxsize=8.0cm\epsfbox{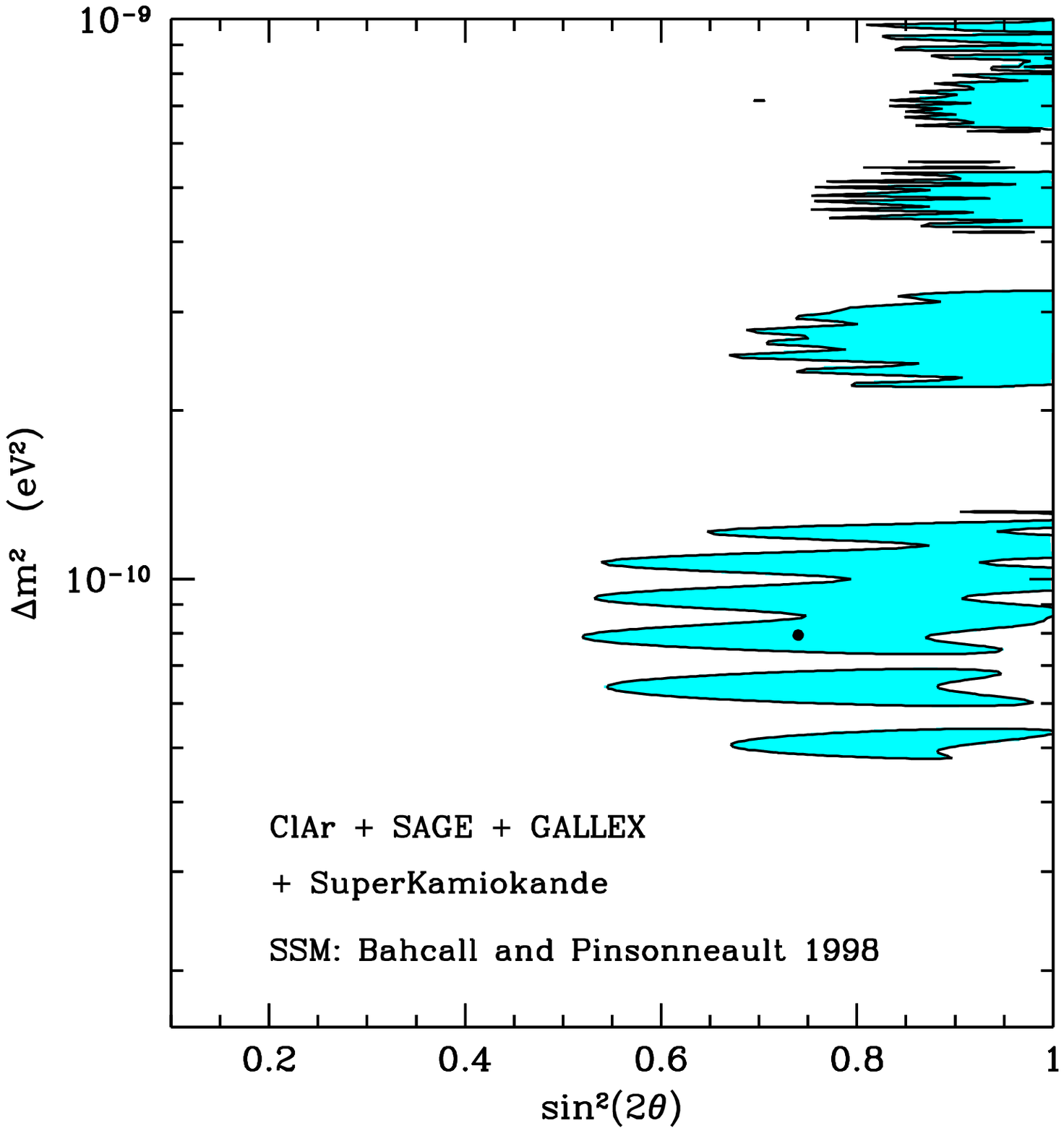}
\hfill}
\caption{\small Left panel: Allowed regions (99 \% c.l.) for the MSW
solutions of the solar neutrino problem for 2-flavour oscillations of
$\nu_e$ into active neutrinos (only rates included). Right panel: The same 
for the VO solution. From ref. \cite{BKS1}. }
\label{rates} 
\end{figure}

The oscillations of solar neutrinos inside the sun can be strongly enhanced 
due to the MSW effect which we discussed in sec. \ref{matter}, and the
solar data can be fitted even with a very small vacuum mixing angle. Solar 
matter can also influence neutrino oscillations if the vacuum mixing
angle is not small. The allowed values of the neutrino oscillation parameters 
$\sin^2 2\theta_0$ and  $\Delta m^2$ which fit the detection rates in the 
chlorine, gallium and water Cherenkov experiments in the 2-flavour scheme are 
shown in fig. \ref{rates} for matter enhanced (left panel) and vacuum (right 
panel) oscillations of $\nu_e$ into active neutrinos. In the case of the 
matter enhanced oscillations, there are three allowed ranges of the 
parameters corresponding to the small mixing angle (SMA), large mixing angle 
(LMA) and low $\Delta m^2$ (LOW) MSW solutions \cite{BKS1,Valencia}. The LOW 
solution has a very low probability and only exists at 99\% c.l.. The vacuum 
oscillation (VO) solution corresponds to very small values of $\Delta m^2$,
for which the neutrino oscillation length (\ref{losc}) for typical solar
neutrino energies ($\sim$ a few MeV) is comparable to the distance between
the sun and the earth. This solution is also known as ``just so'' 
oscillation solution. For $\nu_e \to \nu_{sterile}$ oscillations, there is 
only the SMA solution with the allowed region of parameters similar to
that for oscillations into active neutrinos. 

Solar neutrinos detected during night travel some distance inside the 
earth on their way to the detector, and their oscillations can be affected 
by the matter of the earth. In particular, a fraction of $\nu_\mu$ or
$\nu_\tau$ produced as a result of the solar $\nu_e$ oscillations can 
be reconverted into $\nu_e$ by oscillations enhanced by the  matter of the
earth. This earth ``regeneration'' (or day/night) effect can be appreciable 
in the case of the LMA solution, but is expected to be very small in the 
case of the SMA solution. The day/night effects (and in general, the
zenith angle dependence of the neutrino signal) can in principle be
observed in real-time experiments, such as Super-Kamiokande. 

In the case of the MSW solutions of the solar neutrino problem the
neutrino state arriving at the earth is an incoherent superposition of the 
mass eigenstates $\nu_1$ and $\nu_2$ (see Problem 23 below). The 
probability $P_{SE}$ of finding a solar $\nu_e$ after it traverses the 
earth depends on the average $\nu_e$ survival probability in the 
sun $\bar{P}_S$: 
\be
P_{SE}=\bar{P}_S + \frac{1-2 \bar{P}_S}{\cos 2\theta_0}\,(P_{2e}-\sin^2
\theta_0)\,.
\label{PSE}
\ee
Here $P_{2e}=P(\nu_2\to\nu_e)$ is the probability of oscillations of the 
second mass eigenstate into electron neutrino inside the earth. 
It follows from this expression that the sign of the night-day asymmetry 
$(N-D)/(N+D)$ depends on whether $\bar{P}_S$ is larger or smaller than  1/2. 
For the LMA solution, $\bar{P}_S<1/2$, and the night-day 
asymmetry is expected to be positive. 
For the SMA solution, $\bar{P}_S$ for the Super-Kamiokande experiment is close 
to 1/2, and therefore the expected day/night effect is small. It can in 
principle be of either sign. The value measured in the Super-Kamiokande 
experiment is 
\cite{Totsuka}
\be
\frac{N}{D}-1=0.067\pm0.033\,(stat.)\pm 0.013\,(syst.)\,,
\label{dn}
\ee
i.e. shows an excess of the night-time flux at about 2$\sigma$. The zenith 
angle event dependence measured by the Super-Kamiokande shows a rather flat 
distribution of the excess of events over different night-time zenith angle 
bins. This is rather typical for the LMA solution whereas for the SMA 
solution one normally expects the excess (or deficiency) to be concentrated 
in the vertical upward bin with zenith angles $\theta$ in the range 
$-1<\cos\theta <-0.8$. Thus, the night-time zenith angle dependence and 
the overall night-day asymmetry (\ref{dn}) favour the LMA solution of the
solar neutrino problem. 

As we discussed before, oscillations of solar neutrinos in the 3-flavour 
scheme can be described through the effective two-flavour $\nu_e$ survival 
probability (see eq. (\ref{prob7}) and Problem 20). For a recent detailed 
analysis of the solar neutrino oscillations in the 3-flavour framework, 
see \cite{Lisi1}. 

{\it Problem 22. Discuss effects of the solar and earth's matter on the
oscillations of solar neutrinos in the case of VO solution. Take into account 
that the matter densities $\rho_{sun}$ and $\rho_{earth}$ vary in the
ranges 0 -- 150 and 2.7 -- 12.5 $g/cm^3$ respectively.} 

{\it Problem 23. Show that in the case of the MSW solutions the 
solar neutrino flux arriving at the earth can be considered as an
incoherent superposition of the neutrino mass eigenstates $\nu_1$ and
$\nu_2$. Hint: consider the relative phase between $\nu_1$ and
$\nu_2$ states acquired by neutrinos on their way from the sun to the
earth. Assume that the energy resolution of the neutrino detector
$\Delta E \aprge 100$ keV. 
} 

{\it Problem 24. Derive equation (\ref{PSE}) for the probability of
finding a solar $\nu_e$ after it crosses the earth during night. Hint: 
use the results of the previous problem.}

\begin{figure}[htb] 
\hbox to \hsize{\hfil\epsfxsize=10cm\epsfbox{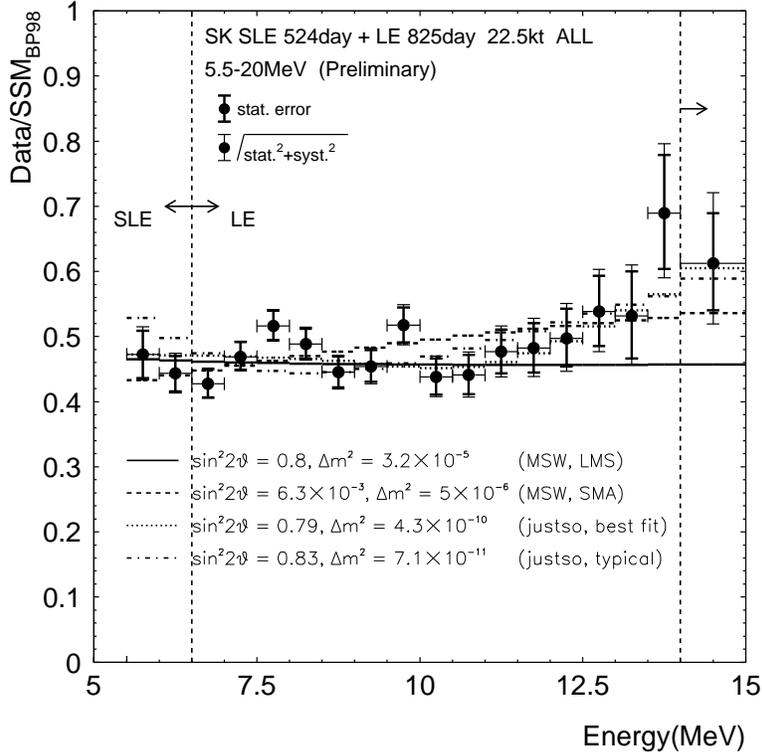}\hfil}
\caption{\small Recoil electron energy spectrum in the Super-Kamiokande 
experiment along with typical predictions of SMA, LMA and VO solutions 
(from ref. \cite{Totsuka}). }
\label{SKspectrum} 
\end{figure}

As we have already discussed, neutrino oscillations (both in matter and in 
vacuum) depend on neutrino energy and therefore should result in certain 
distortions of the spectrum of the detected solar neutrinos. One can obtain 
an information on the spectrum of incoming neutrinos by measuring the recoil 
electron spectrum in the reaction (\ref{nue}). This spectrum has been 
measured in the Super-Kamiokande experiment, and the results (along
with the typical predictions of the LMA, SMA and VO solutions) are shown 
in fig. (\ref{SKspectrum}). In the absence of the neutrino spectrum
distortion, the ratio of measured/expected electron spectrum presented in 
this figure should be a horizontal line. The characteristic feature of 
the measured spectrum is an excess of events in the region of high
energies. The experimental errors are still too large to allow a clean 
discrimination between different solutions of the solar neutrino problem; 
however, the spectrum favours VO over the other solutions. Another possible 
explanation of the excess of the high energy events is that the flux 
of the highest-energy $hep$ neutrinos has been underestimated by about a  
factor of 20. Since the $hep$ flux is very poorly known, this is still a
possibility. 

At present, the situation with the solar neutrino data is somewhat
confusing: different pieces of the data favour different solutions of the 
solar neutrino problem. The total rates are better fitted within the SMA 
solution, the night/day ratio prefers the LMA solution, and the electron
recoil spectrum in Super-Kamiokande favours the VO solution. Clearly, 
more data are needed to clear the situation up. Fortunately, two new
experiments which can potentially 
resolve the problem are now under way or will soon be put into operation.
The SNO (Sudbury Neutrino Observatory) experiment has started taking data,
and its first results are expected some time during 2000. The Borexino 
experiment is scheduled to start data taking in 2001. 

The SNO detector consists of 1000 tons of heavy water, and it detects
solar neutrinos in three different channels:
\begin{eqnarray}
\nu_e+d \to p+p+e^- \quad\quad ({\rm CC}),~~E_{min}=1.44~{\rm MeV}\,, 
\label{CC}\\
\nu_a+d\to p+n+\nu_a \quad\quad ({\rm NC}),~~E_{min}=2.23~{\rm MeV}\,,
\label{NC}
\end{eqnarray}
and $\nu_a e$ scattering process (\ref{nue}) which can proceed through both 
CC and NC channels. The CC reaction (\ref{CC}) is very well suited for 
measuring the solar neutrino spectrum. Unlike in the case of $\nu_a e$ 
scattering (\ref{nue}) in which the energy of incoming neutrino is shared 
between two light particles in the final state, the final state of the 
reaction (\ref{CC}) contains only one light particle -- electron, and a
heavy $2p$ system whose kinetic energy is very small. Therefore by measuring 
the electron energy one can directly measure the spectrum of 
the solar neutrinos. Detecting the solar neutrino spectrum would be 
of great importance since any deviation from the spectrum predicted  by
the nuclear beta decay would be a ``smoking gun'' signature of new 
neutrino physics.

The cross section of the NC reaction (\ref{NC}) is the same for neutrinos 
of all three flavours, and therefore oscillations between $\nu_e$ and 
$\nu_\mu$ or $\nu_\tau$ would not change the NC detection rate in the SNO 
experiment. On the other hand, these oscillations would deplete the solar 
$\nu_e$ flux, reducing the CC event rate. Therefore the CC/NC ratio is 
a sensitive probe of neutrino flavour oscillations. If solar $\nu_e$ 
oscillate into electroweak singlet (sterile) neutrinos, both CC and NC
event rates will be suppressed. 

The Borexino experiment will detect solar neutrinos through the $\nu_a e$
scattering with a very low energy threshold, and will be able to detect
the $^7$Be neutrino line. Different solutions of the solar neutrino problem 
predict different degree of suppression of $^7$Be neutrinos, and their 
detection could help discriminate between these solutions. 
Observation of the $^7$Be neutrino line would be especially important 
in the case of the VO solution. Due to the eccentricity of the earth's
orbit the distance between the sun and the earth varies by about 3.5\%
during the year, and this should lead to varying oscillation phase (and 
therefore varying solar neutrino signal) in the case of vacuum neutrino 
oscillations. This seasonal variation can in principle be separated from 
the trivial 7\% variation due to the $1/L^2$ law which is not related to 
neutrino oscillations. However, the oscillation phase depends on neutrino 
energy [see (\ref{prob1})], and integration over significant energy
intervals may make it difficult to observe the seasonal variations of the 
solar neutrino flux due to the VO. The $^7$Be neutrinos are monochromatic, 
which should facilitate the observation of the seasonal variations. The
oscillation phase in (\ref{prob1}) depends on the ratio $L/E$, therefore 
for neutrinos with continuous spectrum VO should lead to seasonal variations 
of the spectrum distortion which are correlated with the seasonal variations 
of the flux \cite{MS}. Such variations would be a clear signature of the
VO solution. 

Although up to now the sensitivities of the solar neutrino experiments 
have been insufficient to allow a clear-cut discrimination between different 
solutions of the solar neutrino problem, one can hope that the combined data 
of the currently operating and forthcoming experiments will allow to finally 
resolve the solar neutrino problem. 

\section{Atmospheric neutrinos 
\label{atm}}

Atmospheric neutrinos 
%
%
are electron and muon neutrinos and their antineutrinos which are produced 
in the hadronic showers induced by primary cosmic rays in the earth's 
atmosphere. 
%
%
The main mechanism of production of the atmospheric neutrinos is given 
by the following chain of reactions:
\be
\begin{array}{lllll}
p(\alpha, ...)+Air&\rightarrow &\pi^{\pm}(K^{\pm})&+ & X  \\
&   &\pi^{\pm}(K^{\pm})&\rightarrow
&\mu^{\pm}+\nu_{\mu}(\bar{\nu}_{\mu})\\
& & & &\mu^{\pm}\rightarrow e^{\pm}+\nu_{e}(\bar{\nu}_{e})+
\bar{\nu}_{\mu}(\nu_{\mu})
\end{array}
\label{nuprod}
\ee
Atmospheric neutrinos can be observed directly in large mass underground
detectors predominantly by means of their CC interactions: 
\begin{eqnarray}
&  \nu_{e}(\bar{\nu}_{e})+A\rightarrow e^{-}(e^{+})+X\,, 
\nonumber \\
&  \nu_{\mu}(\bar{\nu}_{\mu})+A\rightarrow \mu^{-}(\mu^{+})+X\,.
\label{nudet}
\end{eqnarray} 
Naively, from the reaction chain (\ref{nuprod}) one would expect to have
two atmospheric muon neutrinos or antineutrinos for every electron neutrino 
or antineutrino. In reality, the situation is more complicated: one should
take into account the differences in the lifetimes of $\pi^{\pm}$, $K^{\pm}$ 
and $\mu^{\pm}$ as well as the differences in their spectra. Also, although 
the reaction chain (\ref{nuprod}) is the dominant source of atmospheric
neutrinos, it is not the only one. Calculations of the atmospheric
neutrino fluxes predict the $\nu_\mu/\nu_e$ ratio that depends on neutrino
energy and the zenith angle of neutrino trajectory, approaching 2 for low 
energy neutrinos and horizontal trajectories but exceeding this value for 
higher energy neutrinos and for trajectories close to vertical.  

Accurate calculation of the atmospheric neutrino fluxes
is a difficult job which includes such ingredients as spectra and chemical
composition of cosmic rays (including geomagnetic effects and solar
activity), cross sections of $\pi$ and $K$ production off the nuclear
targets, Monte Carlo simulation of hadronic cascades in the atmosphere and
the calculation of neutrino spectra including muon polarization effects.     
Each step introduces some uncertainty in the calculation. 
The overall uncertainty of the calculated atmospheric neutrino fluxes is
rather large, and the total fluxes calculated by different authors differ
by as much as 20 -- 30\%. At the same time, the ratio of the muon to
electron neutrino fluxes is fairly insensitive to the above uncertainties,  
and different calculations  
%
%
yield the ratios of muon-like to electron-like contained events which 
agree to about 5\%. 
This ratio has been measured in a number of experiments, and the Kamiokande 
and IMB Collaborations reported smaller than expected ratio in their
contained events, with the double ratio $R(\mu/e) \equiv 
[(\nu_\mu+\bar{\nu}_\mu)/(\nu_e+\bar{\nu}_e]_{data}/[(\nu_\mu+\bar{\nu}_\mu)
/(\nu_e+\bar{\nu}_e)]_{MC}
\simeq 0.6$ where MC stands for Monte Carlo simulations. The discrepancy
between the observed and predicted atmospheric neutrino fluxes was called
the atmospheric neutrino anomaly. The existence of this anomaly was 
subsequently confirmed by Soudan 2, MACRO and Super-Kamiokande experiments. 
Most remarkably, the Super-Kamiokande (SK) Collaboration obtained a very 
convincing evidence for the up-down asymmetry and zenith-angle dependent 
deficiency of the flux of muon neutrinos, which has been interpreted as an 
evidence for neutrino oscillations. We shall now discuss the SK data and
their interpretation.  

The SK detector is a 50 kt water Cherenkov detector (22.5 kt fiducial
volume) which is overseen by more than 13,000 photomultiplier tubes. The
charged leptons born in the CC interactions of neutrinos produce the rings 
of the Cherenkov light in the detector which are observed by the phototubes. 
Muons can be distinguished from electrons since their Cherenkov rings  
are sharp whereas those produced by electrons are diffuse. 
%
%
The SK Collaboration subdivided their atmospheric neutrino events into several 
groups, depending on the energy of the charged leptons produced.
Fully contained (FC) events are those for which the neutrino  
interaction vertex is located inside the detector and all final state 
particles do not get out of it. FC events are further subdivided into sub-GeV 
(visible energy $<1.33$ GeV) and multi-GeV (visible energy $>1.33$ GeV) 
events. Partially contained (PC) events are those for which the produced muon
exits the inner detector volume (only muons are penetrating enough). The
average energy of a neutrino producing a PC event in SK is $\sim 15$ GeV. 
Muon neutrinos can also be detected indirectly by observing the muons that
they have produced in the material surrounding the detector. To reduce the 
background from atmospheric muons, only upward--going neutrino-induced muons 
are usually considered. A rough estimate of the energy spectrum of the 
upward--going muons has been obtained dividing them in two categories, 
passing (or through-going) and stopping muons. The latter, which stop inside 
the detector, correspond to the average parent neutrino energy $\sim 10$ 
GeV, whereas for the through-going muons the average neutrino energy is
$\sim 100$ GeV. 

The measurements of the double ratio $R(\mu/e)$ for contained events at SK
(848 live days) give 
\cite{Mann}   
\begin{eqnarray}
& & R=0.68\pm0.02\,(stat.)\pm 0.05\,(syst.)\quad \mbox{(sub-GeV)}\,,  
\nonumber \\
& & R=0.68\pm0.04\,(stat.)\pm 0.08\,(syst.)\quad \mbox{(multi-GeV)}\,.  
\label{R}
\end{eqnarray}
The value of $R$ for sub-GeV events is different from unity (to which it
should be equal in no-oscillation case) by $5.9\sigma$. 

\begin{figure}[h]
\hbox to \hsize{\hfil\epsfxsize=10cm\epsfbox{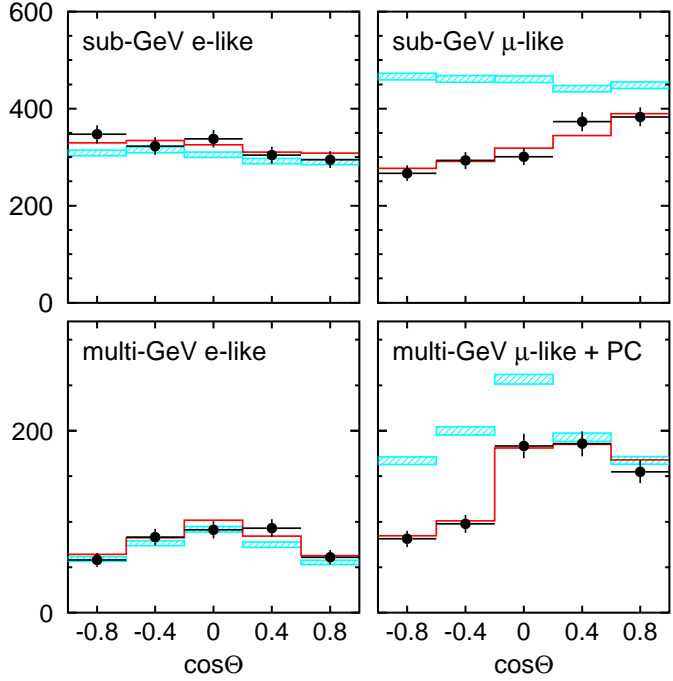}\hfil}
\caption{\small Zenith angle distributions for sub-GeV and
multi-GeV e-like and $\mu$-like events at SK. The bars show the
(no-oscillations) Monte Carlo predictions; the lines show the
predictions for $\nu_\mu \leftrightarrow \nu_\tau$ oscillations
with the best-fit parameters $\Delta m^2=3.5\times 10^{-3}$ eV$^2$, 
$\sin^2 2\theta=1.0$. From \cite{Scho}.} 
\label{zen1}
\end{figure}

We shall now discuss the zenith angle distributions of the atmospheric 
neutrino events. It should be remembered that the zenith angle distributions 
of the charged leptons which are experimentally measured  do not coincide with 
those of their parent neutrinos: for multi-GeV neutrinos the average angle 
between the momenta of neutrinos and charged leptons is about $17^\circ$, 
whereas for sub-GeV neutrinos it is close to $60^\circ$. This is properly 
taken into account in MC simulations. For PC events and upward going muons
the correlation between the directions of momenta of muons and parent 
neutrinos is much better. The distances $L$ traveled
by neutrinos before they reach the detector vary in a wide range: for 
vertically downward going neutrinos (neutrino zenith angle $\theta_\nu=0$)
$L\sim 15$ km; for horizontal neutrino trajectories ($\theta_\nu=90^\circ$) 
$L\sim 500$ km; the vertically up-going neutrinos ($\theta_\nu=180^\circ$)
cross the earth along its diameter and for them $L\sim 13,000$ km. 

In fig. \ref{zen1} the zenith angle distributions of the SK e-like and 
$\mu$-like events are shown separately for sub-GeV and multi-GeV contained 
events. The data correspond to 736 live days. (By the time these lecture 
notes were being written, the results for 848 live days of data taking had 
been reported, but no high-quality figures for zenith angle distributions 
were publicly available). 
One can see that for e-like events, the measured zenith angle distributions 
agree very well with the MC predictions (shown by bars), both in the
sub-GeV and multi-GeV samples, while for $\mu$-like events both samples 
show zenith-angle dependent deficiency of event numbers compared to 
expectations. The deficit of muon neutrinos is stronger for upward going 
neutrinos which have larger pathlengths. In the multi-GeV sample, there is
practically no deficit of events caused by muon neutrinos coming from the
upper hemisphere 
($\cos\theta>0$), whereas in the sub-GeV sample, all $\mu$-like events 
exhibit a deficit which decreases with $\cos\theta$. This pattern is
perfectly consistent with oscillations $\nu_\mu \leftrightarrow \nu_\tau$
or $\nu_\mu \leftrightarrow \nu_{s}$ where $\nu_s$ is a sterile neutrino.
Muon neutrinos responsible for 
the multi-GeV sample are depleted by the oscillations when their pathlength 
is large enough; the depletion becomes less pronounced as the pathlength 
decreases ($\cos\theta$ increases); for neutrinos coming from the upper 
hemisphere, the pathlengths are too short and there are practically no 
oscillations. Neutrinos responsible for the sub-GeV $\mu$-like events have
smaller energies, and so their oscillation lengths are smaller; therefore even 
neutrinos coming from the upper hemisphere experience sizeable depletion 
due to the oscillations. For up-going sub-GeV neutrinos the oscillation 
length is much smaller than the pathlength and they experience averaged 
oscillations. The solid line in fig. \ref{zen1} obtained with the $\nu_\mu 
\leftrightarrow \nu_\tau$ oscillation parameters in the 2-flavour scheme
$\Delta m^2=3.5 \times 10^{-3}$ eV$^2$, $\sin^2 2\theta=1.0$ gives an 
excellent fit of the data.

\begin{figure}[htb]
\hbox to \hsize{\hfil\epsfxsize=16cm\epsfbox{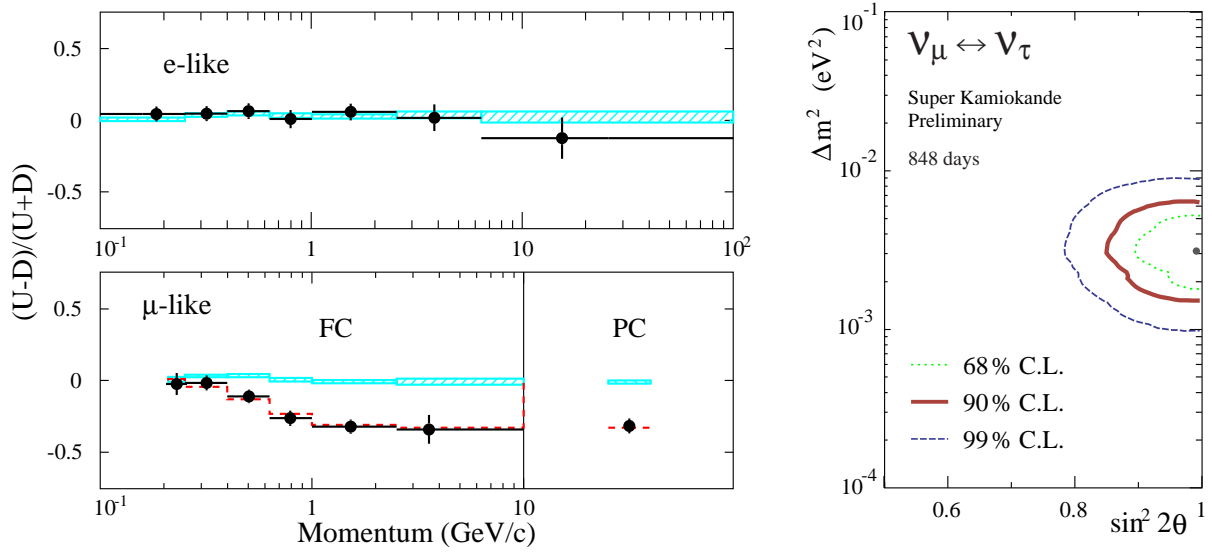}\hfil}
\caption{\small Left panel: up-down asymmetry vs event momentum for single
ring e-like and $\mu$-like events in SK. Right panel: SK allowed regions
for oscillations parameters for $\nu_\mu \leftrightarrow \nu_\tau$ channel
in 2-flavour scheme (FC + PC single ring events). From \cite{Mann}. }
\label{asymm}
\end{figure}

An informative parameter characterizing the distortion of the zenith angle 
distribution is the up-down asymmetry $A$, where up corresponds to the events 
with $\cos\theta <-0.2$ and down to those with $\cos\theta > 0.2$. 
The flux of atmospheric neutrinos is expected to  
be nearly up-down symmetric for neutrino energies $E\aprge 1$ GeV, with 
minor deviations coming from geomagnetic effects 
which are well understood and can be accurately taken into account. 
In particular, at the geographical location of the SK detector 
small positive asymmetry is expected.
%
Any significant deviation of the up-down asymmetry of neutrino induced
events from the asymmetry due to the geomagnetic effects 
is an indication of neutrino oscillations or some other new neutrino
physics. The asymmetry measured for the SK multi-GeV $\mu$-like events is 
\cite{Mann}
\be
A=\frac{U-D}{U+D}=-0.32\pm 0.04\,(stat.) \pm 0.01\,(syst.)\,,
\label{A}
\ee
i.e. is negative and differs from zero by almost $8\sigma$! 
The dependence of the asymmetries 
for e-like FC and $\mu$-like FC+PC events on the event momentum is shown
in fig. \ref{asymm} (left panel). One can see that for e-like events 
$A\simeq 0$ for all momenta. At the same time, for $\mu$-like events the
asymmetry is close to zero at low momenta and decreases with momentum.
This is easily understood in terms of the $\nu_\mu$ oscillations. For very 
small momenta, the oscillation length is small and both up-going and 
down-going neutrino fluxes are depleted by oscillations to about the same 
extent; in addition, loose correlation between the directions of the momenta 
of the charged lepton and of its patent neutrino tends to smear out the 
asymmetry at low energies. 
With increasing momentum the oscillation length increases, and the 
pathlength of down-going neutrinos becomes too small for oscillations to 
develop. The right panel of fig. \ref{asymm} gives the allowed values of the
oscillation parameters following from the SK FC and PC event data. 
The best fit corresponds to $\Delta m^2=3.1 \times 10^{-3}$ eV$^2$,
$\sin^2 2\theta=0.99$ and has a very good $\chi^2/d.o.f.=55/67$. 
In this analysis the poorly known overall normalization
factor of the neutrino flux was considered as a free parameter, and the best 
fit was achieved with 5\% upward renormalization of the flux of Honda et al. 
\cite{Honda}. 
The above value of $\chi^2$ should be compared with that of the no-oscillation 
hypothesis: $\chi^2/d.o.f.=177/69$, which is a very poor fit. 

\begin{figure}[htb]
\hbox to \hsize{\hfil\epsfxsize=16cm\epsfbox{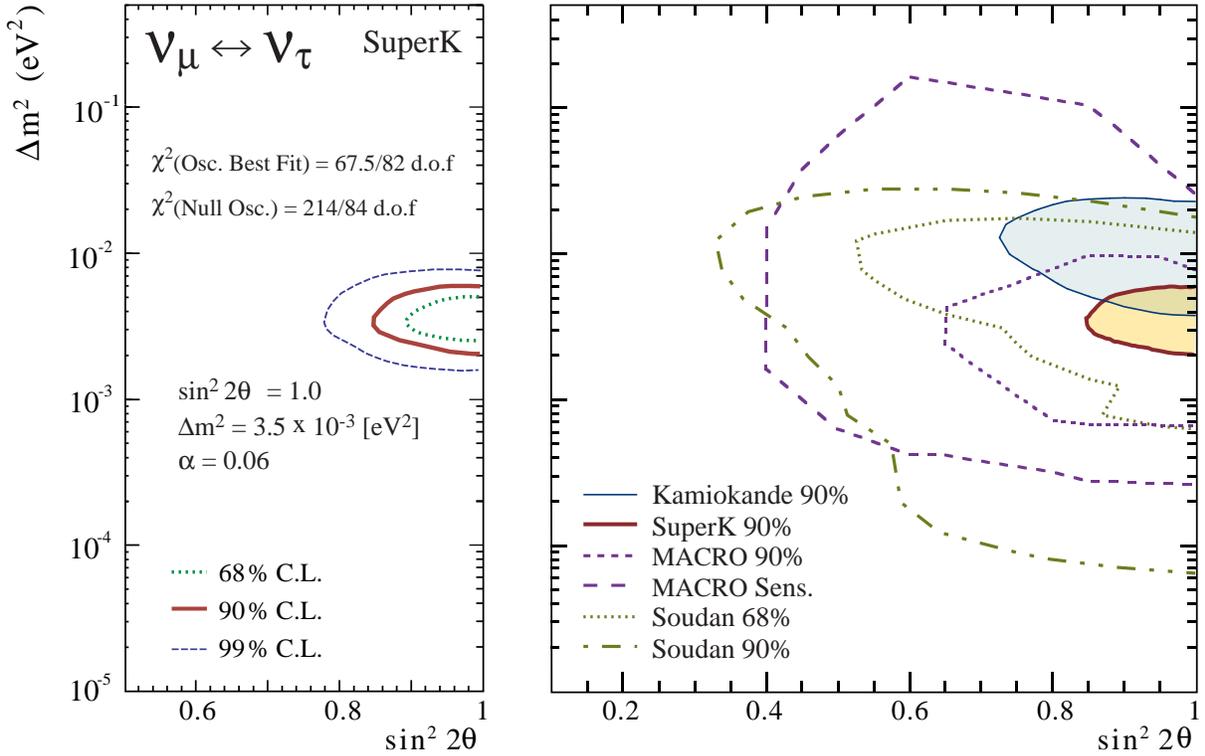}\hfil}
\caption{\small Left panel: SK allowed regions of oscillation parameters
for $\nu_\mu \leftrightarrow \nu_\tau$ channel in 2-flavour scheme 
(FC + PC single ring events + upward through-going + upward stopping events). 
Right panel: comparison of allowed values of oscillation parameters obtained 
in different experiments. From \cite{Mann}.}
\label{sum}
\end{figure}

The SK data show evidence for neutrino oscillations not only in their 
FC and PC $\mu$-like events: upward stopping and upward through-going 
events also demonstrate zenith angle dependent deficiency of muon
neutrinos consistent with neutrino oscillations, although the statistics
for up-going muons is lower than that for contained events. Fig. \ref{sum} 
(left panel) summarizes the allowed region for 2-flavour neutrino oscillations 
obtained from the combined analysis of the SK FC and PC single ring events 
(848 live days), upward through-going (923 live days) and upward stopping
events (902 live days). The best fit ($\chi^2/d.o.f=67.5/82$) was obtained 
for $\Delta m^2=3.5 \times 10^{-3}$ eV$^2$, $\sin^2 2\theta=1.0$ and the 
upward 6\% renormalization of the flux of Honda et al. The no-oscillation  
fit gives $\chi^2/d.o.f.=214/84$. The right panel gives the comparison of
the allowed regions obtained in different atmospheric neutrino experiments. 
The evidence for neutrino oscillations is very impressive. 

Are neutrino oscillations that are responsible for the depletion of the
$\nu_\mu$ flux $\nu_\mu\leftrightarrow \nu_\tau$ or $\nu_\mu\leftrightarrow 
\nu_s$? For contained events, the oscillation probabilities in these two
channels are nearly the same and the data can be fitted equally well in
both cases, with very similar allowed ranges of the oscillation parameters.  
However, for higher energy upward going events there are important 
differences between  these two cases.  
In the 2-flavour scheme, $\nu_\mu\leftrightarrow \nu_\tau$ oscillations are 
not affected by matter because the interactions of $\nu_\mu$ and $\nu_\tau$ 
with matter are identical. However, sterile neutrinos do not interact with 
matter at all, and therefore the $\nu_\mu\leftrightarrow \nu_s$ oscillations 
are affected by the matter-induced potential $V_\mu-V_s=V_\mu$. At low 
energies, the kinetic energy difference $\Delta m^2/2E$ dominates over 
$V_\mu$, and the earth's matter effects are unimportant. They become important 
at higher energies, when $\Delta m^2/2E\sim V_\mu$; at very high energies, when
$\Delta m^2/2E\ll V_\mu$, matter strongly suppresses neutrino oscillations 
both in $\nu_\mu\leftrightarrow \nu_s$ and $\bar{\nu}_\mu\leftrightarrow 
\bar{\nu}_s$ channels (see eq. (\ref{amp}) where $N_e$ has to be replaced  
by $\mp N_n/2$ for these oscillation channels). Therefore the oscillations
of high energy neutrinos travelling significant distances in the earth
should be strongly suppressed in this case. 

\begin{figure}[htb]
\hbox to \hsize{\hfil\epsfxsize=12cm\epsfbox{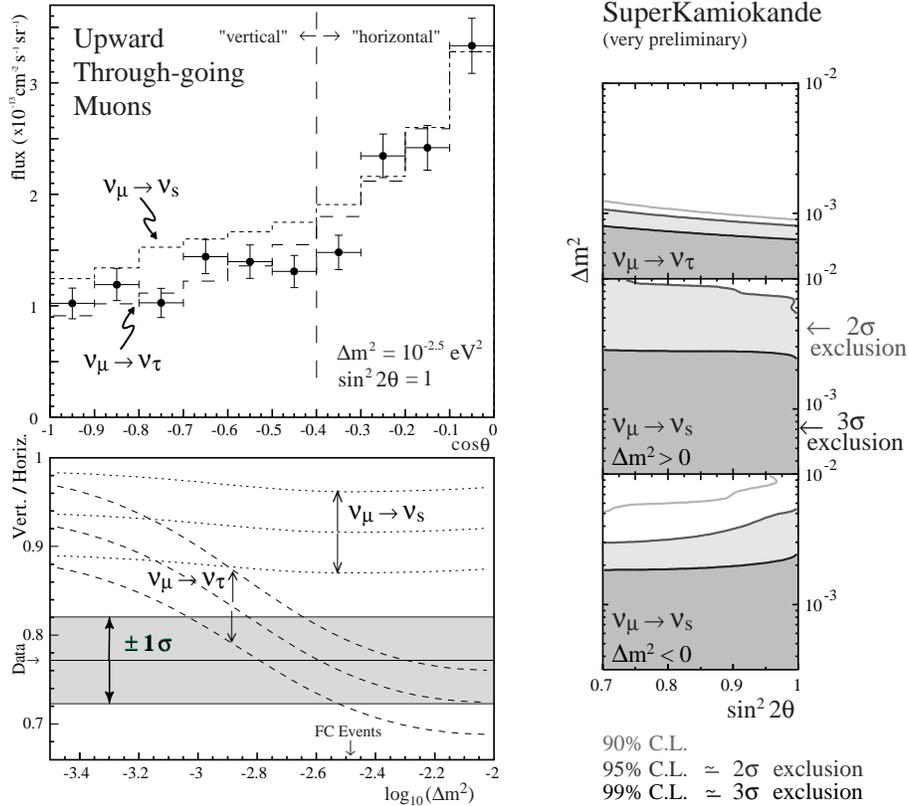}\hfil}
\caption{\small Left top panel: Zenith angle distribution of upward
through-going events at SK. The dashed and short-dashed histograms show the 
$\nu_\mu \leftrightarrow \nu_\tau$ and $\nu_\mu\leftrightarrow \nu_s$ solutions 
respectively. Left bottom: Vertical/Horizontal ratio for upward through-going 
muons along with the predictions of $\nu_\mu \leftrightarrow \nu_\tau$ and 
$\nu_\mu\leftrightarrow \nu_s$ oscillation scenarios.  Right panel: Exclusion 
regions obtained from Up/Down ratio for energetic PC sample ($E_{vis}>5$ GeV, 
$\langle E \rangle \sim 25$ GeV) and Vertical/Horizontal ratio for 
upward through-going events. From \cite{Mann}.}
\label{ster}
\end{figure}

Fig. \ref{ster} (upper left panel) shows the zenith angle distribution of
the upward through-going events ($\langle E \rangle\sim 100$ GeV) along 
with the predictions of $\nu_\mu \leftrightarrow \nu_\tau$ and 
$\nu_\mu\leftrightarrow \nu_s$ scenarios with parameters obtained from the
fits of contained events. It can be seen that the $\nu_\mu\leftrightarrow 
\nu_\tau$ oscillation channel describes the ``vertical'' ($\cos\theta<-0.4$) 
events better than the $\nu_\mu\leftrightarrow \nu_s$ channel. For horizontal 
and nearly horizontal events ($-0.4<\cos\theta<0$) both scenarios give similar 
predictions since the neutrinos do not travel inside the earth or travel only 
small distances there. This is further illustrated by the left bottom panel 
of fig. \ref{ster} in which the Vertical/Horizontal event ratio is plotted. 
The excluded ranges of the oscillation parameters for various oscillation 
channels are shown in the right panel of fig. \ref{ster}. One can conclude 
that the $\nu_\mu\leftrightarrow \nu_s$ oscillations are currently disfavoured 
at the $2\sigma$ level, although not yet completely ruled out as a possible 
solution of the atmospheric neutrino problem.

Another way to discriminate between $\nu_\mu\leftrightarrow \nu_\tau$ and 
$\nu_\mu\leftrightarrow \nu_s$ as the two possible main channels of the 
atmospheric neutrino oscillations is through the detection of single
pions produced in the NC reactions $\nu N\to \nu N \pi^0$. Sterile neutrinos 
do not participate in these reactions and therefore $\nu_\mu\leftrightarrow 
\nu_s$ oscillations would deplete the number of the observed $\pi^0$'s. 
Unfortunately, cross sections of these NC reactions are only known with 
large uncertainties, 
and at the moment a clear discrimination between the $\nu_\mu\leftrightarrow 
\nu_\tau$ and $\nu_\mu\leftrightarrow \nu_s$ oscillations of atmospheric
neutrinos through the detection of single neutral pions does not seem to
be possible. The situation may change in the near future, however, as the 
K2K Collaboration is planning to perform high statistics measurements 
of neutrino interactions in their near detector.

Can $\nu_\mu\leftrightarrow \nu_e$ oscillations be responsible for the 
observed anomalies in the atmospheric neutrino data? 
%
%
The answer is no, at least not as the dominant channel. Explaining the data 
requires oscillations with large mixing equal or close to the maximal one;  
$\nu_\mu\leftrightarrow \nu_e$ oscillations would then certainly lead to 
a significant distortion of the zenith angle distributions of the e-like 
contained events, contrary to observations. In addition, for $\Delta m^2$ in 
the range $\sim 10^{-3}$ eV$^2$ which is required by the atmospheric
neutrino data, $\nu_\mu\leftrightarrow \nu_e$ oscillations are severely 
restricted by the CHOOZ reactor antineutrino experiment, which excludes 
these oscillations as the main channel of the atmospheric neutrino
oscillations (see sec. \ref{reacc}). 
 
However, $\nu_\mu\leftrightarrow \nu_e$ 
and $\nu_e\leftrightarrow \nu_\tau$ 
can be present as 
subdominant channels of the oscillations of atmospheric neutrinos. 
This may lead to interesting matter effects on oscillations of neutrinos
crossing the earth on their way to the detector. 
Matter can strongly affect $\nu_e\leftrightarrow \nu_{\mu,\tau}$ and 
$\bar{\nu}_e\leftrightarrow \bar{\nu}_{\mu,\tau}$ oscillations leading to
an enhancement of the oscillation probabilities for neutrinos and
suppression for antineutrinos or vice versa, depending on the sign of 
the corresponding mass squared difference. For certain neutrino energies, 
one can expect an enhancement due to the MSW effect in the mantle or in the 
core of the earth. 
In addition, there can be a specific enhancement effect, different from the 
MSW one, due to a cumulative build-up of the transition probability for 
neutrinos crossing the earth's mantle, core and then again mantle (parametric 
resonance \cite{ETC,Akh3,LS,ADLS}). The matter enhancement effects are 
expected to take place in a relatively narrow range of neutrino energies, and 
it is rather difficult to observe them experimentally. Yet the possibility
of the  observation of matter effects on oscillations of atmospheric neutrinos 
is very exciting. 

Since three neutrino flavours are known to exist, oscillations of atmospheric 
neutrinos should in general be considered in the full 3-flavour framework 
(assuming that no sterile neutrinos take part in these oscillations). As 
follows from the analyses of solar and atmospheric neutrino data, there 
should be two distinct mass squared difference scales in the three neutrino 
framework, $\Delta m^2_{atm}\sim 10^{-3}\,\mbox{eV}^2$ and $\Delta m^2_\odot 
\aprle 10^{-4}\,\mbox{eV}^2$. The hierarchy $\Delta m^2_{atm}\gg \Delta 
m_\odot^2$ means that one of the three neutrino mass eigenstates (which we 
denote $\nu_3$) is separated by the larger mass gap from the other two.  
The neutrino mass squared differences thus satisfy 
$|\Delta m_{31}|^2\simeq |\Delta m_{32}|^2 \gg |\Delta m_{21}|^2$. 
We know already 
that $\nu_\mu \leftrightarrow \nu_\tau$ should be the main channel of 
oscillations whereas $\nu_e\leftrightarrow \nu_{\mu,\tau}$ oscillations can 
only be present as the subdominant channels. In the 3-flavour framework
this means that the element $|U_{e3}|$ of the lepton mixing matrix 
(\ref{mix2}) is small.
%
The smallness of  $|U_{e3}|$ means that the $\nu_\mu \leftrightarrow \nu_\tau$
and $\nu_\mu \leftrightarrow \nu_e$ oscillations approximately decouple, and 
the 2-flavour analysis gives a good first approximation. In terms of the 
parametrization (\ref{U3}) of the lepton mixing matrix, the mixing angle 
describing the main channel of the atmospheric neutrino oscillations is 
$\theta_{23}$, and its best-fit value following from the SK data is
$45^\circ$. This fact and the smallness of $|U_{e3}|=s_{13}$ mean that the 
mass eigenstate $\nu_3$ mainly consists of the flavour eigenstates $\nu_\mu$ 
and $\nu_\tau$ with approximately equal weights, while the admixture of 
$\nu_e$ to this state is small or zero. 

Can one observe true 3-flavour effects in the atmospheric neutrino 
oscillations? In principle, yes. Although in the 2-flavour scheme
$\nu_\mu\leftrightarrow \nu_\tau$ oscillations are not affected by matter, 
in the full 3-flavour approach these oscillations are modified by matter 
because of the mixing with $\nu_e$. These matter effects are, however,
rather small because of the smallness of $|U_{e3}|$. 
Three-flavour effects can also lead to interesting phenomena in the 
subdominant $\nu_\mu\leftrightarrow \nu_e$ and $\nu_e\leftrightarrow \nu_\tau$  
oscillations. To see this, let us consider the fluxes $F_\mu$ and $F_e$ of 
atmospheric $\nu_\mu$ and $\nu_e$ in the presence of oscillations. 
As we discussed in sec. \ref{3flmatt}, in the case of the hierarchy 
$|\Delta m_{31}|^2\simeq |\Delta m_{32}|^2 \gg |\Delta m_{21}|^2$ the
oscillation 
probabilities in the 3-flavour scheme simplify significantly and can be
expressed through the evolution matrix of the effective 2-flavour problem. 
Using eqs. (\ref{3f3}) - (\ref{3f6}) one can express $F_\mu$ and $F_e$
through the corresponding no-oscillation fluxes $F_\mu^0$ and $F_e^0$ as 
%
\begin{eqnarray}
& F_e = F_e^0 \left[ 1 + P_2(r s_{23}^2 - 1)\right]\,,
\label{fluxe} 
\\
& F_{\mu} = F_{\mu}^0 \left[ 1 - s_{23}^4 \left(1 - \frac{1}{r s_{23}^2}
\right) P_2 + 2s_{23}^2 c_{23}^2 \left(Z \cos\Phi+W_3 \sin\Phi - 1
\right) \right],
\label{fluxmu}
\end{eqnarray}
where
$r(E, \theta_\nu) = F_{\mu}^0(E, \theta_\nu)/ F_e^0(E, \theta_\nu)$. 
It is interesting that the subdominant $\nu_e\leftrightarrow \nu_{\mu,\tau}$ 
oscillations may lead either to an excess or to a deficiency of e-like events 
depending on the values of $r$ and $s_{23}$. Indeed, the effect of 
oscillations on the electron neutrino flux is proportional to the factor
$(r s_{23}^2 - 1)$. If one assumes $r = 2$, there will be an excess of
e-like events for $\theta_{23}>45^\circ$ and a deficiency for $\theta_{23} 
<45^\circ$. The SK best fit is $\theta_{23}=45^\circ$; for this value of 
$\theta_{23}$ there would be no deviation from the prediction for $r=2$.
However, for upward going neutrinos in the multi-GeV range $r$ is typically 
3 -- 3.5 rather than 2, so there should be an excess of e-like events for 
$\theta_{23}\aprge 33^\circ$ and a deficiency for $\theta_{23}\aprle  
33^\circ$. Thus, the distortions of the zenith angle distributions of the 
e-like events due to the subdominant $\nu_e\leftrightarrow \nu_{\mu,\tau}$ 
oscillations should depend on the value of the mixing angle $\theta_{23}$ 
governing the dominant $\nu_\mu\leftrightarrow \nu_\tau$ oscillation channel. 
However, these distortions are expected to be small because of the smallness 
of $|U_{e3}|$ (notice that the probability $P_2$ in (\ref{fluxe}) and
(\ref{fluxmu}) vanishes in the limit $U_{e3}\to 0$). 
For recent analyses of the SK atmospheric neutrino data in the 3-flavour
framework, see \cite{Lisi2,ADLS}.

If the solution of the solar neutrino problem is the LMA MSW effect,
$\Delta m_{21}^2$ can 
be as large as $\sim 10^{-4}$ eV$^2$ and its influence of the atmospheric
neutrino oscillations cannot be neglected. In that case eqs. (\ref{fluxe}) 
and (\ref{fluxmu}) have to be modified. It was shown in \cite{PerSm} that
in this case the subdominant oscillations can lead to a negative up-down 
asymmetry of e-like events at relatively large momenta. If the trend of 
slightly decreasing with momentum asymmetry of the e-like events which 
can be seen in the left top panel of fig. \ref{asymm} is of physical
origin and not just due to statistical fluctuations, it can be
explained by relatively large $\Delta m_{21}^2$. 

Are the standard neutrino oscillations the sole possible explanation of
the observed atmospheric neutrino anomalies? In principle, other explanations 
are possible. Those include exotic types of neutrino oscillations  -- 
matter-induced oscillations due to flavour-changing interactions of neutrinos 
with medium \cite{Jose2}, oscillations due to small violations of the Lorentz 
or CPT invariance or of the gravitational equivalence principle \cite{exotic}, 
and also neutrino decay \cite{Pakvasa}. Exotic oscillations lead to periodic 
variations of the $\nu_\mu$ survival probability with the oscillation lengths
$l_{osc}\propto E^{-n}$ where $n=0$ in the case of flavour-changing neutrino 
interactions or violation of CPT invariance, and $n=1$ for oscillations
due to the violations of the Lorentz invariance or equivalence principle. 
This has to be contrasted with $n=-1$ in the case of the standard neutrino 
oscillations. The energy dependence of the oscillation length can be
tested in the atmospheric neutrino experiments as the energies of detected 
neutrinos span more than 3 orders of magnitude. The analysis was performed 
in \cite{Lisi3}, and the authors found that the fit of the atmospheric 
neutrino data assuming oscillations with $l_{osc}\propto E^{-n}$ gives   
$n=-0.9\pm 0.4$ at 90\% c.l.. This clearly favours the standard oscillations 
over the exotic ones. In contrast to this, the neutrino decay mechanism fits 
the SK data quite well, the quality of the fit being as good as the one for 
the standard neutrino oscillations \cite{Barger1}. However, the assumed 
neutrino decay mode (into a sterile neutrino and a Majoron) requires
particle physics models which look less appealing than those that provide 
neutrino masses required for the standard neutrino oscillations. 

\noindent
{\it Problem 25. Using eqs. (\ref{3f3}) - (\ref{3f6}) of sec. \ref{3flmatt} 
derive expressions (\ref{fluxe}) and (\ref{fluxmu}) for the atmospheric
$\nu_e$ and $\nu_\mu$ fluxes in the presence of oscillations in the case of 
the hierarchy $|\Delta m_{31}|^2 |\simeq \Delta m_{32}|^2 \gg |\Delta
m_{21}|^2$.} 

\section{Reactor and accelerator neutrino experiments
\label{reacc}}

In reactor neutrino experiments oscillations of electron antineutrinos 
into another neutrino species are searched for by studying possible depletion 
of the $\bar{\nu}_e$ flux beyond the usual geometrical one 
\cite{gratta,dilella}. These are the disappearance experiments, because 
the energies of the reactor $\bar{\nu}_e$'s ($\langle E\rangle \simeq$ 3 MeV) 
are too small to allow the detection of muon or tauon antineutrinos in CC 
experiments. Small $\bar{\nu}_e$ energy makes the reactor neutrino experiments 
sensitive to oscillations with rather small values of $\Delta m^2$. 

\begin{figure}[h] 
\setlength{\unitlength}{1cm}
\hbox{\hfill
\epsfxsize=5.5cm\epsfbox{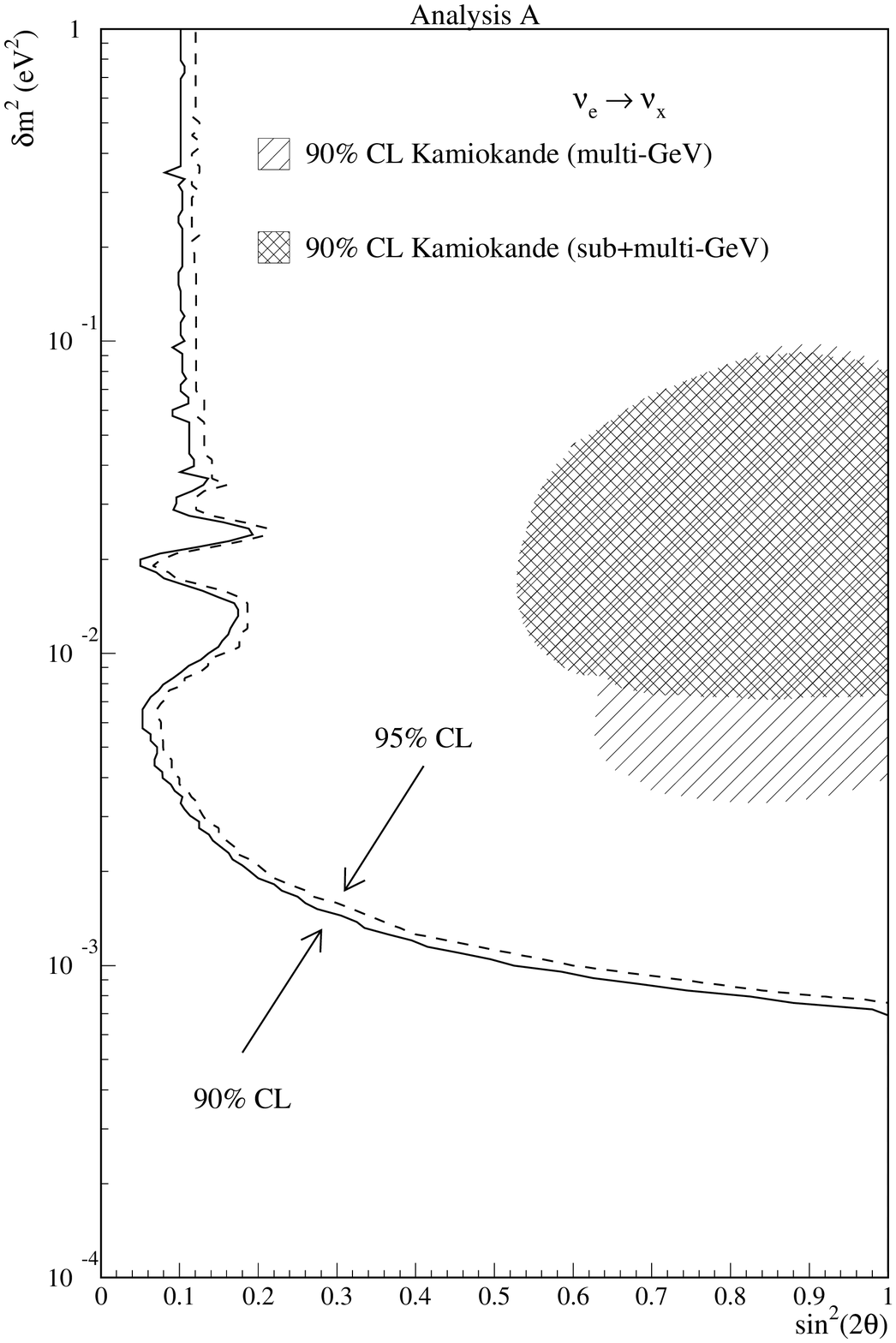}
\hspace{1.0cm}
\epsfxsize=6.2cm\epsfbox{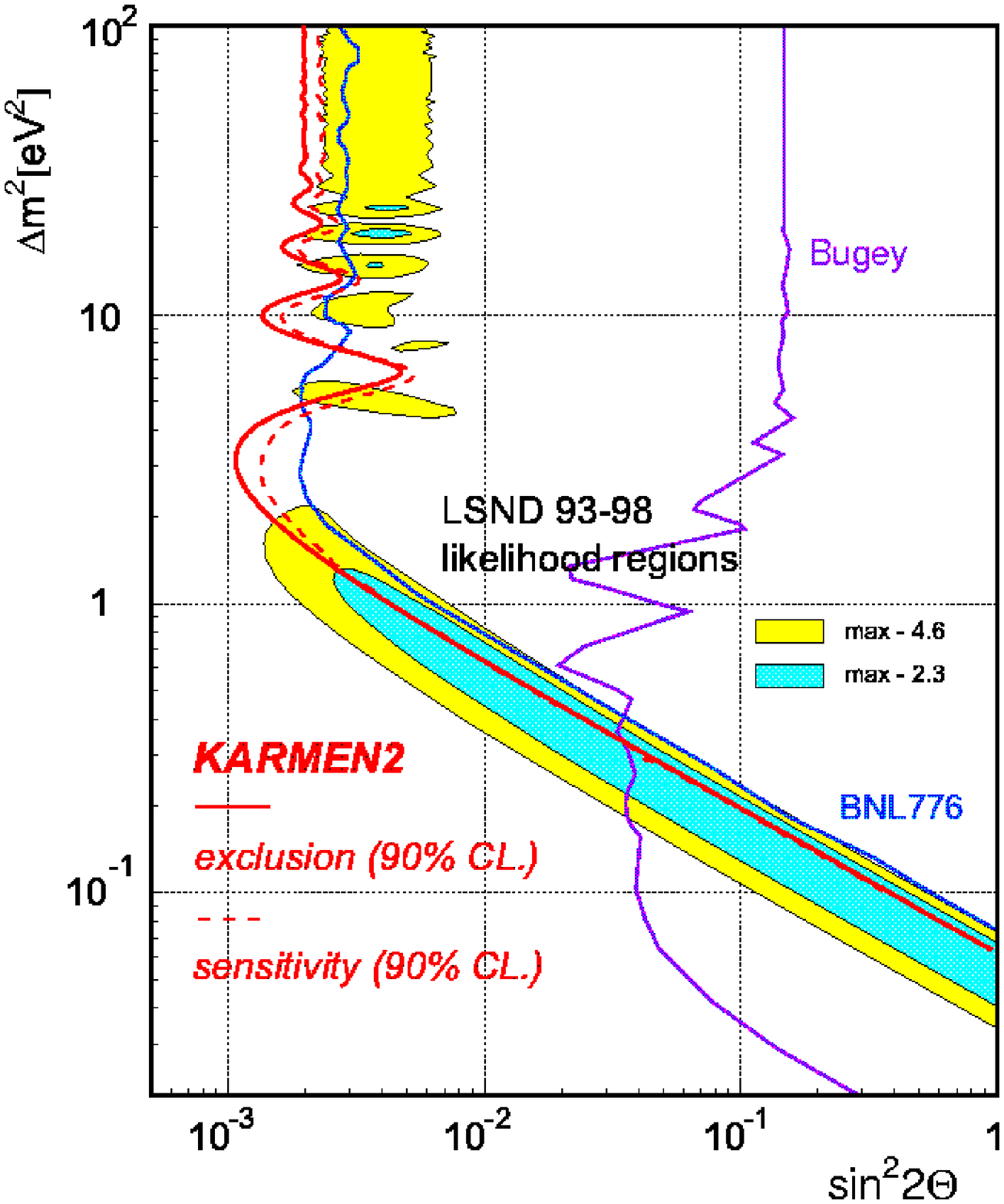}
\hfill}
\caption{\small Left panel: CHOOZ exclusion region for 
$\bar{\nu}_e\to \bar{\nu}_x$ oscillations. The region to the right of the 
curves are excluded. Shaded areas correspond to the Kamiokande allowed region 
for atmospheric $\nu_\mu\to \nu_e$ oscillations \cite{CHOOZ}. 
Right panel: LSND allowed region for $\bar{\nu}_\mu\to \bar{\nu}_e$ 
oscillations (shaded areas) along with KARMEN, BNL776 and Bugey exclusion
regions \cite{dilella}. }
\label{choozLSND} 
\end{figure}

Up to now, no evidence for neutrino oscillations has been 
found in the reactor neutrino experiments, which allowed to exclude
certain regions in the neutrino parameter space. The best 
constraints were obtained by the CHOOZ experiment in France \cite{CHOOZ} 
(see fig. \ref{choozLSND}). As follows from eq. (\ref{prob6}), in the case
of the hierarchy $\Delta m_{31}^2 \simeq \Delta m_{32}^2 \gg \Delta m_{21}^2$, 
the survival probability of $\nu_e$ (which by CPT invariance is equal to that
of $\bar{\nu}_e$) in the 3-flavour scheme has a simple form which coincides
with that in the two-flavour framework. Therefore the excluded regions 
obtained in the 2-flavour analysis apply to the 3-flavour case as well, with 
the identification $\Delta m^2 =\Delta m_{31}^2\simeq \Delta m_{32}^2$, 
$\sin^2 2\theta =\sin^2 2\theta_{13}$.  For the values of $\Delta m_{31}^2 
\equiv \Delta m_{atm}^2$ in the Super-Kamiokande allowed region
$(2 - 6)\cdot 10^{-3}$ eV$^2$, the CHOOZ results give the following 
constraint on the element $U_{e3}$ of the lepton mixing matrix: $|U_{e3}|^2 
(1-|U_{e3}|^2)<0.045 - 0.02$, i.e. $|U_{e3}|$ is either small or close to 
unity. The latter possibility is excluded by solar and atmospheric
neutrino observations, and one finally obtains 
\be
\sin^2 \theta_{13}\equiv |U_{e3}|^2 \le (0.047 - 0.02)\, \quad
{\rm for}\quad\Delta m_{31}^2
=(2 - 6)\cdot 10^{-3}~{\rm eV}^2\,.
\label{choozres}
\ee
This is the most stringent constraint on $|U_{e3}|$ to date.

\begin{figure}[h] 
\setlength{\unitlength}{1cm}
\hbox{\hfill
\epsfxsize=5.0cm\epsfbox{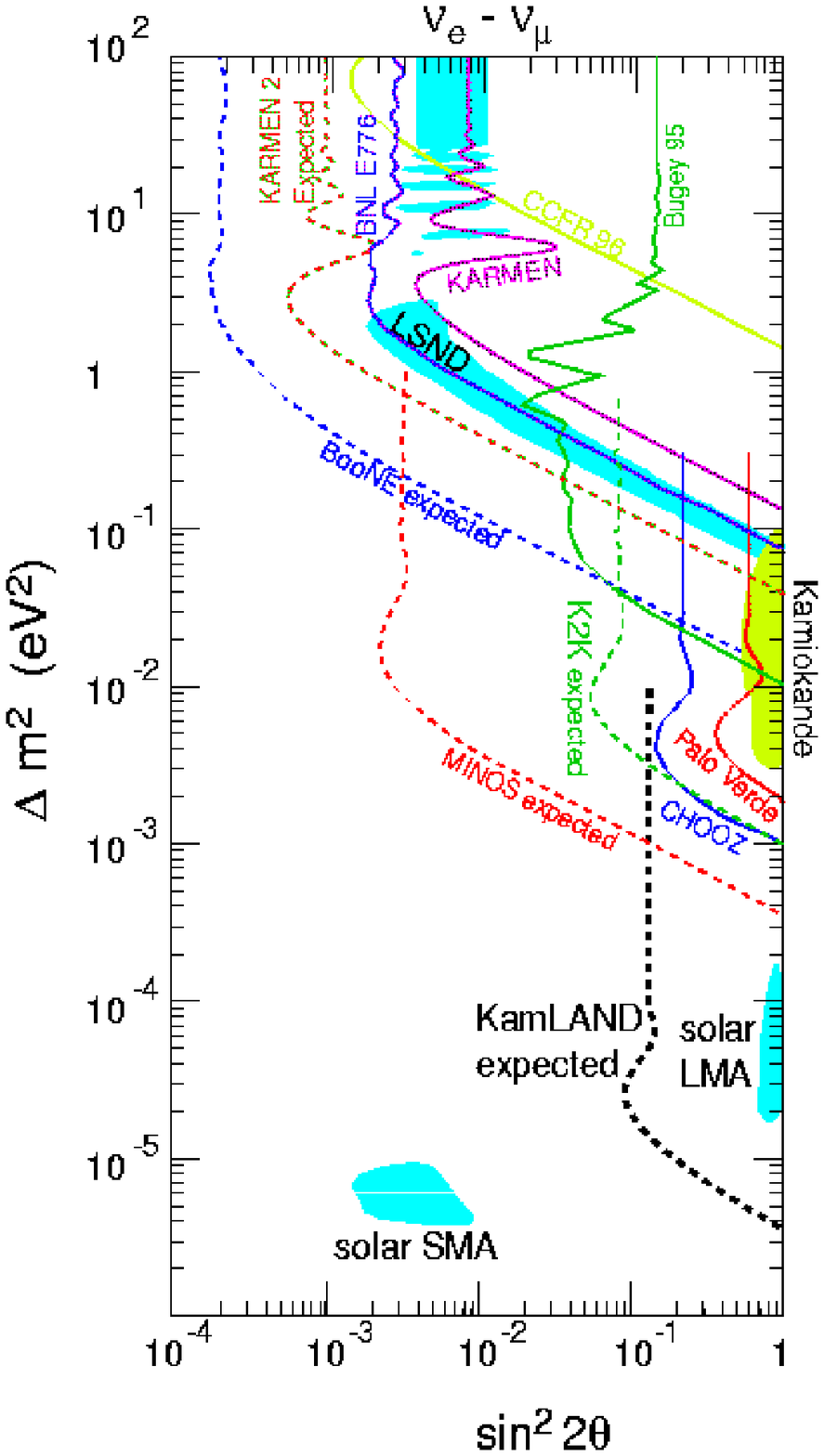}
\hspace{1.0cm}
\epsfxsize=6.7cm\epsfbox{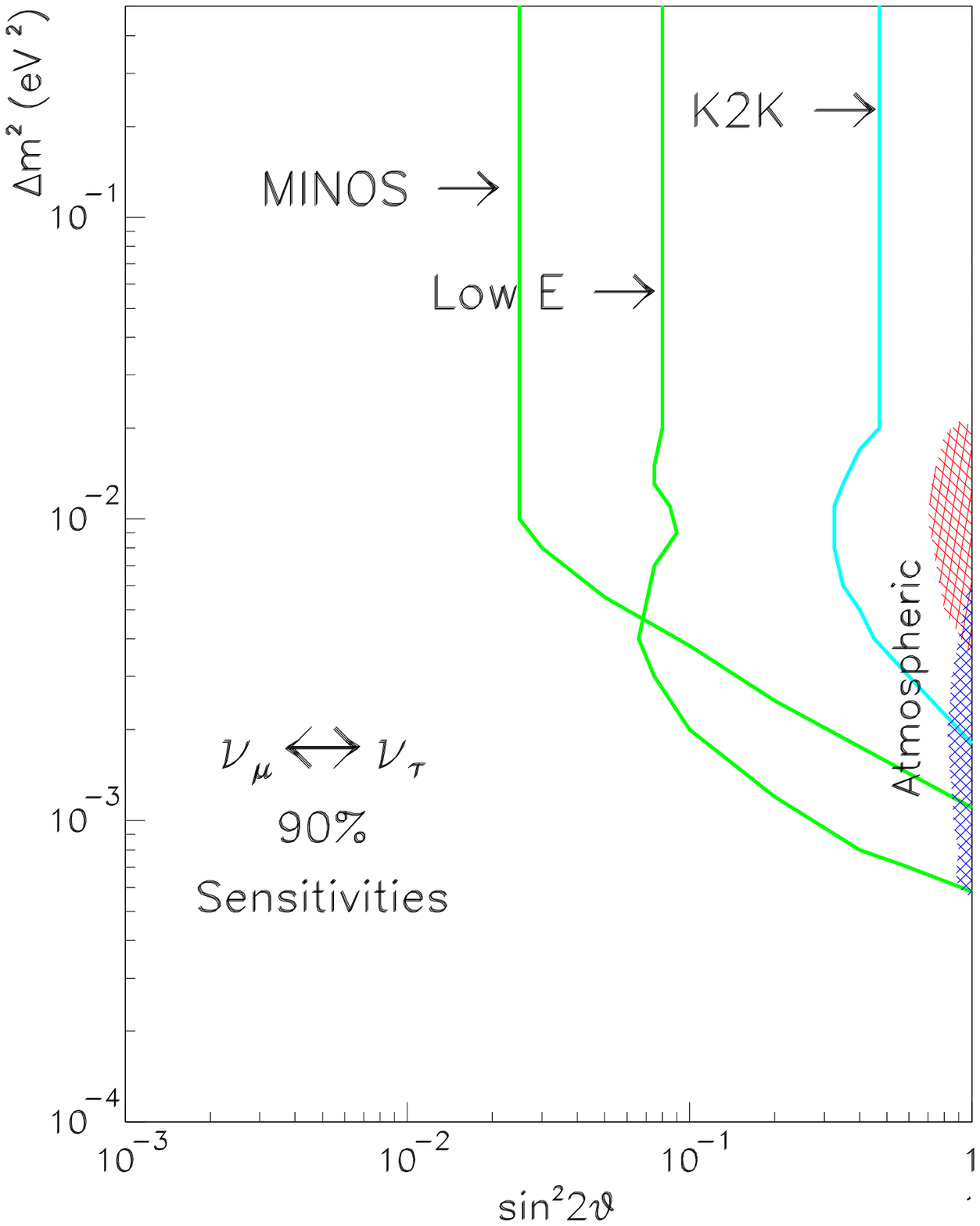}
\hfill}
\caption{\small Left panel: results of the present and sensitivities of 
future $\nu_e\leftrightarrow \nu_\mu$ oscillations searches (90\% c.l.) 
\cite{ASuz}. Right panel: sensitivities of MINOS and K2K long baseline 
experiments \cite{Conrad}. }
\label{future} 
\end{figure}

Presently, a long baseline reactor experiment KamLAND is under construction 
in Japan. This will be a large liquid scintillator detector
experiment using the former Kamiokande site. KamLAND will detect electron 
antineutrinos coming from several Japanese power reactors at an average 
distance of about 180 km. KamLAND is scheduled to start taking data in 2001 
and will be sensitive to values of $\Delta m^2$ as low as $4\times 10^{-6}$ 
eV$^2$, i.e. in the range relevant for the solar neutrino oscillations! It is 
expected to be able to probe the LMA solution of the solar neutrino problem 
(see fig. \ref{future}). It may also be able to directly detect solar 
$^8$B and $^7$Be neutrinos after its liquid scintillator has been purified 
to ultra high purity level by recirculating through purification \cite{ASuz}. 

In the 3-flavour framework, the oscillations due to the large 
$\Delta m_{atm}^2\sim 3\times 10^{-3}$ eV$^2$ should be in the averaging 
regime in the KamLAND experiment 
and therefore the $\bar{\nu}_e$ survival probability should be described by 
eqs. (\ref{prob7}) and (\ref{prob8}) rather than by (\ref{prob6}). In the
case of SMA, LMA and VO solutions of the solar neutrino problem the 
$\bar{\nu}_e$ survival probability at KamLAND is then just $P(\bar{\nu}_e\to 
\bar{\nu}_e)\simeq c_{13}^4$, i.e. the mixing angle $\theta_{13}$ is directly 
measured. The CHOOZ data allow $c_{13}^4 \aprge 0.90$ for $\Delta m_{atm}^2$ 
in the SK range. In the case of the LMA solution, $P(\bar{\nu}_e\to 
\bar{\nu}_e)_{\small 3fl}\simeq c_{13}^4 P(\bar{\nu}_e\to \bar{\nu}_e)_{\small 
2fl}$ where $P(\bar{\nu}_e\to \bar{\nu}_e)_{\small 2fl}$ is the usual 2-flavour
survival probability (\ref{prob8}). Thus, in this case the 2-flavour
analysis 
is modified by a non-vanishing $\theta_{13}$. 


There have been a number of accelerator experiments looking for neutrino 
oscillations. In all but one no evidence for oscillations was found and 
constraints on oscillation parameters were obtained \cite{dilella,Conrad}.
The LSND Collaboration 
have obtained an evidence for $\bar{\nu}_\mu \to \bar{\nu}_e$ and 
$\nu_\mu \to \nu_e$ oscillations \cite{LSND}. The LSND result is the only 
indication for neutrino oscillations that is a signal and not a deficiency. 
The KARMEN experiment is looking for neutrino oscillations in $\bar{\nu}_\mu 
\to \bar{\nu}_e$ channel. No evidence for oscillations has been obtained, and 
part of the LSND allowed region has been excluded. In fig. \ref{choozLSND} 
(right panel) the results from LSND and KARMEN experiments are shown along 
with the relevant constraints from the BNL776 and Bugey experiments. One
can see that the only domain of the LSND allowed region which is presently
not excluded is a narrow strip with $\sin^2 2\theta \simeq 1.5\times 10^{-3} 
- 4\times 10^{-2}$ and $\Delta m^2 \simeq 0.2 - 2$ eV$^2$.

The existing neutrino anomalies (solar neutrino problem, atmospheric neutrino 
anomaly and the LSND result), if all interpreted in terms of neutrino
oscillations, require three different scales of mass squared differences:
$\Delta m^2_\odot \aprle 10^{-4}$ eV$^2$, $\Delta m^2_{atm} \sim 10^{-3}$ 
eV$^2$ and $\Delta m^2_{\rm LSND}\aprge 0.2$ eV$^2$. This is only possible
with four (or more) light neutrino species. The fourth light neutrino cannot 
be just the 4th generation neutrino similar to $\nu_e$, $\nu_\mu$ and 
$\nu_\tau$ because this would be in conflict with the experimentally measured 
width of $Z^0$ boson [see eq. (\ref{nnu})]. It can only be an electroweak 
singlet (sterile) neutrino. Therefore the LSND result, if correct, would
imply the existence of a light sterile neutrino. 

Out of all experimental evidences for neutrino oscillations, the LSND result 
is the only one that has not yet been confirmed by other experiments. It is 
therefore very important to have it independently checked. This will be
done by the MiniBooNE (first phase of BooNE) experiment at Fermilab. 
MiniBooNE will be capable of observing both $\nu_\mu \to \nu_e$ appearance
and $\nu_\mu$ disappearance. If the LSND signal is due to $\nu_\mu \to \nu_e$ 
oscillations, MiniBooNE is expected to detect an excess of $\sim$1500 $\nu_e$ 
events during its first year of operation, establishing the oscillation signal 
at $\sim 8\sigma$ level (see fig. \ref{future}). If this happens, the second 
detector will be installed, with the goal to accurately measure the 
oscillation parameters. MiniBooNE will begin taking data in 2002.

A number of long baseline accelerator neutrino experiments have been proposed 
to date. They are designed to independently test the oscillation 
interpretation of the results of the atmospheric neutrino experiments, 
accurately measure the oscillation parameters and to (possibly) identify 
the oscillation channel. The first of these experiments, K2K (KEK to 
Super-Kamiokande) already started taking data in 1999. It has a baseline 
of 250 km and is looking for $\nu_\mu$ disappearance. K2K should be able 
to test practically the whole region of oscillation parameters allowed
by the SK atmospheric neutrino data except perhaps the lowest-$\Delta m^2$ 
part of it (see fig. \ref{future}). Two other long baseline projects, 
NuMI - MINOS (Fermilab to Soudan mine in the US) and CNGS (CERN to Gran 
Sasso in Europe), each with the baseline of about 730 km, will be sensitive 
to smaller values of $\Delta m^2$ and should be able to test the whole 
allowed region of SK (fig. \ref{future}). MINOS will look for $\nu_\mu$ 
disappearance and spectrum distortions due to $\nu_\mu \to \nu_x$ 
oscillations. It may run in three different energy regimes -- high,
medium and low energy. MINOS is scheduled to start data taking 
in the end of 2002. CERN to Gran Sasso will be an appearance experiment
looking specifically for $\nu_\mu \to \nu_\tau$ oscillations. It will also  
probe $\nu_\mu \to \nu_e$ appearance. 
At the moment, two detectors have been approved for participation in the 
experiment -- OPERA and ICANOE. The whole project was approved in December
of 1999 and the data taking is planned to begin on May 15, 2005
\cite{nuoscind}. 

Among widely discussed now future projects are neutrino factories -- muon
storage rings producing intense beams of high energy neutrinos. 
In addition to high statistics studies of neutrino interactions, 
experiments at neutrino factories should be capable of measuring neutrino 
oscillation parameters with high precision and probing  
the subdominant neutrino oscillation channels, matter effects and CP
violation effects in neutrino oscillations \cite{nufact}.

\section{Phenomenological neutrino mass matrices
\label{phen}}

The information on neutrino masses and lepton mixing obtained in solar
and atmospheric the neutrino experiments can be summarized as follows:
\begin{eqnarray}
{\rm SMA:} \quad\quad &\Delta m_\odot^2\simeq (4 - 10)\cdot 10^{-6}
~{\rm eV}^2\,,\quad\quad &\sin^2 2\theta_\odot\simeq (0.1 - 1.0)\cdot   
10^{-2} \nonumber \\
{\rm LMA:} \quad\quad &\Delta m_\odot^2\simeq (2 - 20)\cdot 10^{-5}
~{\rm eV}^2\,,\quad\quad &\sin^2 2\theta_\odot \simeq 0.65 -
0.97\nonumber\\
{\rm VO:} \quad\quad &~~\Delta m_\odot^2 \simeq (0.5 - 5)\cdot 10^{-10}
~{\rm eV}^2\,,\quad\quad &\sin^2 2\theta_\odot \simeq 0.6 - 1.0 \nonumber
\\
{\rm Atm:} \quad\quad & ~\Delta m_{atm}^2 \simeq (2 - 6)\cdot 10^{-3}
~{\rm eV}^2\,,\quad\quad &\sin^2 2\theta_{atm} \simeq 0.82 - 1.0
\label{constr1} 
\end{eqnarray}
In the 3-flavour framework (neglecting the LSND result), $\Delta m_\odot$
has to be identified with 
$\Delta m_{21}^2$, $\Delta m_{atm}$ with $\Delta m_{31}^2\simeq 
\Delta m_{32}^2$,  $\theta_\odot$ with $\theta_{12}$, and $\theta_{atm}$
with $\theta_{23}$. The 2-flavour analysis gives a good first approximation 
to the results of the 3-flavour studies \cite{Lisi1,Lisi2} because the
mixing angle $\theta_{13}$ is constrained to be rather small by the CHOOZ 
data (\ref{choozres}). 

\begin{figure}[h] 
\setlength{\unitlength}{1cm}
\epsfig{file=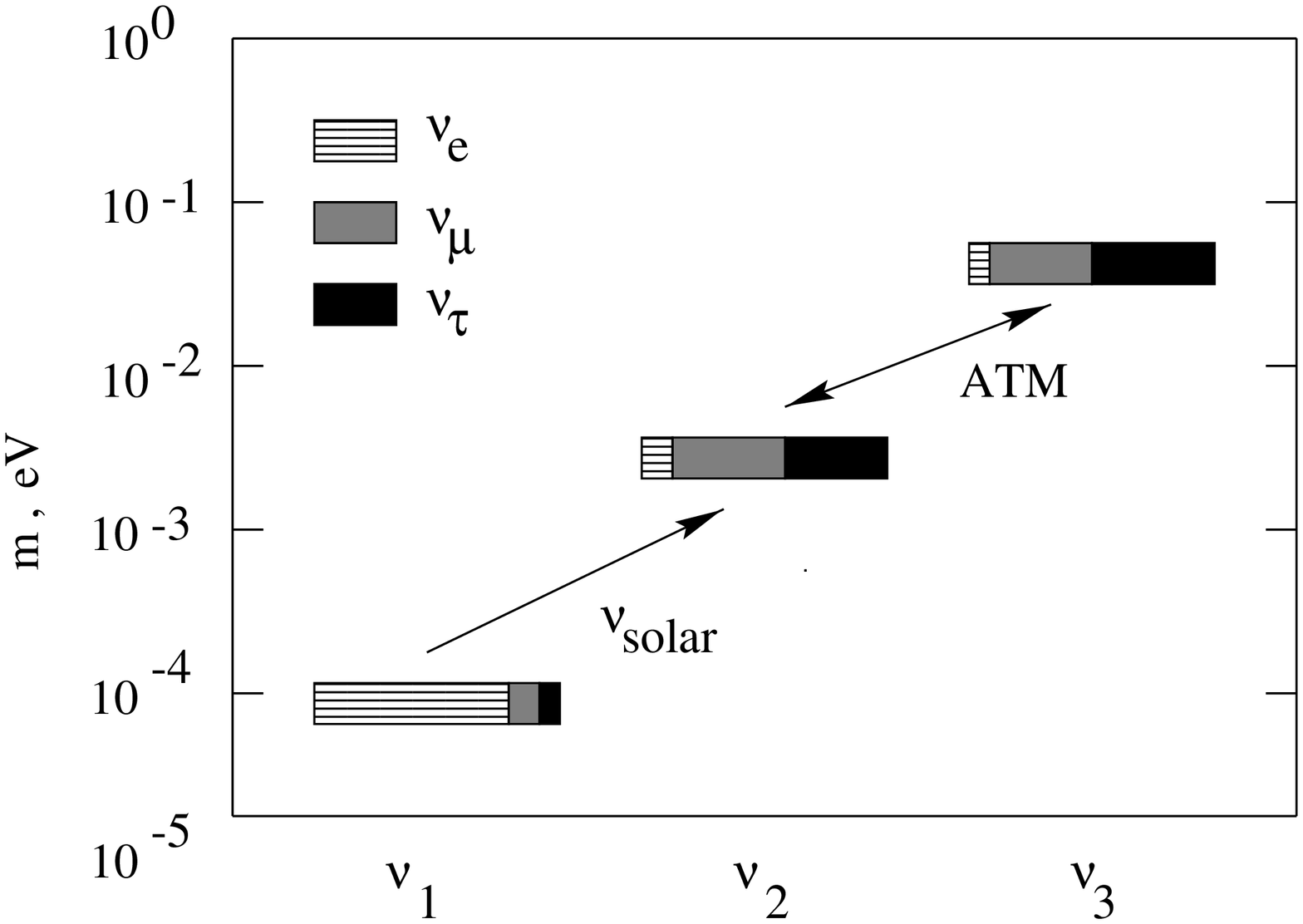,height=5.5cm}
\hspace{1.0cm}
\epsfig{file=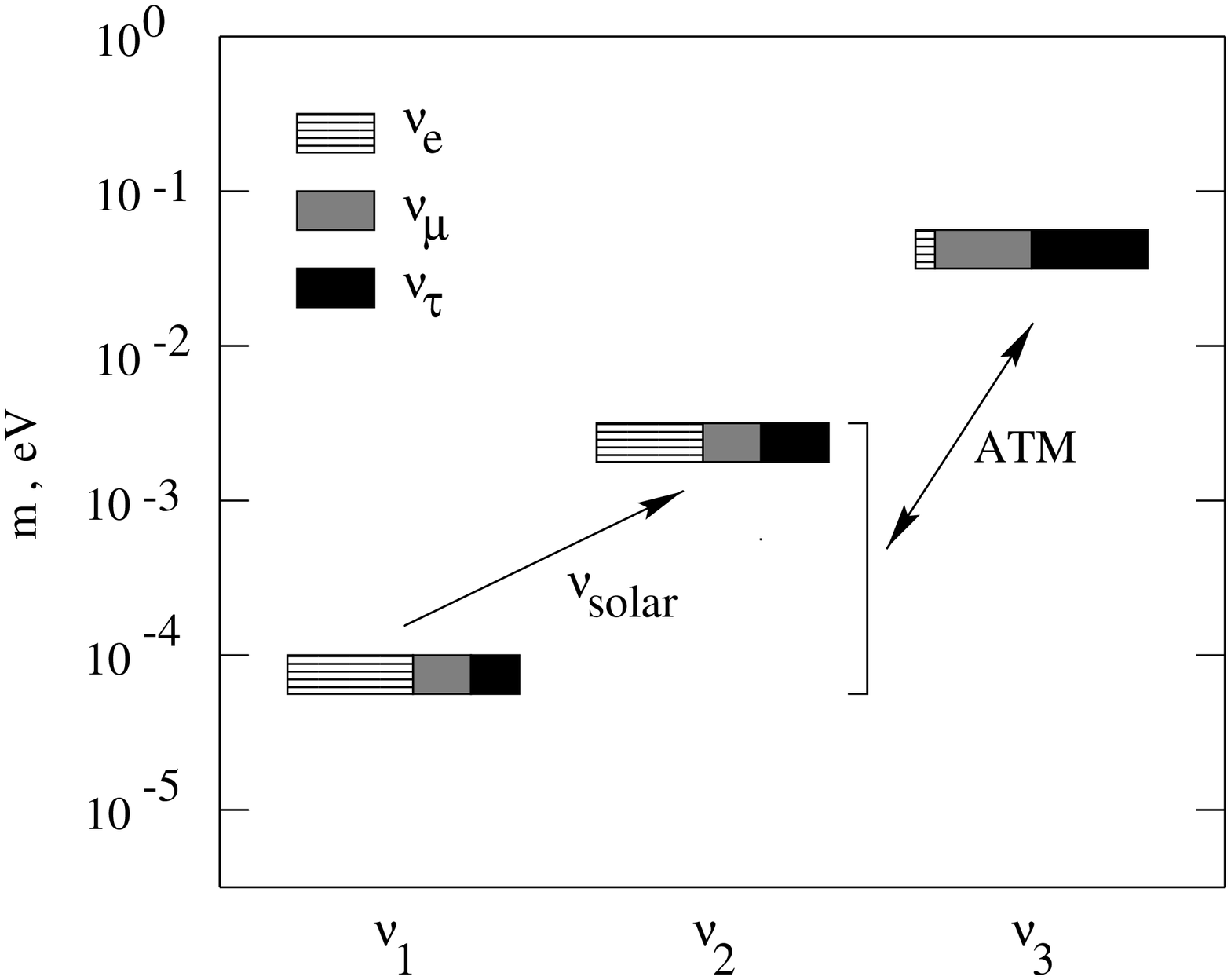,height=5.5cm}
\caption{\small 3-flavour schemes of neutrino masses and mixing. 
Solutions of the solar neutrino problem are SMA (left panel) and LMA right 
panel) \cite{SmMor}}
\label{schemes1} 
\end{figure}

\begin{figure}[h] 
\setlength{\unitlength}{1cm}
\epsfig{file=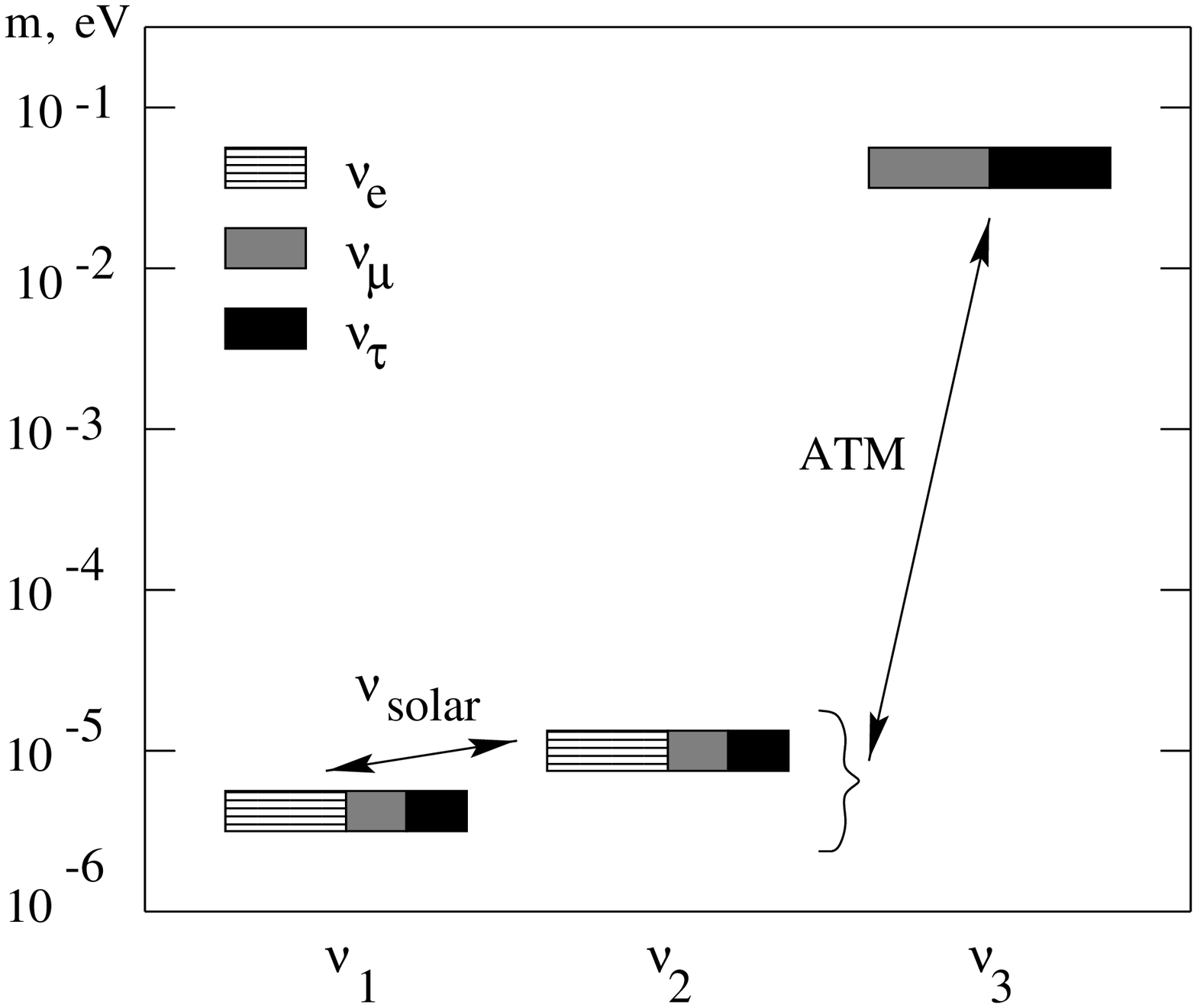,height=5.5cm}
\hspace{0.6cm}
\epsfig{file=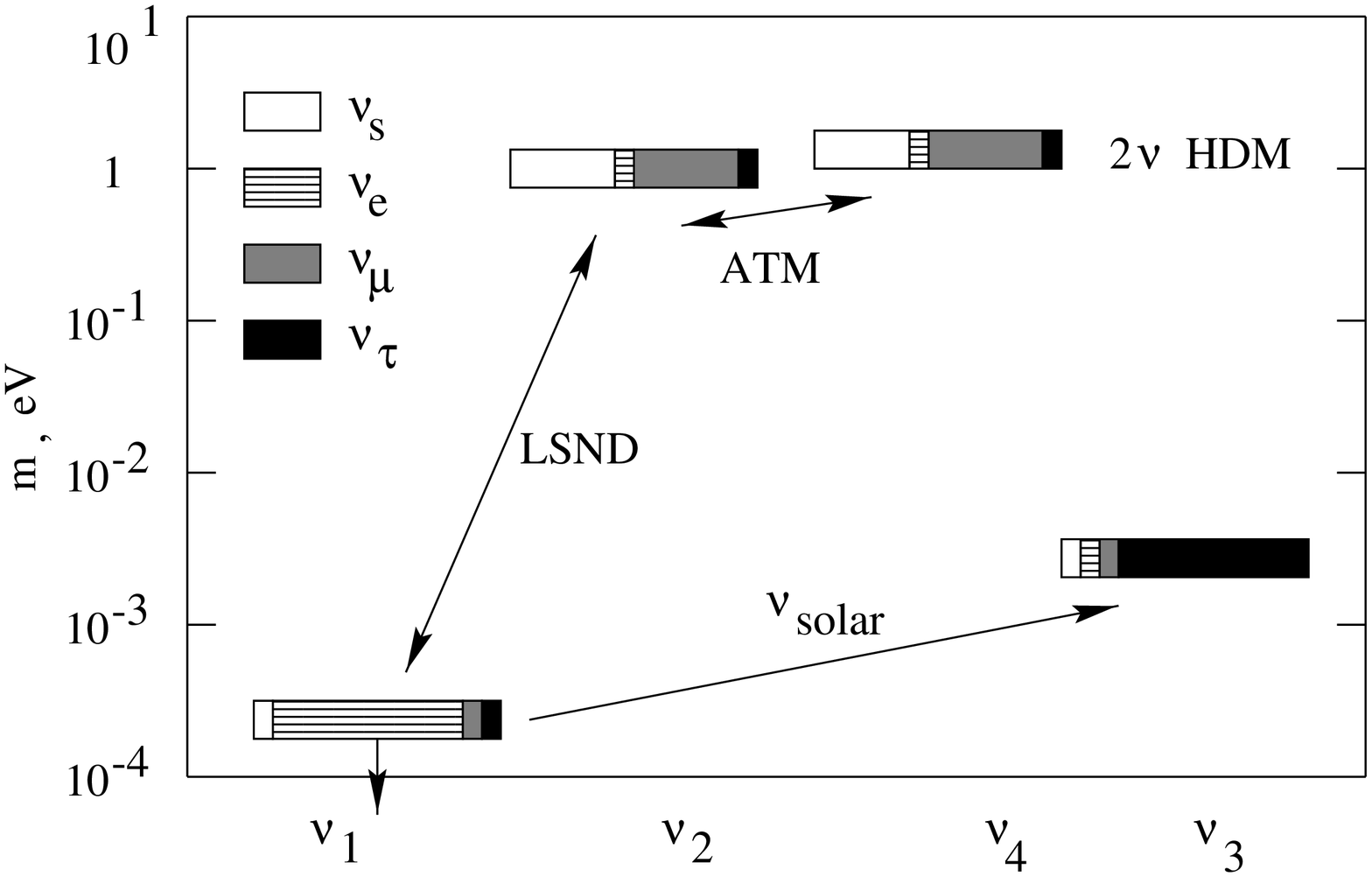,height=5.5cm}
\caption{\small Schemes of neutrino masses and mixing. 
Left panel: 3-flavour scheme with the VO solution of the solar neutrino
problem. Right panel:  SMA solution in the 4-flavour scheme \cite{SmMor}. 
}
\label{schemes2} 
\end{figure}

A very concise and illuminating way of summarizing the neutrino data is to 
present the neutrino mass eigenstates graphically as rectangles of unit area, 
the position of which on the vertical scale reflects their mass whereas
the areas of differently marked parts are equal to $|U_{ai}|^2$, i.e. to
the weights with which the $a$th flavor eigenstate is present in the $i$th 
mass eigenstate \cite{SmMor}. 
In the 3-flavour framework, the present-day data allow three main 
possibilities corresponding to the three main solutions of the solar neutrino 
problem. They are shown in fig. \ref{schemes1} and in fig. \ref{schemes2} 
(left panel). The schemes have different contributions of $\nu_e$, $\nu_\mu$ 
and $\nu_\tau$ in the two low-lying mass eigenstates. The weight of $\nu_e$ 
in the mass eigenstate $\nu_3$ which is separated from $\nu_1$ and $\nu_2$ by 
a large mass gap is $|U_{e3}|^2$; it is either small or zero. The weights of 
$\nu_\mu$ and $\nu_\tau$ within each mass eigenstate are approximately equal 
to each other as a consequence of $\theta_{23}\simeq 45^\circ$ and $|U_{e3}|^2 
=s_{13}^2\ll 1$. This, in particular, means that the solar neutrinos oscillate 
into an almost equal superposition of $\nu_\mu$ and $\nu_\tau$,
independently of the type of the solution of the solar neutrino problem 
(SMA, LMA, LOW or VO).  
The scheme of the left panel of fig. \ref{schemes1}
corresponds to single large mixing, whereas those of the right panel of 
fig. \ref{schemes1} and left panel of fig. \ref{schemes2} describe bi-large 
or bi-maximal (in the case $U_{e3}=0$) mixing. The 3-flavour schemes of 
figs \ref{schemes1} and \ref{schemes2} correspond to the so-called normal  
neutrino mass hierarchy, $m_1 \ll m_2 \ll m_3$. In addition, there are 
schemes (not shown in the figures) in which the position of the mass
eigenstate $\nu_3$ and of the pair ($\nu_1, \nu_2$) are interchanged 
(inverted mass hierarchy,  $m_3 \ll m_1 \simeq m_2$). The present-day data 
do not discriminate between the normal and inverted hierarchies; such a 
discrimination may become possible in future if the earth's matter effects
in atmospheric or long baseline $\nu_e \leftrightarrow \nu_\mu$ or $\nu_e 
\leftrightarrow \nu_\tau$ oscillations are observed.  

The normal and inverted neutrino mass hierarchies discussed above do not 
exhaust all possible schemes of 3-flavour neutrino mass schemes. Since
neutrino oscillation experiments only allow the determination of neutrino
mass squared differences and not of the masses themselves, it is also possible 
that neutrinos are quasi-degenerate in mass, and only their mass squared 
differences satisfy the hierarchies shown in figs. \ref{schemes1} and 
\ref{schemes2}. Direct neutrino mass measurements allow the average neutrino 
mass as large as a few eV (provided that the $2\beta 0\nu$ constraint 
(\ref{meff}) is satisfied). In that case neutrinos could constitute a
significant fraction of the dark matter of the universe (hot dark matter).  

If the LSND result is correct, there should be at least four
light neutrino species. Four-flavour neutrino schemes can be analyzed
similarly to the 3-flavour case. The data allow essentially two schemes. 
In each of these schemes there are two pairs of nearly degenerate mass
eigenstates separated by large $\Delta m_{LSND}^2$. The mass splittings
between the components of the quasi-degenerate pairs are $\Delta m_\odot^2$
and $\Delta m_{atm}^2$. An example of a 4-flavour scheme is shown in 
fig. \ref{schemes2} (right panel).  More detailed discussions of the 4-flavour 
schemes can be found in \cite{4fl}.

\begin{table}
\caption{\small Zeroth order form of the neutrino mass matrix  for double
and single maximal mixing. A~ -- normal mass hierarchy, B -- inverted 
hierarchy, C -- quasi-degenerate neutrinos. From \cite{AF}.}
\small
\begin{center}
\begin{tabular}{|c|c|c|c|}
\hline                         & & double & single \\ & $m_{diag}$ &
maximal &
maximal \\ & & mixing & mixing \\
\hline & & & \\ A & Diag[0,0,1] &
$\left[ 
\matrix{ 0&0&0\cr 0&\um&-\um\cr 0&-\um&\um}
\right]$ &
$\left[
\matrix{ 0&0&0\cr 0&\um&-\um\cr 0&-\um&\um}
\right]$ \\ & & & \\
\hline & & & \\ B1 & Diag[1,-1,0] &
$\left[
\matrix{ 0&\sq&\sq\cr
\sq&0&0\cr
\sq&0&0}
\right]$ &
$\left[   
\matrix{ 1&0&0\cr 0&-\um&-\um\cr 0&-\um&-\um}
\right]$ \\ & & & \\
\hline & & & \\ B2 & Diag[1,1,0] &
$\left[
\matrix{ 1&0&0\cr 0&\um&\um\cr 0&\um&\um}
\right]$ &
$\left[   
\matrix{ 1&0&0\cr 0&\um&\um\cr 0&\um&\um}
\right]$ \\ & & & \\
\hline & & & \\ C0 & Diag[1,1,1] &
$\left[
\matrix{ 1&0&0\cr 0&1&0\cr 0&0&1}
\right]$ &
$\left[
\matrix{ 1&0&0\cr 0&1&0\cr 0&0&1}
\right]$ \\ & & & \\
\hline & & & \\ C1 & Diag[-1,1,1] &
$\left[
\matrix{ 0&-\sq&-\sq\cr -\sq&\um&-\um\cr -\sq&-\um&\um}
\right]$ &
$\left[
\matrix{ -1&0&0\cr 0&1&0\cr 0&0&1}
\right]$ \\ & & & \\
\hline & & & \\ C2 & Diag[1,-1,1] &
$\left[
\matrix{ 0&\sq&\sq\cr
\sq&\um&-\um\cr
\sq&-\um&\um}
\right]$ &
$\left[   
\matrix{ 1&0&0\cr 0&0&-1\cr 0&-1&0}
\right]$ \\ & & & \\
\hline & & & \\ C3 & Diag[1,1,-1] &
$\left[
\matrix{ 1&0&0\cr 0&0&1\cr 0&1&0}
\right]$ &
$\left[
\matrix{ 1&0&0\cr 0&0&1\cr 0&1&0}
\right]$ \\ & & & \\
\hline
\end{tabular}
\end{center}
\label{textures}
\end{table}
The experimental information on the neutrino masses and lepton mixing
angles allows one to reconstruct the phenomenologically allowed forms of the 
neutrino mass matrix. This can be done by inverting the relation $(m_L)_{diag}
=diag(m_1,\,m_2,\,m_3)=U^T m_L\, U$. The detailed structure of the neutrino 
mass matrix $m_L$ is not yet known since some of the neutrino parameters
are still rather uncertain. However, one can derive the zeroth-approximation 
structures (textures) of the neutrino mass matrices just using the already
known gross features of neutrino spectrum and mixing angles. In particular, 
in the zeroth approximation one can assume $m_1=m_2$, $s_{13}=0$, 
$\theta_{23}=45^\circ$. Some of the neutrino mass matrix textures that can
be obtained in this way are summarized in Table \ref{textures} taken from 
\cite{AF}. Notice that these textures are defined only up to the trivial 
sign changes due to the rephasing of the neutrino fields. 

In realistic models, zeros in neutrino mass textures must be filled 
with small elements and in addition the large entries can be perturbed
by small corrections. The small entries of the neutrino mass matrices 
should satisfy certain constraints; one of them follows from the requirement 
that for the LMA and SMA MSW solutions of the solar neutrino problem 
the lower-lying of the mass eigenstates $\nu_1$ and 
$\nu_2$ have a larger $\nu_e$ component. As was discussed in sec. 
\ref{matter}, if this condition is not satisfied, the MSW resonance 
is only possible for antineutrinos and not for neutrinos. This requirement 
on the small entries of the neutrino mass matrices disfavours the inverted 
mass hierarchy and quasi-degenerate neutrinos and favours the normal mass 
hierarchy \cite{Akh4}. 

The neutrino mass matrix textures can provide us with a hint of the symmetries 
or dynamics underlying the theory of neutrino mass. With the forthcoming data 
from future neutrino experiments, it may eventually become possible to unravel 
the mechanism of the neutrino mass generation, which may hold the clue to the 
general fermion mass problem.

\vspace{0.5cm}
We live in a very fascinating time for neutrino physics. It is very
likely that in a few years from now new solar and long baseline neutrino
experiments will bring us an important knowledge about neutrinos allowing 
to answer many questions and solving the solar neutrino problem. 
Neutrinos may also bring us new surprises, as they did many times in the
past. 

\section*{Acknowledgements}
I am grateful to the organizers for a very pleasant atmosphere 
of the school, to Frank Kr\"uger for his help with figures and to 
Orlando Peres for a comment. 
This work was supported by Funda\c{c}\~ao para a Ci\^encia e a Tecnologia
through the grant PRAXIS XXI/BCC/16414/98 and also in part by the TMR
network grant ERBFMRX-CT960090 of the European Union.


\begin{thebibliography}{99}
\bibitem{SK1}  Super-Kamiokande Collaboration, Y. Fukuda {\it et al.}, 
Phys. Rev. Lett. 81 (1998) 1562; ibid. 82 (1999) 2644; ibid. 82 (1999) 
5194; hep-ex/9908049.

\bibitem{SNP} For a recent review see, e.g., Y. Suzuki, Talk at the 
XIX International Symposium on Lepton and Photon Intaractions at High 
Energies, Stanford University, August 914, 1999. 

\bibitem{Mann} For a recent review see, e.g., W.A. Mann, Talk at the 
XIX International Symposium on Lepton and Photon Intaractions at High 
Energies, Stanford University, August 914, 1999; hep-ex/9912007. 

\bibitem{LSND} LSND Collaboration, C. Athanassopoulos {\it et al.}, Phys.
Rev. Lett. 81 (1998) 1774; 
Phys. Rev. C 58 (1998) 2489.

\bibitem{nuoscind} For numerous links to various solar, atmospheric,
reactor, accelerator and other neutrino experiments see the Neutrino
Oscillation Industry Web page maintained by M. Goodman, 
http://www.hep.anl.gov/ndk/hypertext/nu\_industry.html 

\bibitem{MSW} L. Wolfenstein, Phys. Rev. D17 (1978) 2369; S.P. Mikheyev, 
A. Yu. Smirnov, Sov. J. Nucl. Phys. 42 (1985) 913. 

\bibitem{history} For history of neutrino physics see, e.g., P. Ramond,
hep-ph/9809401; S.M. Bilenky, hep-ph/9710251, hep-ph/9908335; W.C. Haxton, 
B.R. Holstein, hep-ph/9905257; M. Nakagawa, hep-ph/9902413. 

\bibitem{Babu} K.S. Babu, lectures at this school, to be published in 
the Proceedings. 

\bibitem{numass} A.Yu. Smirnov, hep-ph/9901208; R.N. Mohapatra, 
hep-ph/9910365; J.W.F. Valle, hep-ph/9907222. 

\bibitem{Walker} T.P. Walker, lectures at this school, to be published in  
the Proceedings. 
\bibitem{raffelt} G.G. Raffelt, {\it Stars as Laboratories for Fundamental
Physics}, Chicago Univ. Press, 1996; hep-ph/9902271; hep-ph/9903472.  
\bibitem{sarkar} S. Sarkar, Rep. Prog. Phys.59 (1996) 1493; hep-ph/9602260. 
\bibitem{raffelt2} G.G. Raffelt, W. Rodejohann, Lectures given at 4th
National Summer School for German-speaking Graduate Students of 
Theoretical Physics, Saalburg, Germany, 31 Aug - 11 Sep 1998, 
hep-ph/9912397. 
\bibitem{BilPe} S.M. Bilenky, S.T. Petcov, Rev. Mod. Phys. 59
(1987) 671; Erratum: ibid. 61 (1989) 169.  

\bibitem{PM} P.B. Pal, R.N. Mohapatra, {\it Massive Neutrinos in 
Physics and Astrophysics}, World Scientific, Singapore, 1991. 

\bibitem{Akh1} E.Kh. Akhmedov, hep-ph/9705451; J. Pulido, Phys. Rep. 211
(1992) 167. 

\bibitem{GR}  G. Gelmini, E. Roulet, Rep. Prog. Phys. 58 (1995) 1207,
hep-ph/9412278. 

\bibitem{BGG} S.M. Bilenky, C. Giunti, W. Grimus, Prog. Part. Nucl. Phys.
43 (1999) 1. 


\bibitem{KP} C.W. Kim, A. Pevsner, {\it Neutrinos in Physics and
Astrophysics}, Harwood Academic, 1993. 


\bibitem{BV} F. Boehm, P. Vogel, {\it Physics of Massive Neutrinos},
Cambridge Univ. Press, 1992.  


\bibitem{PDG} Particle Data Group, C. Caso et al., Eur.
Phys. J. C3 (1998) 1.

\bibitem{Kayser} B. Kayser, Phys. Rev. D30 (1984) 1023. 
 
\bibitem{RS} 
V.A. Rubakov, M.E. Shaposhnikov, Phys. Usp. 39 (1996) 461; hep-ph/9603208. 

\bibitem{Planckscale} S. Weinberg, Phys. Rev. Lett. 43 (1979) 1566; 
R. Barbieri, J. Ellis, M.K. Gaillard, Phys. Lett. B90 (1980) 249;
E. Akhmedov, Z. Berezhiani, G. Senjanovic, Phys. Rev. Lett. 69 (1992) 3013. 

\bibitem{tripl} G.B. Gelmini, M. Roncadelli, Phys. Lett. B99 (1981) 411.  

\bibitem{LR} J.C. Pati, A. Salam, Phys. Rev. D10 (1974) 275; 
R.N. Mohapatra, J.C. Pati, Phys. Rev. D11 (1975) 2558; 
G. Senjanovi\'c, R.N. Mohapatra, Phys. Rev. D12 (1975) 1502. 

\bibitem{seesaw} M. Gell-Mann, P. Ramond and R. Slansky, in 
{\it Supergravity}, ed. by P. van Nieuwenhuizen and D. Z. Freedman (North
Holland, Amsterdam, 1979), p. 315; T. Yanagida, in Proc. of the {\it Workshop 
on the Unified Theory and Baryon Number in the Universe}, ed. by O. Sawada
and A. Sugamoto (KEK report 79-18, 1979), p.95, Tsukuba, Japan; 
R.N. Mohapatra and G. Senjanovi\'c, Phys. Rev. Lett. 44 (1980) 912.  

\bibitem{Pontecorvo} B. Pontecorvo, Sov. Phys. JETP 6 (1958) 429; 
Sov. Phys. JETP 7 (1958) 172.  
 
\bibitem{MNS} Z. Maki, M. Nakagawa, S. Sakata, Progr. Theor. Phys. 28
(1962) 870. 

\bibitem{MS-Uspekhi} S.P. Mikheyev, A.Yu. Smirnov, Sov. Phys. 
Usp. 30 (1987) 759; Prog. Part. Nucl. Phys. 23 (1989) 41. 

\bibitem{SZ} A.Yu. Smirnov, G.T. Zatsepin, Mod. Phys. Lett. A7 (1992) 1272.

\bibitem{Majorosc} S.M. Bilenky, J. Ho\v{s}ek, S.T. Petcov, Phys. Lett. 
B94 (1989) 495; I.Yu. Kobzarev {\it et al.}, Sov. J. Nucl. Phys. 32 (1980) 
823; M. Doi {it et al.}, Phys. Lett. B102 (1981) 323; 
P. Langacker {\it et al.}, Nucl. Phys. B282 (1987) 589.  

\bibitem{KuoPa} For reviews see, e.g., \cite{MS-Uspekhi,BGG} and T.K. Kuo,
J. Pantaleone, Rev. Mod. Phys. 61 (1989) 937. 

\bibitem{SmCec} C. Lunardini, A.Yu. Smirnov, in preparation. 

\bibitem{Giunti} C. Giunti, C.W. Kim, M. Monteno, Nucl. Phys. B521 (1998) 3. 
\bibitem{Akh2} E.Kh. Akhmedov, Nucl. Phys. B538 (1999) 25.  
\bibitem{ADLS} E.Kh. Akhmedov {\it et al.}, Nucl. Phys. B542 (1999) 3. 
\bibitem{Lim} C.-S. Lim, preprint BNL-39675, 1987.
\bibitem{CP} For recent discussion of CP violation in neutrino oscillations 
see, e.g., A. De R\'ujula, M. B. Gavela and P. Hern\'andez, Nucl. Phys. B547 
(1999) 21; 
K. Dick {\it et al.}, Nucl. Phys. B562 (1999) 29-56;
M. Tanimoto, Phys. Lett. B462 (1999) 115;
A. Donini {\it et al.}, hep-ph/9909254; 
M. Koike, J. Sato, hep-ph/9909469. 

\bibitem{decay} Z.G. Berezhiani, M.I. Vysotsky, Phys. Lett. B199
(1987) 281; Z.G. Berezhiani, A.Yu. Smirnov, Phys. Lett. B220 (1989) 279. 

\bibitem{prec} L.B. Okun, M.B. Voloshin, M.I. Vysotsky, Sov. Phys. JETP
64 (1986) 446, Erratum: ibid. 65 (1987) 209;  
C.-S. Lim, W.J. Marciano, Phys. Rev. D37 (1988) 1368; 
E.Kh. Akhmedov, Phys. Lett. B213 (1988) 64. 

\bibitem{ETC} V.K. Ermilova, V.A. Tsarev, V.A. Chechin,
Short Notices of the Lebedev Institute 5 (1986) 26.
\bibitem{Akh3} E. Kh. Akhmedov, Yad. Fiz. 47 (1988) 475.

\bibitem{LS} 
Q.Y. Liu, A. Yu. Smirnov, Nucl. Phys. B524 (1998) 505; Q.Y. Liu, 
S.P. Mikheyev, A. Yu. Smirnov, Phys. Lett. B440 (1998) 319. 

\bibitem{P} S.T. Petcov, Phys. Lett. B434 (1998) 321.

\bibitem{Klapdor} Heidelberg-Moscow Collaboration, L. Baudis {\it et al.}, 
Phys. Rev. Lett. 83 (1999) 41; H.V. Klapdor-Kleingrothaus, Int. J. Mod. Phys. 
13 (1998) 3953.  

\bibitem{Jose1} J. Schechter, J.W.F. Valle, Phys. Rev. D25 (1982) 2951.  

\bibitem{jnb} J.N. Bahcall's home page, http://www.sns.ias.edu/$\sim$jnb/
\bibitem{jnb1} J.N. Bahcall, astro-ph/9808162.
\bibitem{BKS1} J.N. Bahcall, P.I. Krastev, A.Yu. Smirnov, Phys. Rev. D58 
(1998) 096016.
\bibitem{Valencia} M. C. Gonzalez-Garcia, P. C. de Holanda, C. Pe\~na-Garay,   
J. W. F. Valle, hep-ph/9906469.

\bibitem{Totsuka} Y. Totsuka, Invited talk given at the XV Int. Conf. 
Particle and Nuclei ({\it PANIC99}), Uppsala, Sweden, June 10-19, 1999. 
\bibitem{Lisi1} G. Fogli {\it et al.}, hep-ph/9912231;
M.C. Gonzalez-Garcia, C. Pe\~na-Garay, hep-ph/0001129. 
\bibitem{MS} S.P. Mikheyev, A.Yu. Smirnov, Phys. Lett. B429 (1998) 343. 
\bibitem{Scho} K. Scholberg, hep-ex/9905016.
\bibitem{Honda} M. Honda {\it et al.}, Phys. Rev. D52 (1995) 4985. 
\bibitem{Lisi2} G.L. Fogli {\it et al.}, Phys. Rev. D59 (1999) 033001, 
hep-ph/9808205; hep-ph/9904465.  
\bibitem{PerSm} O.L.G. Peres, A.Yu. Smirnov, Phys. Lett. B456 (1999) 204.
\bibitem{Jose2}  M.C. Gonzalez-Garcia {\it et al.}, 
Phys. Rev. Lett. 82 (1999) 3202.
\bibitem{exotic} S. Coleman, S.L. Glashow, Phys. Lett. B405 (1997) 249; 
S.L. Glashow {\it et al.}, Phys. Rev. D56 (1997) 2433; S. Coleman and 
S.L. Glashow, Phys. Rev. D59 (1999) 116008;  M. Gasperini, Phys. Rev. D38 
(1988) 2635; Phys. Rev. D39 (1989) 3606. 
\bibitem{Pakvasa} V. Barger {\it et al.}, Phys. Rev. Lett. 82 (1999) 2640. 
\bibitem{Lisi3} G.L. Fogli {\it et al.},  Phys.Rev. D60 (1999) 053006. 
\bibitem{Barger1} V. Barger {\it et al.}, Phys. Lett. B462 (1999) 109.
\bibitem{gratta} G. Gratta, hep-ex/9905011.
\bibitem{dilella} L. DiLella, hep-ex/9912010. 
\bibitem{CHOOZ} CHOOZ Collaboration, M. Apollonio {\it et al.}, 
hep-ex/9907037. 

\bibitem{ASuz} A. Suzuki, Talk given at 8th International Workshop on 
Neutrino Telescopes, Venice, Italy, February 23 - 26, 1999.  

\bibitem{Conrad} J. Conrad, hep-ex/9811009. 

\bibitem{nufact} For discussions of neutrino oscillation experiments at 
muon storage rings see, e.g., 
A. De R\'ujula, M. B. Gavela and P. Hern\'andez, ref. \cite{CP}; 
M. Campanelli, A. Bueno and A. Rubbia, hep-ph/9905240;
V. Barger, S. Geer and K. Whisnant, hep-ph/9906487;
O. Yasuda, hep-ph/9910428; I. Mocioiu and R. Shrock, hep-ph/9910554;
V. Barger, S. Geer, R. Raja and K. Whisnant, hep-ph/9911524.
M. Freund {\it et al.}, hep-ph/9912457. 

\bibitem{SmMor} A.Yu. Smirnov, Invited talk at 34th Rencontres de Moriond:
Electroweak Interactions and Unified Theories, Les Arcs, France,
March 13-20, 1999;  hep-ph/9907296.  

\bibitem{4fl} For recent discussions of neutrino mass and mixing schemes 
and data analyses in the 4-flavour framework, see e.g. C. Giunti,
hep-ph/9909395; D. Dooling {\it et al.}, hep-ph/9908513; 
C. Giunti, M.C. Gonzalez-Garcia, C. Pe\~na-Garay, hep-ph/0001101. 

\bibitem{AF}  G. Altarelli, F. Feruglio, hep-ph/9905536. 
\bibitem{Akh4} E.Kh. Akhmedov, Phys. Lett. B467 (1999) 95. 


\end{thebibliography}
\end{document}